\documentclass[11pt]{article}
\newcounter{ourcount}
\usepackage{amsmath,amsfonts,amssymb,graphicx,epsfig,pdflscape,multirow,theorem,gensymb}
\usepackage[matrix,arrow,curve]{xy} 
\numberwithin{equation}{section}
\graphicspath{{./eps/}}
\usepackage[nosort]{cite}
\usepackage[usenames]{color}
\usepackage{pstricks}
\usepackage[backref=false]{hyperref}
\usepackage[capitalise]{cleveref}
\hypersetup{
colorlinks=true,
citecolor=red,
linkcolor=darkblue,
urlcolor=darkblue
}
\definecolor{darkblue}{rgb}{0,0,.8}
\definecolor{red}{rgb}{1,0,0}

\theorembodyfont{\itshape} 
\theoremheaderfont{\scshape}
\theoremstyle{plain}

\newtheorem{Proposition}{Proposition}[section]

\numberwithin{equation}{section}

\newcommand{\nc}{\newcommand}
\def\arxiv#1#2{\href{http://arxiv.org/abs/#1}{\textsf{arXiv:#1 #2}}}

\nc{\fh}{\hat{f}}
\nc{\muh}{\hat{\mu}}
\nc{\nuh}{\hat{\nu}}
\nc{\disp}{\displaystyle}
\nc{\cosec}{\mathop{\mbox{cosec}}}
\nc{\ir}{\mathrm{i}}

\nc{\calT}{{\cal Q}}
\nc{\calD}{{\cal Q}}
\nc{\calQ}{{\cal Q}}
\nc{\bib}{\bibitem}
\nc{\al}{\alpha}
\nc{\g}{\gamma}
\nc{\G}{\Gamma}
\nc{\D}{\Delta}
\nc{\eps}{\epsilon}
\nc{\la}{\lambda}
\nc{\La}{\Lambda}
\nc{\var}{\varphi}
\nc{\pa}{\partial}
\nc{\nn}{\nonumber \\ }
\nc{\hf}{\frac{1}{2}}
\nc{\dz}{\frac{dz}{2\pi i}}
\nc{\bin}[2]{\left(\!\!\!\begin{array}{c} {#1}\\ {#2} \end{array}\!\!\!\right)}
\nc{\be}{\begin{equation}}
\nc{\ee}{\end{equation}}
\nc{\bea}{\begin{eqnarray}}
\nc{\eea}{\end{eqnarray}}
\nc{\bra}[1]{\langle {#1}|}
\nc{\ket}[1]{|{#1}\rangle}
\nc{\chit}{\raisebox{0.25ex}{$\chi$}}
\nc{\Dbh}{\mbox{\boldmath $\hat D$}}
\nc{\Dh}{\mbox{$\hat D$}}
\nc{\Dbb}{\mbox{\boldmath $\bar D$}}
\nc{\Dbm}{\mbox{\boldmath $\mathcal Q$}}
\nc{\Dm}{\mbox{$\mathcal Q$}}
\nc{\Dbmh}{\mbox{\boldmath $\mathcal {\hat  Q}$}}
\nc{\Dmh}{\mbox{$\mathcal {\hat  Q}$}}
\nc{\Dbt}{\mbox{\boldmath $\tilde{D}$}}
\nc{\Tbt}{\mbox{\boldmath $\tilde{T}$}}

\nc{\tl}{\mathsf{TL}}
\nc{\eptl}{\mathsf{\mathcal EPTL}}

\nc{\db}{\mbox{\boldmath $d$}}
\nc{\Ab}{\mbox{\boldmath $A$}}
\nc{\Bb}{\mbox{\boldmath $B$}}
\nc{\Cb}{\mbox{\boldmath $C$}}
\nc{\Db}{\mbox{\boldmath $D$}}
\nc{\eb}{\mbox{\boldmath $e$}}
\nc{\Fb}{\mbox{\boldmath $F$}}
\nc{\Fbt}{\mbox{\boldmath $\tilde{F}$}}
\nc{\fb}{\mbox{\boldmath $f$}}
\nc{\fbt}{\mbox{\boldmath $\tilde{f}$}}
\nc{\Gb}{\mbox{\boldmath $G$}}
\nc{\Hb}{\mbox{\boldmath $H$}}
\nc{\Jb}{\mbox{\boldmath $J$}}
\nc{\Kb}{\mbox{\boldmath $K$}}
\nc{\Lb}{\mbox{\boldmath $L$}}
\nc{\Mb}{\mbox{\boldmath $M$}}
\nc{\Pb}{\mbox{\boldmath $P$}}
\nc{\Qb}{\mbox{\boldmath $Q$}}
\nc{\Rb}{\mbox{\boldmath $R$}}
\nc{\Tb}{\mbox{\boldmath $T$}}
\nc{\Tbb}{\mbox{\boldmath $\bar T$}}
\nc{\Tbm}{\mbox{\boldmath $\mathcal T$}}
\nc{\tb}{\mbox{\boldmath $t$}}
\nc{\Ub}{\mbox{\boldmath $U$}}
\nc{\Vb}{\mbox{\boldmath $V$}}
\nc{\Wb}{\mbox{\boldmath $W$}}
\nc{\Xb}{\mbox{\boldmath $X$}}
\nc{\yb}{\mbox{\boldmath $y$}}
\nc{\Zb}{\mbox{\boldmath $Z$}}
\nc{\Hc}{{\cal H}}
\nc{\Rc}{{\cal R}}
\nc{\Lc}{{\cal L}}
\nc{\Vc}{{\cal V}}
\nc{\Ib}{\mbox{\boldmath $I$}}
\nc{\qb}{\bar{q}}
\nc{\oN}{\mathbb{N}}
\nc{\oZ}{\mathbb{Z}}
\nc{\oR}{\mathbb{R}}
\newrgbcolor{darkgreen}{0., 0.733333, 0.0621042}
\def\vvdots{\mathinner{\mkern1mu\raise1pt\vbox{\kern7pt\hbox{.}}\mkern2mu
  \raise4pt\hbox{.}\mkern2mu\raise7pt\hbox{.}\mkern1mu}}
\nc{\gauss}[2]{\left[\!\!\begin{array}{c} {#1}\\ {#2} \end{array}\!\!\right]}
\nc{\sbin}[2]{\left\{\!\!\!\begin{array}{c} {#1}\\ {#2} 
\end{array}\!\!\!\right\}}
\nc{\sbinlr}[2]{\Big\langle\!\!\begin{array}{c} {#1}\\ {#2} 
\end{array}\!\!\Big\rangle}
\nc{\bino}[2]{\left(\!\!\begin{array}{c} {#1}\\ {#2} \end{array}\!\!\right)}
\def\half {\mbox{$\textstyle \frac{1}{2}$}}
\def\vec#1{\mbox {\boldmath $#1$}}
\def\svec#1{\mbox {\scriptsize\boldmath $#1$}}
\definecolor{lightblue}{rgb}{.7,.7,1}
\definecolor{lightestblue}{rgb}{.95,.95,1}
\definecolor{lightlightblue}{rgb}{.85,.85,1}
\definecolor{midblue}{rgb}{.7,.7,1}

\def\loopa{
\psframe[linewidth=.25pt](0,0)(1,1)
\psarc[linewidth=1.5pt,linecolor=blue](1,0){.5}{90}{180}
\psarc[linewidth=1.5pt,linecolor=blue](0,1){.5}{-90}{0}
}
\def\loopb{
\psframe[linewidth=.25pt](0,0)(1,1)
\psarc[linewidth=1.5pt,linecolor=blue](0,0){.5}{0}{90}
\psarc[linewidth=1.5pt,linecolor=blue](1,1){.5}{180}{270}
}

\def\facegrid#1#2{
\psframe[fillstyle=solid,fillcolor=lightlightblue,linewidth=0pt]#1#2
\psgrid[gridlabels=0pt,subgriddiv=1]#1#2}

\newcommand{\statei}{{\ket{-\frac{3}{32}}}\ar@{};[0,0];}
\newcommand{\stateii}{{\ket{\frac{5}{32}}}\ar@{};[0,0];}
\newcommand{\stateiii}{{\ket{\frac{21}{32}}}\ar@{};[0,0];}
\newcommand{\stateiv}{{\ket{\frac{45}{32}}}\ar@{};[0,0];}
\newcommand{\statev}{{\ket{\frac{77}{32}}}\ar@{};[0,0];}
\newcommand{\statevi}{{\ket{\frac{117}{32}}}\ar@{};[0,0];}
\newcommand{\stated}{{\dots}\ar@{};[0,0];}
\newcommand{\statee}{{}\ar@{};[0,0];}

\renewcommand{\ge}{\geqslant}
\renewcommand{\le}{\leqslant}
\nc{\eE}{\mathsf{e}}

\nc\drtm{{\vec D}}      		
\nc\face{\mathbb{X}} 		
\nc\faceK{\mathbb{K}} 		
\nc\genface[1]{\mathbb{X}^{(#1)}} 
\nc \ham{{\mathcal H}}			

\nc \nface{\mathbb{\hat X}}	

\nc\TLw{\omega} 			
\nc\TLb{\beta} 				

\nc\BMWw{\omega_2}		
\nc\BMWb{\beta_2}			

\nc\fTLw{\omega_2} 			
\nc\fTLb{\beta_2} 			

\nc{\genw}[1]{\omega_{#1}}  	
\nc{\Genw}[2]{\omega_{#1}^{(#2)}}
\nc{\genb}[1]{\beta_{#1}}       	

\def \trinomial[#1][#2][#3][#4]{\left[{#1\atop #2,#3,#4}\right]}

\def \superTrinomial[#1][#2]{
\left({#1 \atop #2} \right)_{\!2}
}

\def \qTrinomial[#1][#2][#3][#4]{\left[{#1\atop #2,#3,#4}\right]_{\!q}}

\nc{\XiShift}{\overline{\xi}_\rho}

\nc{\fws}{\small}

\def\dddots{\mathinner{\mkern1mu\raise9pt\vbox{\kern7pt\hbox{.}}\mkern2mu
  \raise5pt\hbox{.}\mkern2mu\raise1pt\hbox{.}\mkern1mu}}

\nc{\smbin}[2]{\Big(\!\!\!\begin{array}{c} {#1}\\[-3pt] {#2} \end{array}\!\!\!\Big)}
  
\begin{document}

\topmargin -5mm
\oddsidemargin 5mm

\vspace*{-2cm}

\setcounter{page}{1}

\vspace{22mm}
\begin{center}
{\huge {\bf Extended $\boldsymbol T$-systems, $\boldsymbol Q$ matrices and\\[0.15cm] $\boldsymbol T$-$\boldsymbol Q$ relations for $\boldsymbol{s\ell(2)}$ models at roots of unity}}

\vspace{14mm}
{\Large Holger Frahm$^\dagger$, Alexi Morin-Duchesne$^\ddagger$, Paul A. Pearce$^\ast$}
\\[.4cm]
{\em {}$^\dagger$Institut f\"ur Theoretische Physik, Leibniz Universit\"at Hannover}\\
{\em Appelstra\ss e 2, 30167 Hannover, Germany}
\\[.4cm]
{\em {}$^\ddagger$Universit\'e catholique de Louvain, Institut de Recherche en Math\'ematique et Physique}\\
{\em Chemin du Cyclotron 2, 1348 Louvain-la-Neuve, Belgium}
\\[.4cm]
{\em {}$^\ast$School of Mathematics and Statistics, University of Melbourne}\\
{\em Parkville, Victoria 3010, Australia}
\\[.4cm]
{\tt frahm\,@\,itp.uni-hannover.de,\quad \tt alexi.morin-duchesne\,@\,uclouvain.be,\quad \tt papearce\,@\,unimelb.edu.au}
\end{center}


\vspace{8mm}
\centerline{{\bf{Abstract}}}
\vskip.4cm
\noindent 
The mutually commuting $1\times n$ fused single and double-row transfer matrices of the critical six-vertex model are considered at roots of unity $q=\eE^{\ir\lambda}$ with crossing parameter $\lambda=\frac{(p'-p)\pi}{p'}$ a rational fraction of $\pi$. 
The $1\times n$ transfer matrices of the dense loop model analogs, namely the logarithmic minimal models ${\cal LM}(p,p')$, are similarly considered. 
For these $s\ell(2)$ models, we find explicit closure relations for the
$T$-system functional equations and obtain extended sets of bilinear
$T$-system identities. We also define  extended $Q$ matrices as linear
combinations of the fused transfer matrices and obtain extended matrix $T$-$Q$
relations.
These results hold for diagonal 
twisted boundary conditions on the cylinder as well as {$U_q(s\ell(2))$ invariant/Kac vacuum} and off-diagonal/Robin vacuum boundary conditions on the strip.
Using our extended $T$-system and extended $T$-$Q$ relations for eigenvalues, we deduce the usual scalar Baxter $T$-$Q$ relation and the Bazhanov-Lukyanov-Zamolodchikov decomposition of the fused transfer matrices $T^{n}(u+\lambda)$ and $D^{n}(u+\lambda)$, at fusion level $n=p'-1$, in terms of the product $Q^+(u)Q^-(u)$ or $Q(u)^2$. It follows that the zeros of $T^{p'-1}(u+\lambda)$  and $D^{p'-1}(u+\lambda)$ are comprised of the Bethe roots and complete $p'$ strings. 
We also clarify the formal observations of Pronko and Yang-Nepomechie-Zhang and establish, under favourable conditions, the existence of an infinite fusion limit $n\to\infty$ in the auxiliary space of the fused transfer matrices. 
Despite this connection, the infinite-dimensional oscillator representations are not needed at roots of unity due to finite closure of the functional equations.

\vspace{.5cm}
\noindent\textbf{Keywords:} Exactly solvable models, vertex models, dense loop models\\

\newpage
\tableofcontents

\newpage
\hyphenpenalty=30000

\setcounter{footnote}{0}

\section{Introduction}

The six-vertex model~\cite{Pauling,Lieb,Sutherland,LiebWu} is a Yang-Baxter integrable model~\cite{BaxterBook} of a ferroelectric on the square lattice. 
The critical six-vertex model coincides with the critical line of Baxter's symmetric eight-vertex model~\cite{BaxterQ,Baxter73,BaxterBook}. The six- and eight-vertex models are among the most studied models in integrable two-dimensional lattice statistics. Integrability stems from the Yang-Baxter equation and the existence of a one-parameter family of commuting transfer matrices as functions of the spectral parameter $u$. The logarithmic derivative of these transfer matrices at $u=0$ yields~\cite{Sutherland1970,Baxter73} the Hamiltonians of the XXZ and XYZ quantum spin chains. The XYZ quantum spin chain and its specializations to the XXZ and XXX (or Heisenberg) models are central in the study of integrable one-dimensional quantum systems. 

A number of important algebraic and analytic tools have been developed to study the six-vertex, eight-vertex and related loop and RSOS models. These include the Bethe Ansatz, Functional Equations, Quantum Inverse Scattering Methods (QISM) and $T$- and $Y$-systems. The Bethe Ansatz~\cite{Bethe1931} extends to various forms such as the coordinate Bethe ansatz, functional Bethe ansatz (including separation of variables) and the algebraic Bethe ansatz. The Functional Equation approach was initiated by Baxter~\cite{BaxterQ} and includes Baxter's $T$-$Q$ relation and the fusion hierarchies of functional equations satisfied by the fused transfer matrices. The Quantum Inverse Scattering Method~\cite{QISM} has its origins in the inverse problem methods~\cite{ClassInvMeth} for solving classical nonlinear equations. Lastly, the $T$- and $Y$-system methods~\cite{Zamolodchikov,KlumperP,KNS} are a generalization~\cite{WiegmannEtAl} of Hirota's bilinear equations~\cite{Hirota} and extend to the analysis of the associated Thermodynamic Bethe Ansatz (TBA) and Non-Linear Integral Equations (NLIE).

In this paper, we are interested in critical $s\ell(2)$ vertex and dense loop models for which the quantum group parameter $q=\eE^{\ir\lambda}$ is a root of unity, characterized by rational values of the crossing parameter $\lambda$:
\be
\lambda=\lambda_{p,p'}=\frac{(p'-p)\pi}{p'},\qquad 1\le p<p,\qquad \mbox{$p,p'$ coprime}.
\label{rational}
\ee 
The generic $s\ell(2)$ dense loop model, restricted to roots of unity, coincides with the logarithmic minimal models ${\cal LM}(p,p')$~\cite{PRZ2006,MDPR}. At roots of unity on the critical line, these systems exhibit an extended $s\ell(2)$ loop algebra symmetry~\cite{loopAlg} not present at generic points. The extra symmetry at the rational points accounts for additional degeneracies between the eigenvalues and allows for the occurrence of nontrivial Jordan blocks in the transfer matrices. It is therefore apparent that, at roots of unity, these critical systems are described, in the continuum scaling limit, by logarithmic Conformal Field Theories (CFTs)~\cite{specialissue}. Each rational point on the critical line is described by a different logarithmic CFT. 
Our focus is on the interplay between the two approaches to these roots-of-unity models based on $T$-systems and $T$-$Q$ relations.
Since the early work, there is an extensive literature on these topics so let us just cite a selective but far from exhaustive list~\cite{AlcarazEtAl87,Sklyanin,MezNepomechie1992,BLZ97,Pronko2000,RossiWeston2002,Nepomechie2002,Doikou2002,Caoetal03,Nepo04,deGierPyatov2004,FabMcCoy2004,Korff2005,YNZ2006,YZ2006,Roan2006,BazhMang2007,BLMS2010,Niccoli2010,BFLMS2011,FGSW2011,KazakovTsuboi2012,HaoNepomechieS2013,HaoNepomechieS2014,NepomechieWang2014,GHNS2015,GainutdinovNepomechie2016,YupengBook}. 
Perhaps the most extensive overview of these topics is to be found in the Bazhanov-Mangazeev article~\cite{BazhMang2007}. 

The goal of this paper is to propose a general and systematic framework extending the key precepts of $T$-systems, $Q$ matrices and $T$-$Q$ relations to derive the  Bethe ansatz equations of $s\ell(2)$ models at roots of unity. More specifically, we (i) derive an extended set of $T$-system bilinear equations, 
(ii)~define extended $Q$ matrices, as explicit linear combinations of fused transfer matrices, and deduce that they satisfy extended $T$-$Q$ relations and (iii) show that, for eigenvalues, the extended $T$-$Q$ relations imply the usual Baxter $T$-$Q$ relations, Bethe ansatz equations and Bazhanov-Lukyanov-Zamolodchikov
decompositions \cite{BLZ97,BazhMang2007}. This program is carried out to completion for (i) diagonal twist boundary conditions on the cylinder, (ii) $U_q(s\ell(2))$ invariant/Kac vacuum boundary conditions on the strip and (iii) off-diagonal/Robin vacuum boundary conditions on the strip. The treatment holds simultaneously for the six-vertex and dense loop models. 

The layout of the paper is as follows. In \cref{sec:the.models},
we review the definitions of the fundamental transfer matrices of the six-vertex and dense loop models for twisted boundary conditions on the cylinder and open boundary conditions on the strip. 
We also define separately the fused transfer matrices for the related $s\ell(2)$ vertex and loop models. These transfer matrices satisfy the same functional equations as a consequence of the common underlying Temperley-Lieb algebras elaborated upon in \cref{app:diag.algebras}. These common structures enable a uniform treatment of both models. 
In \cref{sec:periodic,sec:strip,sec:Robin}, we present our results as listed above for the $s\ell(2)$ models at roots of unity under the three 
respective boundary conditions. The proofs of the extended $T$-systems of bilinear functional equations are relegated to \cref{AppA}. 
\cref{sec:Robin}, on off-diagonal/Robin boundaries, is special since our final analysis splits into two cases depending on whether the extended $Q$ matrices are analytic or non-analytic functions of the spectral parameter $u$. Concluding remarks are given in \cref{sec:conclusion}. 

\section{Fused transfer matrices of the six-vertex and dense loop models}
\label{sec:the.models}

\subsection[$R$-matrices and transfer matrices of the six-vertex model]{$\boldsymbol R$-matrices and transfer matrices of the six-vertex model}
\label{sec:vertex.model}

In this section, we briefly review the six-vertex model on
the square lattice and its commuting transfer matrices for
various boundary conditions.  In the natural basis $\{++,+-,-+,--\}$, the Boltzmann weights of the local
vertex configurations can be arranged in an operator valued $R$-matrix
$R:\mathbb{C}\to\mathrm{End}(\mathbb{C}^2\otimes \mathbb{C}^2)$:
\begin{equation}
  \label{rmatrix}
  R(u) = \left(
    \begin{array}{cccc}
      \frac{\sin(\lambda-u)}{\sin\lambda}&0&0&0 \\
      0& \frac{\sin u}{\sin\lambda} & g^{-1}&0\\
      0& g & \frac{\sin u}{\sin\lambda}&0\\
      0&0&0& \frac{\sin(\lambda-u)}{\sin\lambda}
    \end{array}\right),\quad 
    \check R(u) = P_{12} R(u)=\left(
    \begin{array}{cccc}
      \frac{\sin(\lambda-u)}{\sin\lambda}&0&0&0 \\
      0& g &\frac{\sin u}{\sin\lambda} &0\\
      0& \frac{\sin u}{\sin\lambda} & g^{-1}&0\\
      0&0&0& \frac{\sin(\lambda-u)}{\sin\lambda}
    \end{array}\right),
\end{equation}
where $P_{12}$ permutes the order of the two copies of $\mathbb{C}^2$.
The more common convention with weights $\sin(\lambda+u)$ instead of $\sin(\lambda-u)$
is obtained by the simple involution $\lambda \leftrightarrow \pi-\lambda$.
The gauge factor $g$ is arbitrary. 
Choosing $g=z=\eE^{\ir u}$ allows us to express $\check R_{j,j+1}(u)$ as
\begin{equation}
  \check R_{j,j+1}(u) = \frac{\sin(\lambda-u)}{\sin\lambda}\,I +
  \frac{\sin u}{\sin\lambda}\, e_j
\end{equation}
in terms of the generators of the Temperley-Lieb algebras of
\cref{app:diag.algebras} with $\beta=2\cos\lambda$. In the following, however, we set $g=1$.

The $R$-matrix satisfies the Yang-Baxter equation
\begin{equation}
  \label{ybe}
  R_{12}(u-v)\,R_{13}(u)\,R_{23}(v) = R_{23}(v)\,R_{13}(u)\,R_{12}(u-v)\,
\end{equation}
on the space $V_1\otimes V_2\otimes V_3$, $V_j\simeq \mathbb{C}^2$,
$R_{jk}\in\mathrm{End}(V_j\otimes V_k)$. The monodromy matrix, defined as
\begin{equation}
  \label{monodromy}
  \stackrel{a}{\mathcal{T}}(u)
  = \ir^N\,R_{a1}(u+\zeta_1)\,R_{a2}(u+\zeta_2)\,\cdots
  \,R_{aN}(u+\zeta_N)
  = \left(
    \begin{array}{cc}
      A(u) & B(u) \\ C(u) & D(u)
    \end{array} \right)\,,
\end{equation}
can be represented as a matrix in the auxiliary space $V_a$ with operator
valued entries $A$, $B$, $C$, $D$ acting on the quantum space $V^{\otimes N}$.
The normalization factor $\ir^N$ is unimportant and is chosen for later
notational convenience.  As a consequence of the Yang-Baxter equation
\eqref{ybe} the monodromy matrix is a representation of the Yang-Baxter
algebra:
\begin{equation}
  \label{yba}
  R_{12}(u-v) \stackrel{1}{\mathcal{T}}(u)  \stackrel{2}{\mathcal{T}}(v)
  =  \stackrel{2}{\mathcal{T}}(v)  \stackrel{1}{\mathcal{T}}(u) R_{12}(u-v).
\end{equation}
The single-row transfer matrix of the inhomogeneous six-vertex
model on the cylinder with periodic boundary conditions,
\begin{equation}
  \label{transfer6v}
  \boldsymbol{T}(u) = 
  \mathrm{tr}_a \stackrel{a}{\mathcal{T}}(u)\,,
\end{equation}
generates a family of commuting operators on the quantum space.  
For the homogeneous model with $\zeta_j= 0$, this family includes the
Hamiltonian of the spin-$\frac12$ XXZ quantum chain.

Integrability is preserved for certain twisted boundary conditions:
the $c$-number 
twist matrices
\begin{equation}
  \label{twist6v}
  \mathbf{\Theta}=\left(
    \begin{array}{cc}
      \omega & 0 \\ 0 & \omega^{-1}\end{array}\right)
\end{equation}
are representations of the Yang-Baxter algebra \eqref{yba} with
one-dimensional quantum space \cite{Vega84}.  Therefore, the commutativity of
the transfer matrices obtained from
${\mathcal{T}}_\Theta(u) = \mathbf{\Theta}\, {\mathcal{T}}(u)$ is retained.
The resulting transfer matrices commute with the total magnetization $S^z = \frac12 \sum_{j=1}^{N} \sigma_j^z$.

The introduction of $K$-matrices~\cite{Cherednik} and commuting double-row
transfer matrices~\cite{Sklyanin} describing integrable boundary conditions
for the six-vertex model on the strip is by now standard.  A boundary
condition on the strip encoded in a $K$-matrix is integrable if the right- and
left-reflection or boundary Yang-Baxter equations~\cite{Cherednik} are
satisfied:
\begin{subequations} \label{reflalg}
\begin{align}
  \label{reflalgR}
  R_{12}(u-v) \stackrel{1}{K}_R\!(u)\,R_{21}(u+v) \stackrel{2}{K}_R\!(v)
  &=\,\stackrel{2}{K}_R\!(v)\, R_{12}(u+v) \stackrel{1}{K}_R\!(u)\, R_{21}(u-v),\\
   \label{reflalgL}
  R_{12}(v-u) \stackrel{1}{K}{\!}^{\textrm{t}_1}_L\!(u)\, R_{21}(2\lambda-u-v) \stackrel{2}{K}{\!}^{\textrm{t}_2}_L\!(v)
  &=\, \stackrel{2}{K}{\!}^{\textrm{t}_2}_L\!(v)\, R_{12}(2\lambda-u-v) \stackrel{1}{K}{\!}^{\textrm{t}_1}_L\!(u)\, R_{21}(v-u).
\end{align}
\end{subequations}
For diagonal $U_q(s\ell(2))$ invariant boundary conditions, the $K$-matrices
are \cite{KuSk91}
\begin{equation}
\label{eq:Ksl2q}
  K_R 
  (u) = 
    \begin{pmatrix}
      \mathsf{e}^{\ir u} & 0 \\ 0 & \mathsf{e}^{-\ir u}\end{pmatrix},\qquad
      K_L
      (u) =  \begin{pmatrix}
      \mathsf{e}^{\ir (\lambda-u)} & 0 \\ 0 & \mathsf{e}^{-\ir (\lambda-u)}\end{pmatrix}.
\end{equation}
The more general three-parameter solutions for the off-diagonal $K$-matrices are given in \cite{VeGo93,GhZa94}. Here we parameterize the solution for $K_R(u)$ in terms of {three variables $\nu_1, \nu_2$ and $\xi$:}
\be
K_R(u) = 
\begin{pmatrix}
\eE^{\ir \xi}\nu_1 \frac{\sin(\lambda + \xi + u)}{\sin\lambda} + \eE^{-\ir \xi}\nu_1^{-1} \frac{\sin(\lambda + \xi - u)}{\sin\lambda}
& 
\nu_2 \frac{\sin (2u)}{\sin \lambda}
\\[0.1cm]
\nu_2^{-1}\frac{\sin (2u)}{\sin \lambda}
& 
\eE^{\ir \xi}\nu_1 \frac{\sin(\lambda + \xi - u)}{\sin\lambda} + \eE^{-\ir \xi}\nu_1^{-1} \frac{\sin(\lambda + \xi + u)}{\sin\lambda}
\end{pmatrix}.
\ee
Likewise, the solution for $K_L(u)$ is parameterized in terms of {$\mu_1, \mu_2$ and $\eta$:}
\be
K_L(u) = 
\begin{pmatrix}
\eE^{\ir \eta}\mu_1 \frac{\sin(2\lambda + \eta - u)}{\sin\lambda} + \eE^{-\ir \eta}\mu_1^{-1} \frac{\sin(\eta +u)}{\sin\lambda}
& 
\mu_2^{-1} \frac{\sin (2\lambda-2u)}{\sin \lambda}
\\[0.1cm]
\mu_2 \frac{\sin (2\lambda-2u)}{\sin \lambda}
& 
\eE^{\ir \eta}\mu_1 \frac{\sin(\eta +u)}{\sin\lambda} + \eE^{-\ir \eta}\mu_1^{-1} \frac{\sin(2\lambda + \eta - u)}{\sin\lambda}
\end{pmatrix}.
\ee
We note that $K_L(u) = K^{\textrm t}_R(\lambda-u)|_{\nu\to\mu,\,\xi\to\eta}$. The $U_q(s\ell(2))$ invariant 
 boundary conditions are recovered in the limits $\xi,\eta\to\ir \infty$:
\begin{subequations}\label{eq:Gamma.6V}
\begin{alignat}{2}
\lim_{\xi \to \ir \infty} \frac{K_R(u)}{\Gamma_R(u)} &=  
\begin{pmatrix}\mathsf{e}^{\ir u} & 0 \\ 0 & \mathsf{e}^{-\ir u}\end{pmatrix}, 
\quad 
&&\Gamma_R(u) = \frac{\sin(\lambda+\xi-u)}{\sin{\lambda}}(\eE^{\ir(u+\xi)}\nu_1+\eE^{-\ir(u+\xi)}\nu_1^{-1}),\\
\lim_{\eta \to \ir \infty} \frac{K_L(u)}{\Gamma_L(\lambda-u)} &=
\begin{pmatrix}\mathsf{e}^{\ir (\lambda-u)} & 0 \\ 0 & \mathsf{e}^{-\ir (\lambda-u)}\end{pmatrix}, 
\quad
&&\Gamma_L(u) = \frac{\sin(\lambda+\eta-u)}{\sin{\lambda}}(\eE^{\ir(u+\eta)}\mu_1+\eE^{-\ir(u+\eta)}\mu_1^{-1}).
\end{alignat}
\end{subequations}
The double-row transfer matrix on the strip is then
\begin{equation}
  \boldsymbol{D}(u) = \mathrm{tr}_a \Big(\!
  \stackrel{a}{K}_L\!(u) \stackrel{a}{\mathcal{T}}(u)
    \stackrel{a}{K}_R\!(u)  \stackrel{a}{\mathcal{T}}\!{'}(u)\Big)
\end{equation}
where
\be 
\stackrel{a}{\mathcal{T}}\!{'}(u) = 
  \ir^{-N}\,R_{aN}(u-\zeta_N)\,\cdots\,R_{a2}(u-\zeta_2)\,R_{a1}(u-\zeta_1)\,.
\ee

\subsection{Fused transfer matrices of the six-vertex model}
\label{app:fusedVertex}

Using the fusion procedure introduced in \cite{KRS81}, $R$-matrices acting on
spaces $\mathbb{C}^{n+1}\otimes \mathbb{C}^{m+1}$ can be constructed starting
from \eqref{rmatrix}.  This construction relies on the fact that the
elementary $R$-matrix degenerates for a particular choice of the spectral
parameter. Here $R(\lambda)$ is a projector onto a one-dimensional subspace of
$\mathbb{C}^2\otimes\mathbb{C}^2$.  Explicit expressions for the $m\times n$
fused vertex weights for the six-vertex model have been given, for example, in
\cite{KirillovResh87}.  For $m>1$, these allow for the construction of
integrable higher spin generalizations of the XXZ spin chain
\cite{ZaFa80,Sogo84}. {Here we consider the $1\times n$ fused transfer
matrices $\boldsymbol{T}^n(u)$ on the cylinder
and double-row transfer matrices $\boldsymbol{D}^n(u)$ on the strip for the six-vertex
model. These are}
associated with a spin-$\tfrac{n}{2}$ representation of $s\ell(2)$ in the
auxiliary space.  Fused transfer matrices are thus labelled by nodes of a
Dynkin diagram which is one-sided $A_\infty$. 

The $1\times n$ fused transfer matrix for the vertex model on the
cylinder is obtained by taking the trace of a product of $n$ monodromy
matrices ${\mathcal{T}}_{\mathbf{\Theta}}(u)$, with $\mathbf{\Theta}$ corresponding to the
twist \eqref{twist6v}, acting on different auxiliary spaces
$V_j$ over the $U_q(s\ell(2))$ spin-$\tfrac{n}{2}$ component of the product
$V_1\otimes V_2\otimes\cdots\otimes V_n$:
\begin{equation}
  \label{transfer6vfused}
  \Tb^{n}(u) = \Bigg(\prod_{j=1}^{n-1}\frac1{f_{j-1}}\Bigg)\,\,
  \mathrm{tr}_{1\cdots n} \Big(
        \boldsymbol{P}_{1\cdots n}\!\! 
        \stackrel{1\cdots n}{\mathcal{T}}_{\!\!\!\mathbf{\Theta}}(u)
    \Big)
\end{equation}
where
\be
	\stackrel{1\cdots n}{\mathcal{T}}_{\!\!\!\mathbf{\Theta}}(u)\, =\,\,
	\stackrel{1}{\mathcal{T}}_{\mathbf{\Theta}}(u)\,
	\stackrel{2}{\mathcal{T}}_{\mathbf{\Theta}}(u+\lambda)\,\cdots
	\stackrel{n}{\mathcal{T}}_{\mathbf{\Theta}}(u+(n-1)\lambda).
\ee
For the six-vertex model defined in terms of the $R$-matrix \eqref{rmatrix} with $g=1$,
the deformed projectors $\boldsymbol{P}_{1\cdots n}$ are defined recursively
as
\begin{equation*}
  \boldsymbol{P}_{1\cdots n} =
  \boldsymbol{P}_{1\cdots(n-1)} - \frac{n-1}{n} \,
  \boldsymbol{P}_{1\cdots(n-1)}\,R_{n-1,n}(\lambda)\,\boldsymbol{P}_{1\cdots
    (n-1)}\,, \qquad
  \boldsymbol{P}_{1} = 
    \begin{pmatrix}
      1 & 0 \\ 0 & 1\end{pmatrix}\,.
\end{equation*}
The fused transfer matrices \eqref{transfer6vfused} have been normalized by
\begin{equation}
  f(u) = \prod_{j=1}^N \frac{\sin(u+\zeta_j)}{\sin \lambda}\,,
  \qquad f_k = f(u+k\lambda)\,.
\end{equation}
With this normalization, the matrix entries of $\boldsymbol{T}^{n}(u)$ are Laurent
polynomials in $z=\mathsf{e}^{\ir u}$ and thus are entire functions of
the spectral parameter $u$.  
The minimal and maximal powers of $z$ are $\pm N$.  
The transfer matrices are mutually
commuting,
\begin{equation}
  \left[ \boldsymbol{T}^{m}(u), \boldsymbol{T}^{n}(v) \right] =0,
\end{equation}
and satisfy the periodicity relations
\begin{equation}
  \boldsymbol{T}^{n}(u+\pi) =(-1)^N \boldsymbol{T}^{n}(u).
\end{equation}

Similarly, the $1\times n$ fused transfer matrices
$\boldsymbol{D}^n(u)$ can be constructed for the
six-vertex model on the strip as in \cite{MezNepomechie1992,Zhou96b}. {These are written in terms of the following two products of monodromy matrices: 
\begin{subequations}
\begin{alignat}{2}
	\stackrel{1\cdots n}{\mathcal{T}}\!(u)\, &=\,\,
	\stackrel{1}{\mathcal{T}}(u)
	\stackrel{2}{\mathcal{T}}(u+\lambda)\cdots
	\stackrel{n}{\mathcal{T}}(u+(n-1)\lambda),
	\\
	\stackrel{1\cdots n}{\mathcal{T}}\!{'}(u)\, &=\,\,
	\stackrel{1}{\mathcal{T}}\!{'}(u)
	\stackrel{2}{\mathcal{T}}\!{'}(u+\lambda)\cdots
	\stackrel{n}{\mathcal{T}}\!{'}(u+(n-1)\lambda).
\end{alignat}
\end{subequations}
The fused $K$ matrices are given by
\begin{subequations}
\begin{alignat}{2}
\label{eq:fusedKR}
	\stackrel{1\cdots n}{K}_{\!\!\!R}\!(u)\, &= 
	\bigg(\prod_{1\le i<j\le n}\frac{(-1)}{s_{i+j-3}}\bigg) 
	\prod_{i=1}^n\bigg[ \stackrel{i}{K}_{\!R}\!\big(u+(i-1)\lambda\big)
	\prod_{j=i+1}^n R_{ji}\big(2u+(i+j-2)\lambda\big)\bigg],
\\
  \label{eq:fusedKL}
  \stackrel{1\cdots n}{K}_{\!\!\!L}\!(u)\,&=
  \bigg(\prod_{1\le i<j\le n}\frac{1}{s_{i+j-3}}\bigg)\!\!\!
  \prod_{\substack{i=n\\[0.1cm] \textrm{step}=-1}}^1\!\!\!\!\bigg[
  \stackrel{i}{K}_{\!L}\! \big(u+(i-1)\lambda\big)\!\!
  \prod_{\substack{j=i-1\\[0.1cm] \textrm{step}=-1}}^1 \!\!R_{ji}\big(\!-2u-(i+j-4)\lambda\big)\bigg].
\end{alignat}
\end{subequations}
The fused double-row transfer matrix at level $n$ is then defined as
\be
\Db^n(u) = \Bigg(\prod_{j=1}^{n-1}\frac1{f_{j-1}}\Bigg)\,\,
  \mathrm{tr}_{1\cdots n} \Big(
        \boldsymbol{P}_{1\cdots n}\!\! 
        \stackrel{1\cdots n}{K}_{\!\!\!L}\!(u)\!
        \stackrel{1\cdots n}{\mathcal{T}}\!(u)\!
        \stackrel{1\cdots n}{K}_{\!\!\!R}\!(u)\!
        \stackrel{1\cdots n}{\mathcal{T}}\!{'}(u)
    \Big)
\ee
where 
\be
f(u) = \prod_{i=1}^N \frac{\sin (u + \zeta_i)\sin (u - \zeta_i)}{\sin^2 \lambda}, \qquad f_k = f(u+k \lambda).
\ee
With this definition, the matrix entries of $\Db^{n}(u)$ are Laurent polynomials in $z=\mathsf{e}^{\ir u}$ 
with minimal and maximal powers $\pm 2N$ for the diagonal quantum group invariant
  boundary conditions \eqref{eq:Ksl2q} and $\pm (2N+4n)$ for the
  generic $K$-matrices. The transfer matrices are mutually
commuting,
\begin{equation}
  \left[ \Db^{m}(u), \Db^{n}(v) \right] =0,
\end{equation}
and satisfy the periodicity and crossing relations
\begin{equation}
  \Db^{n}(u+\pi) = \Db^{n}(u), \qquad \Db^n(u) = \Db^n\big((2-n)\lambda-u\big).
\end{equation}
}

\subsection{Face operators and transfer tangles of the dense loop model}
\label{sec:loop.model}

In this section, we review the definitions of the dense loop model on the square lattice. This model is defined for generic values of the crossing parameter $\lambda$ but, when specialized to the roots of unity values $\lambda=\lambda_{p,p'}$ \eqref{rational} of interest here, it coincides with the logarithmic minimal model ${\cal LM}(p,p')$~\cite{PRZ2006}. 

We give the construction of the dense loop transfer tangles with twisted boundary conditions on the cylinder and Kac/Robin vacuum boundary conditions on the strip. These tangles are elements of the diagrammatic Temperley-Lieb algebras $\tl_N(\beta)$, $\eptl_N(\alpha,\beta)$ and $\tl^{\textrm{\tiny$(2)$}}_N(\beta, \alpha_1, \alpha_2, \beta_1, \beta_2, \gamma)$ defined in \cref{app:diag.algebras}. In later sections, we use the terms {\it transfer matrix} and {\it $Q$ matrix} in both the vertex and loop settings with the understanding that, in the loop setting, these objects should be interpreted as tangles in the corresponding diagrammatic algebras. 

The elementary face operator for the dense loop model is defined as
\be
\psset{unit=.8cm}
\begin{pspicture}[shift=-.42](1,1)
\facegrid{(0,0)}{(1,1)}
\psarc[linewidth=0.025]{-}(0,0){0.16}{0}{90}
\rput(.5,.5){$u$}
\end{pspicture}
\ \ = \frac{\sin (\lambda-u)}{\sin \lambda}\ \ \begin{pspicture}[shift=-.45](1,1)
\facegrid{(0,0)}{(1,1)}
\rput[bl](0,0){\loopa}
\end{pspicture}
\ \ +\frac{\sin u}{\sin \lambda}\ \
\begin{pspicture}[shift=-.42](1,1)
\facegrid{(0,0)}{(1,1)}
\rput[bl](0,0){\loopb}
\end{pspicture}
\ee
where $u$ is the spectral parameter and $\lambda$ is the crossing parameter, related to the fugacity of bulk loops via the relation $\beta = 2 \cos \lambda$. 
The elementary face operators satisfy the Yang-Baxter equation which ensures integrability. For periodic boundary conditions, the single-row transfer tangle is an element of $\eptl_N(\alpha,\beta)$ defined by
\be
\Tb (u)= \ir^{N}\ 
\psset{unit=0.9}
\begin{pspicture}[shift=-0.4](-0.2,0)(5.2,1.0)
\facegrid{(0,0)}{(5,1)}
\psarc[linewidth=0.025]{-}(0,0){0.16}{0}{90}
\psarc[linewidth=0.025]{-}(1,0){0.16}{0}{90}
\psarc[linewidth=0.025]{-}(4,0){0.16}{0}{90}
\psline[linewidth=1.5pt,linecolor=blue]{-}(0,0.5)(-0.2,0.5)
\psline[linewidth=1.5pt,linecolor=blue]{-}(5,0.5)(5.2,0.5)
\rput(2.5,0.5){$\ldots$}
\rput(3.5,0.5){$\ldots$}
\rput(0.5,.5){$_{u+\zeta_1}$}
\rput(1.5,.5){$_{u+\zeta_2}$}
\rput(4.5,.5){$_{u+\zeta_N}$}
\end{pspicture}\ \ ,
\label{eq:Tu}
\ee
where the $\zeta_i$ are bulk inhomogeneity parameters. The periodic transfer tangles satisfy the commutativity and periodicity relations
\be
[\Tb(u), \Tb(v)] = 0, \qquad \Tb(u+\pi) = (-1)^N \Tb(u).
\ee
The normalization factor $\ir^N$ in \eqref{eq:Tu} is unimportant and is chosen for later notational convenience.

{For integrability in the presence of boundaries, the left and right boundaries are decorated~\cite{BPO} with triangle face operators that must satisfy the boundary Yang-Baxter equation.}
For Kac vacuum boundary conditions on the strip, the double-row transfer tangle \cite{PRZ2006} is defined, as an element of $\tl_N(\beta)$, by 
\be
\Db(u) = \ \
\psset{unit=0.9cm}
\begin{pspicture}[shift=-0.9](-1,0)(5.5,2)
\facegrid{(0,0)}{(5,2)}
\rput(2.5,0.5){$\ldots$}
\rput(2.5,1.5){$\ldots$}
\rput(3.5,0.5){$\ldots$}
\rput(3.5,1.5){$\ldots$}
\pspolygon[fillstyle=solid,fillcolor=lightlightblue](5,1)(6,2)(6,0)(5,1)
\pspolygon[fillstyle=solid,fillcolor=lightlightblue](0,1)(-1,2)(-1,0)(0,1)
\rput(0,0){\rput(0.5,0.5){$_{u+\zeta_1}$}\psarc[linewidth=0.025]{-}(0,0){0.16}{0}{90}}
\rput(1,0){\rput(0.5,0.5){$_{u+\zeta_2}$}\psarc[linewidth=0.025]{-}(0,0){0.16}{0}{90}}
\rput(4,0){\rput(0.5,0.5){$_{u+\zeta_N}$}\psarc[linewidth=0.025]{-}(0,0){0.16}{0}{90}}
\rput(0,1){\rput(0.5,0.5){$_{u-\zeta_1}$}\psarc[linewidth=0.025]{-}(1,0){0.16}{90}{180}}
\rput(1,1){\rput(0.5,0.5){$_{u-\zeta_2}$}\psarc[linewidth=0.025]{-}(1,0){0.16}{90}{180}}
\rput(4,1){\rput(0.5,0.5){$_{u-\zeta_N}$}\psarc[linewidth=0.025]{-}(1,0){0.16}{90}{180}}
\psarc[linecolor=blue,linewidth=1.5pt](0,1){0.5}{90}{-90}
\psarc[linecolor=blue,linewidth=1.5pt](5,1){0.5}{-90}{90}
\end{pspicture} \qquad .
\ee
In this case, the boundary conditions take the form of simple arcs.

For Robin vacuum boundary conditions on the strip, loop segments may connect to the boundary 
and the transfer tangle $\vec D(u)$ is
an element of $\tl^{\textrm{\tiny$(2)$}}_N(\beta, \alpha_1, \alpha_2, \beta_1, \beta_2, \gamma)$. 
Here $\beta$ is the fugacity of bulk loops and, as discussed in \cref{app:diag.algebras}, $\alpha_1, \alpha_2, \beta_1, \beta_2$ are fugacities of boundary loops and $\gamma$ is the fugacity of loop segments spanning between the left and right edges of the strip. 
The boundary face operators in this case are defined as
\be\label{eq:b.face}
\psset{unit=.8cm}
\begin{pspicture}[shift=-0.9](-1,0)(0,2)
\pspolygon[fillstyle=solid,fillcolor=lightlightblue](0,1)(-1,2)(-1,0)(0,1)
\rput(-0.65,1){$_{u,\eta}$}
\end{pspicture}
\,=\Gamma_L(u) \
\begin{pspicture}[shift=-0.89](-1,0)(0,2)
\pspolygon[fillstyle=solid,fillcolor=lightlightblue](0,1)(-1,2)(-1,0)(0,1)
\psarc[linewidth=1.5pt,linecolor=blue](0,1){.7}{135}{-135}
\end{pspicture}
\ +\frac{\sin 2u}{\sin \lambda} \
\begin{pspicture}[shift=-0.89](-1,0)(0,2)
\pspolygon[fillstyle=solid,fillcolor=lightlightblue](0,1)(-1,2)(-1,0)(0,1)
\psline[linecolor=blue,linewidth=1.5pt]{-}(-0.4,0.6)(-1,0.6)
\psline[linecolor=blue,linewidth=1.5pt]{-}(-0.4,1.4)(-1,1.4)
\end{pspicture}\ \ , 
\qquad \ \ 
\begin{pspicture}[shift=-0.9](0,0)(1,2)
\pspolygon[fillstyle=solid,fillcolor=lightlightblue](0,1)(1,2)(1,0)(0,1)
\rput(0.65,1){$_{u,\xi}$}
\end{pspicture}
\,=\Gamma_R(u) \
\begin{pspicture}[shift=-0.89](0,0)(1,2)
\pspolygon[fillstyle=solid,fillcolor=lightlightblue](0,1)(1,2)(1,0)(0,1)
\psarc[linewidth=1.5pt,linecolor=blue](0,1){.7}{-45}{45}
\end{pspicture}
\ +\frac{\sin 2u}{\sin \lambda} \
\begin{pspicture}[shift=-0.89](0,0)(1,2)
\pspolygon[fillstyle=solid,fillcolor=lightlightblue](0,1)(1,2)(1,0)(0,1)
\psline[linecolor=blue,linewidth=1.5pt]{-}(0.4,0.6)(1,0.6)
\psline[linecolor=blue,linewidth=1.5pt]{-}(0.4,1.4)(1,1.4)
\end{pspicture}\ \ ,
\ee
where the functions $\Gamma_L(u)$ and $\Gamma_R(u)$ are given by \cite{PRT14}:
\begin{subequations}\label{eq:Gamma.loops}
\begin{alignat}{2}
\Gamma_L(u) &= \frac{\sin(\lambda + \eta - u)}{\sin \lambda} \Big(\alpha_1  \frac{\sin(\lambda + \eta + u)}{\sin \lambda} - \alpha_2 \frac{\sin(\eta+u)}{\sin \lambda}\Big),\\[0.15cm]
\Gamma_R(u) &= \frac{\sin(\lambda + \xi - u)}{\sin \lambda} \Big(\beta_1 \frac{\sin(\lambda + \xi + u)}{\sin \lambda} - \beta_2 \frac{\sin(\xi+u)}{\sin \lambda}\Big).
\end{alignat}
\end{subequations}
Here, $\eta$ and $\xi$ are left and right boundary fields respectively. We note that the right boundary triangle is obtained from the left boundary triangle by performing a rotation of $180^\circ$ and substituting $(\beta_1, \beta_2, \xi)$ for $(\alpha_1, \alpha_2, \eta)$. 
The double-row transfer tangle with Robin boundary conditions is then defined diagrammatically by 
\be
\Db(u) = \ \
\psset{unit=1.1cm}
\begin{pspicture}[shift=-0.9](-1,0)(6,2)
\facegrid{(0,0)}{(5,2)}
\rput(2.5,0.5){$\ldots$}
\rput(2.5,1.5){$\ldots$}
\rput(3.5,0.5){$\ldots$}
\rput(3.5,1.5){$\ldots$}
\rput(0,0){\rput(0.5,0.5){$_{u+\zeta_1}$}\psarc[linewidth=0.025]{-}(0,0){0.16}{0}{90}}
\rput(1,0){\rput(0.5,0.5){$_{u+\zeta_2}$}\psarc[linewidth=0.025]{-}(0,0){0.16}{0}{90}}
\rput(4,0){\rput(0.5,0.5){$_{u+\zeta_N}$}\psarc[linewidth=0.025]{-}(0,0){0.16}{0}{90}}
\rput(0,1){\rput(0.5,0.5){$_{u-\zeta_1}$}\psarc[linewidth=0.025]{-}(1,0){0.16}{90}{180}}
\rput(1,1){\rput(0.5,0.5){$_{u-\zeta_2}$}\psarc[linewidth=0.025]{-}(1,0){0.16}{90}{180}}
\rput(4,1){\rput(0.5,0.5){$_{u-\zeta_N}$}\psarc[linewidth=0.025]{-}(1,0){0.16}{90}{180}}
\pspolygon[fillstyle=solid,fillcolor=lightlightblue](5,1)(6,2)(6,0)(5,1)\rput(5.6,1){$_{u,\xi}$}
\pspolygon[fillstyle=solid,fillcolor=lightlightblue](0,1)(-1,2)(-1,0)(0,1)\rput(-0.55,1){$_{\lambda-u,\eta}$}
\psline[linecolor=blue,linewidth=1.5pt](0,0.5)(-0.5,0.5)\psline[linecolor=blue,linewidth=1.5pt](0,1.5)(-0.5,1.5)
\psline[linecolor=blue,linewidth=1.5pt](5,0.5)(5.5,0.5)\psline[linecolor=blue,linewidth=1.5pt](5,1.5)(5.5,1.5)
\end{pspicture} \ \ .
\ee

The transfer tangles for Kac/Robin vacuum boundary conditions commute, have period $\pi$ and satisfy crossing symmetry:
\be
[\Db(u), \Db(v)] = 0, \qquad \Db(u+\pi) = \Db(u), \qquad \Db(\lambda - u) = \Db(u).
\ee
The Kac vacuum boundary triangles are in fact obtained as a simple limit of the Robin boundary triangles
\be
\psset{unit=.8cm} 
\lim_{\eta \to \pm \ir \infty}\frac1{\Gamma_L(u)} \ \ 
\begin{pspicture}[shift=-0.9](-1,0)(0,2)
\pspolygon[fillstyle=solid,fillcolor=lightlightblue](0,1)(-1,2)(-1,0)(0,1)
\rput(-0.60,1){$_{u,\eta}$}
\end{pspicture} \ \ = \ \ 
\begin{pspicture}[shift=-0.89](-1,0)(0,2)
\pspolygon[fillstyle=solid,fillcolor=lightlightblue](0,1)(-1,2)(-1,0)(0,1)
\psarc[linewidth=1.5pt,linecolor=blue](0,1){.7}{135}{-135}
\end{pspicture}
\ \ ,
\qquad
\lim_{\xi \to \pm \ir \infty}\frac1{\Gamma_R(u)} \ \
\begin{pspicture}[shift=-0.9](0,0)(1,2)
\pspolygon[fillstyle=solid,fillcolor=lightlightblue](0,1)(1,2)(1,0)(0,1)
\rput(0.65,1){$_{u,\xi}$}
\end{pspicture} \ \ = \ \ 
\begin{pspicture}[shift=-0.89](0,0)(1,2)
\pspolygon[fillstyle=solid,fillcolor=lightlightblue](0,1)(1,2)(1,0)(0,1)
\psarc[linewidth=1.5pt,linecolor=blue](0,1){.7}{-45}{45}
\end{pspicture}\ \ .\label{triLimit}
\ee
This observation justifies our notation $\Db(u)$ used, here and in the sequel, for both the Kac and Robin transfer tangles. It will be clear from context which boundary condition is being discussed.
This observation also implies that {many of the results of \cref{sec:strip} for the $U_q(s\ell(2))$ invariant/Kac vacuum boundary conditions} can be recovered from the off-diagonal/Robin results of \cref{sec:Robin} by taking the limit $\eta, \xi \to \pm \ir \infty$. Notwithstanding, we include \cref{sec:strip} to introduce our approach in the simpler and more familiar context of $U_q(s\ell(2))$ invariant/Kac boundary conditions without all of the additional complications of the off-diagonal/Robin boundary conditions in \cref{sec:Robin}.

\subsection{Fused transfer tangles of the dense loop model}\label{sec:fused.loops}
\label{app:fused.loops}

In this section, we define the fused face operators and the fused transfer tangles $\Tb^n(u)$ and $\Db^n(u)$ of the dense loop models. In this notation, the fundamental transfer tangles are given by $\Tb(u) = \Tb^1(u)$ and $\Db(u) = \Db^1(u)$. 

The $m\times n$ fused face operator is diagrammatically defined as
\be
\psset{unit=.9cm} 
\begin{pspicture}[shift=-.4](1,1)
\label{eq:fusedface}
\facegrid{(0,0)}{(1,1)}
\psarc[linewidth=0.025]{-}(0,0){0.16}{0}{90}
\rput(.5,.45){$u$}
\rput(0.5,0.7){\tiny{$_{(m,n)}$}}
\end{pspicture} \ \  = \prod_{i=1}^{m-1}\prod_{j=1}^{n-1}\frac{1}{s_{j-i}(u)}\ \
\begin{pspicture}[shift=-2.6](-0.3,-0.3)(5.3,5.3)
\pspolygon[fillstyle=solid,fillcolor=pink](0.1,0)(4.9,0)(4.9,-0.3)(0.1,-0.3)(0.1,0)
\pspolygon[fillstyle=solid,fillcolor=pink](0,0.1)(0,4.9)(-0.3,4.9)(-0.3,0.1)(0,0.1)
\pspolygon[fillstyle=solid,fillcolor=pink](5,0.1)(5,4.9)(5.3,4.9)(5.3,0.1)(5,0.1)
\pspolygon[fillstyle=solid,fillcolor=pink](0.1,5)(4.9,5)(4.9,5.3)(0.1,5.3)(0.1,5)
\rput(2.5,-0.15){$_m$}
\rput(2.5,5.15){$_m$}
\rput(-0.15,2.5){$_n$}
\rput(5.15,2.5){$_n$}
\facegrid{(0,0)}{(5,5)}
\psarc[linewidth=0.025]{-}(0,0){0.16}{0}{90}
\psarc[linewidth=0.025]{-}(3,0){0.16}{0}{90}
\psarc[linewidth=0.025]{-}(4,0){0.16}{0}{90}
\psarc[linewidth=0.025]{-}(0,1){0.16}{0}{90}
\psarc[linewidth=0.025]{-}(3,1){0.16}{0}{90}
\psarc[linewidth=0.025]{-}(4,1){0.16}{0}{90}
\psarc[linewidth=0.025]{-}(0,4){0.16}{0}{90}
\psarc[linewidth=0.025]{-}(3,4){0.16}{0}{90}
\psarc[linewidth=0.025]{-}(4,4){0.16}{0}{90}
\rput(4.5,.5){\small $u_0$}
\rput(4.5,1.5){\small $u_1$}
\rput(4.5,2.6){\small $\vdots$}
\rput(4.5,3.6){\small $\vdots$}
\rput(4.5,4.5){\small $u_{n\!-\!1}$}
\rput(3.5,.5){\small $u_{-\!1}$}
\rput(3.5,1.5){\small $u_0$}
\rput(3.5,2.6){\small $\vdots$}
\rput(3.5,3.6){\small $\vdots$}
\rput(3.5,4.5){\small $u_{n\!-\!2}$}
\rput(.5,.5){\small $u_{1\!-\!m}$}
\rput(.5,1.5){\small $u_{2\!-\!m}$}
\rput(.5,2.6){\small $\vdots$}
\rput(.5,3.6){\small $\vdots$}
\rput(.5,4.5){\small $u_{n\!-\!m}$}
\multiput(0,0)(1,0){2}{\rput(1.5,.5){\small $\ldots$}}
\multiput(0,1)(1,0){2}{\rput(1.5,.5){\small $\ldots$}}
\multiput(0,4)(1,0){2}{\rput(1.5,.5){\small $\ldots$}}
\end{pspicture} \ \ , \qquad u_k = u + k \lambda.
\ee
It is a Laurent polynomial in $z = \eE^{\ir u}$ with minimal and maximal powers $\pm 1$. Here we have used the conventions of \cite{MDPR} for depicting the 
Wenzl-Jones projectors as pink rectangles.
{The construction above holds for the generic cases, where all the projectors are well defined. We refer to \cite{MDPR,rsBoundaries1,PRTsuper2014} for discussions of the construction of fusion in the roots of unity cases where some projectors are not well defined.}

For twisted periodic boundary conditions on the cylinder, the $1\times n$ fused transfer tangle is an element of $\eptl_N(\alpha,\beta)$ defined diagrammatically by\footnote{In this paper, we only consider fused transfer tangles with fusion in the vertical direction. We do not consider fused transfer tangles involving fusion in the horizontal direction. In the notation of \cite{MDPR}, our transfer tangles $\Tb^{n}(u)$ and $\Db^{n}(u)$ are denoted by $\Tb^{1,n}(u)$ and $\Db^{1,n}(u)$ respectively.}
\be 
\Tb^n(u)= \ir^{Nn}\ \  
\psset{unit=1.03}
\begin{pspicture}[shift=-0.4](-0.2,0)(5.2,1)
\psline[linewidth=4pt,linecolor=blue]{-}(-0.2,0.5)(0,0.5)\psline[linewidth=2pt,linecolor=white]{-}(-0.2,0.5)(0,0.5)
\psline[linewidth=4pt,linecolor=blue]{-}(5.0,0.5)(5.2,0.5)\psline[linewidth=2pt,linecolor=white]{-}(5.0,0.5)(5.2,0.5)
\facegrid{(0,0)}{(5,1)}
\psarc[linewidth=0.020]{-}(0,0){0.13}{0}{90}
\psarc[linewidth=0.020]{-}(1,0){0.13}{0}{90}
\psarc[linewidth=0.020]{-}(4,0){0.13}{0}{90}
\rput(0.5,0.75){\tiny{$_{(1,n)}$}}
\rput(1.5,0.75){\tiny{$_{(1,n)}$}}
\rput(4.5,0.75){\tiny{$_{(1,n)}$}}
\rput(2.5,0.5){$\ldots$}
\rput(3.5,0.5){$\ldots$}
\rput(0.5,.5){$_{u-\zeta_1}$}
\rput(1.5,.5){$_{u-\zeta_2}$}
\rput(4.5,.5){$_{u-\zeta_N}$}
\end{pspicture}\ \ ,
\ee  
where the double strand indicates $n$ {\it cabled\/} loop segments. It is a Laurent polynomial in $z = \eE^{\ir u}$ of minimal and maximal degree $\pm N$. It satisfes the commutativity and periodicity relations
\be
[\Tb^m(u),\Tb^n(v)]=0, \qquad \Tb^n(u+\pi) = (-1)^N \Tb(u).
\ee

For Kac vacuum boundary conditions on the strip, the $1\times n$ fused transfer tangle is an element of $\tl_N(\beta)$ defined diagrammatically by
\be 
\Db^n(u)= \ \  
\psset{unit=1.3}
\begin{pspicture}[shift=-0.9](-1,0)(6,2)
\pspolygon[fillstyle=solid,fillcolor=lightlightblue](5,1)(6,2)(6,0)(5,1)
\pspolygon[fillstyle=solid,fillcolor=lightlightblue](0,1)(-1,2)(-1,0)(0,1)
\psarc[linewidth=1pt,linecolor=blue]{-}(5,1){0.54}{-90}{90}\psarc[linewidth=1pt,linecolor=blue]{-}(5,1){0.46}{-90}{90}
\psarc[linewidth=1pt,linecolor=blue]{-}(0,1){0.54}{90}{-90}\psarc[linewidth=1pt,linecolor=blue]{-}(0,1){0.46}{90}{-90}
\facegrid{(0,0)}{(5,2)}
\psarc[linewidth=0.020]{-}(0,0){0.13}{0}{90}
\psarc[linewidth=0.020]{-}(1,1){0.13}{90}{180}
\psarc[linewidth=0.020]{-}(1,0){0.13}{0}{90}
\psarc[linewidth=0.020]{-}(2,1){0.13}{90}{180}
\psarc[linewidth=0.020]{-}(4,0){0.13}{0}{90}
\psarc[linewidth=0.020]{-}(5,1){0.13}{90}{180}
\rput(0.5,0.75){\tiny{$_{(1,n)}$}}\rput(0.5,1.75){\tiny{$_{(n,1)}$}}
\rput(1.5,0.75){\tiny{$_{(1,n)}$}}\rput(1.5,1.75){\tiny{$_{(n,1)}$}}
\rput(4.5,0.75){\tiny{$_{(1,n)}$}}\rput(4.5,1.75){\tiny{$_{(n,1)}$}}
\rput(2.5,0.5){$\ldots$}
\rput(2.5,1.5){$\ldots$}
\rput(3.5,0.5){$\ldots$}
\rput(3.5,1.5){$\ldots$}
\rput(0.5,.5){$_{u-\zeta_1}$}
\rput(0.52,1.45){$_{u_{n\!-\!1}-\zeta_1}$}
\rput(1.5,.5){$_{u-\zeta_2}$}
\rput(1.52,1.45){$_{u_{n\!-\!1}-\zeta_2}$}
\rput(4.5,.5){$_{u-\zeta_N}$}
\rput(4.52,1.45){$_{u_{n\!-\!1}-\zeta_N}$}
\end{pspicture}\ \ .\label{KacD}
\ee 
It is a Laurent polynomial in $\eE^{\ir u}$ with minimal and maximal powers $\pm 2N$.

For the Robin boundary conditions, the fused boundary triangle at fusion level $n$ for the right boundary is defined by
\be
\label{eq:fused.bdy.face}
\begin{pspicture}[shift=-.9](0,-1)(1,1)
\pspolygon[fillstyle=solid,fillcolor=lightlightblue](0,0)(1,1)(1,-1)
\rput(0.65,-0.05){\scriptsize $u,\xi$}
\rput(0.65,0.20){\tiny{$_{(n)}$}}
\end{pspicture}
\ \ = \bigg(\prod_{1\le i<j\le n}\frac{(-1)}{s_{i+j-3}}\bigg) \ \
\psset{unit=0.9cm}
\begin{pspicture}[shift=-4.9](0,-5)(5,5)
\multiput(4,-4)(0,2){5}{\pspolygon[fillstyle=solid,fillcolor=lightlightblue](0,0)(1,1)(1,-1)}
\rput(4,-4){\rput(0.65,-0.05){\small{$u_0$}}}
\rput(4,-2){\rput(0.65,-0.05){\small{$u_1$}}}
\rput(4,0){\rput(0.65,-0.05){\small{$u_2$}}}
\rput(4,2){\rput(0.65,-0.05){\small{$...$}}}
\rput(4,4){\rput(0.65,-0.05){$_{u_{n\!-\!1}}$}}
\pspolygon[fillstyle=solid,fillcolor=lightlightblue](0,0)(1,1)(2,0)(1,-1)\psarc[linewidth=0.025]{-}(1,-1){0.21}{45}{135}
\multiput(1,-1)(0,2){2}{\pspolygon[fillstyle=solid,fillcolor=lightlightblue](0,0)(1,1)(2,0)(1,-1)\psarc[linewidth=0.025]{-}(1,-1){0.21}{45}{135}}
\multiput(2,-2)(0,2){3}{\pspolygon[fillstyle=solid,fillcolor=lightlightblue](0,0)(1,1)(2,0)(1,-1)\psarc[linewidth=0.025]{-}(1,-1){0.21}{45}{135}}
\multiput(3,-3)(0,2){4}{\pspolygon[fillstyle=solid,fillcolor=lightlightblue](0,0)(1,1)(2,0)(1,-1)\psarc[linewidth=0.025]{-}(1,-1){0.21}{45}{135}}
\rput(4,-3){\scriptsize$(2u)_1$}
\rput(3,-2){\scriptsize$(2u)_2$}
\rput(4,-1){\scriptsize$(2u)_3$}\rput(2,-1){\scriptsize$(2u)_3$}
\rput(3,0){\scriptsize$(2u)_4$}\rput(1,0){\scriptsize$(2u)_4$}
\rput(4,1){\scriptsize$...$}\rput(2,1){\scriptsize$...$}
\rput(3,2){\scriptsize$...$}
\rput(4,3){\scriptsize$(2u)_{2n-3}$}
\pspolygon[fillstyle=solid,fillcolor=pink](0.2,0.2)(4.8,4.8)(4.6,5)(0,0.4)\rput{0}(2.4,2.6){\scriptsize$_{n}$}
\pspolygon[fillstyle=solid,fillcolor=pink](0.2,-0.2)(4.8,-4.8)(4.6,-5)(0,-0.4)\rput{0}(2.4,-2.6){\scriptsize$_{n}$}
\end{pspicture}\ \ ,
\ee
where $(2u)_k = 2u + k \lambda$. The same operator for the left-boundary is obtained by rotating \eqref{eq:fused.bdy.face} by $180^\circ$ and substituting $(\beta_1, \beta_2, \xi)$ for $(\alpha_1, \alpha_2, \eta)$.
The $1\times n$ fused transfer tangle is then an element of $\tl^{\textrm{\tiny$(2)$}}_N(\beta, \alpha_1, \alpha_2, \beta_1, \beta_2, \gamma)$ defined as
\be 
\Db^n(u)= \ \  
\psset{unit=1.4}
\begin{pspicture}[shift=-0.9](-1,0)(6,2)
\psline[linewidth=4pt,linecolor=blue]{-}(0,0.5)(-0.7,0.5)\psline[linewidth=2pt,linecolor=white]{-}(5,0.5)(-0.7,0.5)
\psline[linewidth=4pt,linecolor=blue]{-}(0,1.5)(-0.7,1.5)\psline[linewidth=2pt,linecolor=white]{-}(5,1.5)(-0.7,1.5)
\psline[linewidth=4pt,linecolor=blue]{-}(5,0.5)(5.7,0.5)\psline[linewidth=2pt,linecolor=white]{-}(5,0.5)(5.7,0.5)
\psline[linewidth=4pt,linecolor=blue]{-}(5,1.5)(5.7,1.5)\psline[linewidth=2pt,linecolor=white]{-}(5,1.5)(5.7,1.5)
\facegrid{(0,0)}{(5,2)}
\psarc[linewidth=0.020]{-}(0,0){0.13}{0}{90}
\psarc[linewidth=0.020]{-}(1,1){0.13}{90}{180}
\psarc[linewidth=0.020]{-}(1,0){0.13}{0}{90}
\psarc[linewidth=0.020]{-}(2,1){0.13}{90}{180}
\psarc[linewidth=0.020]{-}(4,0){0.13}{0}{90}
\psarc[linewidth=0.020]{-}(5,1){0.13}{90}{180}
\rput(0.5,0.75){\tiny{$_{(1,n)}$}}\rput(0.5,1.75){\tiny{$_{(n,1)}$}}
\rput(1.5,0.75){\tiny{$_{(1,n)}$}}\rput(1.5,1.75){\tiny{$_{(n,1)}$}}
\rput(4.5,0.75){\tiny{$_{(1,n)}$}}\rput(4.5,1.75){\tiny{$_{(n,1)}$}}
\rput(2.5,0.5){$\ldots$}
\rput(2.5,1.5){$\ldots$}
\rput(3.5,0.5){$\ldots$}
\rput(3.5,1.5){$\ldots$}
\rput(0.5,.5){$_{u-\zeta_1}$}
\rput(0.52,1.45){$_{u_{n\!-\!1}-\zeta_1}$}
\rput(1.5,.5){$_{u-\zeta_2}$}
\rput(1.52,1.45){$_{u_{n\!-\!1}-\zeta_2}$}
\rput(4.5,.5){$_{u-\zeta_N}$}
\rput(4.52,1.45){$_{u_{n\!-\!1}-\zeta_N}$}
\pspolygon[fillstyle=solid,fillcolor=lightlightblue](5,1)(6,2)(6,0)(5,1)
\pspolygon[fillstyle=solid,fillcolor=lightlightblue](0,1)(-1,2)(-1,0)(0,1)
\rput(5.65,0.95){\scriptsize $u,\xi$}\rput(5.65,1.15){\tiny{$_{(n)}$}}
\rput(-0.55,0.98){\scriptsize $-u_{n-2},\eta$}\rput(-0.55,1.18){\tiny{$_{(n)}$}}
\end{pspicture}\ \ .\label{RobinD}
\ee  
It is a Laurent polynomial in $z = \eE^{\ir u}$ with minimal and maximal degrees $\pm(2N+4n)$. 

Using the limit \eqref{triLimit}, the suitably normalized Robin transfer tangles \eqref{RobinD} reduce to the Kac transfer tangles \eqref{KacD}. 
In this limit, the fusion projectors kill off all of the boundary triangle states except the diagonal cabled state shown in \eqref{KacD}.
For both Kac and Robin boundary conditions, the fused transfer tangles $\Db^n(u)$ satisfy the relations
\be
[\Db^m(u),\Db^n(v)]=0, \qquad \Db^n(u+\pi) = \Db^n(u),\qquad \Db^n(u) = \Db^n\big((2-n)\lambda-u\big).
\ee

\section{Diagonal twist boundary conditions on the cylinder}
\label{sec:periodic}

In this section, we consider the $s\ell(2)$ models at roots of unity with twisted boundary conditions.

\subsection{Fusion hierarchies}

The fusion hierarchies are known~\cite{KirillovResh87,BazResh1989,MDPR} for the $s\ell(2)$ vertex, RSOS and loop models. 
The functional equations, including the fusion hierarchy, for the symmetric eight-vertex model are discussed extensively in \cite{FabMcCoy2004,BazhMang2007}. 
The critical six-vertex model is just a special case since it coincides with the critical line of the symmetric eight-vertex model~\cite{Baxter73}. 
The fusion hierarchy for the logarithmic minimal models was derived in \cite{MDPR}.
The fusion hierarchy equations are
\begin{subequations}
\label{eq:hierarchy}
\begin{alignat}{2}
\vec T^n(u)\vec T^1(u+n\lambda)&=f(u+n\lambda)\vec T^{n-1}(u)+f(u+(n-1)\lambda)\vec T^{n+1}(u),\label{eq:hierarchy1}\\[0.15cm]
\vec T^1(u)\vec T^n(u+\lambda)&=f(u-\lambda)\vec T^{n-1}(u+2\lambda)+f(u)\vec T^{n+1}(u),\label{eq:hierarchy2}
\end{alignat}
\end{subequations}
subject to
\bea
\vec T^{-1}(u)=0,\qquad \vec T^0(u)=f(u-\lambda) \vec I, \qquad f(u) = \prod_{j=1}^N \frac{\sin (u+\zeta_j)}{\sin \lambda}.
\eea
In the sequel, we sometimes use the more compact notations
\be
\label{eq:compact.cyl}
\Tb^n_k = \Tb^n(u+k\lambda), \qquad f_k = f(u+k\lambda).
\ee
The relations \eqref{eq:hierarchy} hold for $n \ge 0$ and recursively define the fused transfer matrices $\vec T^n(u)$ as polynomials in the fundamental transfer matrix $\Tb(u)=\Tb^1(u)$. This holds both for generic values of $\lambda$ and for roots of unity. 

The fusion hierarchy can be extended to $n<0$ with the convention
\be
\label{eq:neg}
\Tb^n(u) := -\Tb^{-2-n}(u+(n+1)\lambda),\qquad n<0.
\ee
By applying \eqref{eq:neg} to \eqref{eq:hierarchy1} with $n<0$, one indeed reproduces \eqref{eq:hierarchy2} for the positive values of the fusion indices.

\subsection[Baxter's $T$-$Q$ relation]{Baxter's $\boldsymbol T$-$\boldsymbol Q$ relation}

Baxter's $T$-$Q$ relation~\cite{BaxterQ,Baxter73,BaxterBook} is 
\bea
\vec T(u)\vec Q(u)=f(u)\vec Q(u-\lambda)+f(u-\lambda)\vec Q(u+\lambda),\qquad [\vec T(u),\vec Q(v)]=0,\label{BaxterTQ}
\eea
{where $\Tb(u)$ is a commuting family of transfer matrices and $\vec Q(u)$ is an auxiliary matrix family.}
The eigenvalues of $\vec Q(u)$ take the form
\bea
Q(u)=Q(u+2\pi)=
\prod_{m=1}^M \sin(u-u_m)\label{Qroots}
\eea
where 
$M$ is a nonnegative integer and $u_m$ are the Bethe roots. In principle, the Bethe roots are determined by solving the Bethe ansatz equations in the form of non-linear equations obtained by substituting \eqref{Qroots} into the eigenvalue equation \eqref{BaxterTQ} and setting $u=u_m$ so that the left-side vanishes.

The matrix $\vec Q(u)$ is in fact not unique. On the cylinder, there are the ``Bloch wave" solutions~\cite{BazhMang2007} denoted by $\vec Q^\pm(u)$. Moreover, if $\vec Q(u)$ satisfies Baxter's $T$-$Q$ relation, then any matrix given by
\bea
\vec Q'(u)=\vec Q(u) \vec Q_0(u),\qquad [\vec Q_0(u),\vec T(u)]=0,\qquad [\vec Q_0(u),\vec Q(u)]=0,\qquad \vec Q_0(u+\lambda)=\vec Q_0(u),
\eea
also satisfies the $T$-$Q$ relation. At roots of unity, this observation is tied to the freedom of adding or removing complete $p'$-strings in the eigenvalues $Q(u)$. 
In fact, working in the vertex setting, Baxter himself constructed two such $\vec Q(u)$ matrices, one $\vec Q_{p,p'}(u)$~\cite{BaxterQ} in 1972 for roots of unity cases and arbitrary system size $N$ and another $\vec Q_\lambda(u)$~\cite{Baxter73} in 1973 for the general case with $N$ even. At roots of unity with $N$ even, these two $\vec Q(u)$ matrices are not equal (see the discussion in Appendix~A of \cite{BazhMang2007}). More specifically, the $Q$ matrix at roots of unity, corresponding to logarithmic theories, is not given by specializing the generic $Q$ matrix:
\bea
\lim_{\lambda\to \lambda_{p,p}} \vec Q_\lambda(u) \ne \vec Q_{p,p'}(u).
\eea

It was observed by Pronko~\cite{Pronko2000} that the fusion hierarchy \eqref{eq:hierarchy} bears a striking similarity to Baxter's $T$-$Q$ equation \eqref{BaxterTQ}.
Applying a shift of $-n\lambda$ to the arguments, the relation \eqref{eq:hierarchy1} takes the form
\begin{align}
\vec T^1(u)\vec T^n(u-n\lambda)&=f(u)\vec T^{n-1}(u-n\lambda)+f(u-\lambda)\vec T^{n+1}(u-n\lambda)\nonumber\\[4pt]
&=f(u)\vec T^{n-1}(u-\lambda-(n-1)\lambda)+f(u-\lambda)\vec T^{n+1}(u+\lambda-(n+1)\lambda).\label{PronkoEqn}
\end{align}
This can be written as a {\it generalized} $T$-$Q$ relation
\bea
\vec T(u)\vec Q^n(u)=f(u)\vec Q^{n-1}(u-\lambda)+f(u-\lambda)\vec Q^{n+1}(u+\lambda),\qquad [\vec T(u),\vec Q^j(v)]=0,
\eea
where 
\bea
\vec Q^n(u)=\vec T^n(u-n\lambda).
\eea

Although it may not be needed for $\lambda=\lambda_{p,p'}$, it was shown in \cite{BLZ97} that, for generic $\lambda$, Baxter's $Q$-operator is associated with an infinite dimensional (spin-$\tfrac{n}{2}\to\infty$) oscillator representation. Indeed, in \cite{YNZ2006,YZ2006}, it was suggested that formally setting
\bea
\vec Q(u)=\lim_{n\to\infty} \vec T^n(u-n\lambda)\label{fuselimit}
\eea
in \eqref{PronkoEqn} leads to a functional equation identical to Baxter's $T$-$Q$ relation \eqref{BaxterTQ}:
\bea
\vec T(u)\vec Q(u)=f(u)\vec Q(u-\lambda)+f(u-\lambda)\vec Q(u+\lambda),\qquad [\vec T(u),\vec Q(v)]=0.\label{TQ}
\eea
However, no arguments were put forward for the convergence of the limit \eqref{fuselimit} and, assuming that the limit actually exists, no relation was proposed between the $\vec Q(u)$ in \eqref{fuselimit} and Baxter's $\vec Q(u)$. Indeed, closer inspection shows there are problems to make sense of this simple limiting procedure. Nevertheless, as shown in \cite{YNZ2006,YZ2006,FGSW2011}, it seems that the $T$-$Q$ relations arising from this procedure make good physical sense and that they lead to correct results. We will make proper sense of this limit, under certain circumstances, in subsequent sections.

\subsection[Closure, the infinite fusion limit and extended $Q$ matrices]{Closure, the infinite fusion limit and extended $\boldsymbol Q$ matrices}
\label{InfiniteFusion}

For the RSOS models, the fusion hierarchies truncate at a finite level. In contrast, the fusion hierarchies \eqref{eq:hierarchy} of the vertex and loop models at roots of unity do not truncate and instead satisfy a closure relation. Indeed, for the roots of unity cases $\lambda = \lambda_{p,p'}$, the fused transfer matrix $\Tb^{p'}(u)$ satisfies the closure relation
\begin{equation}\label{eq:closure.cyl}
\Tb^{p'}(u)=\Tb^{p'-2}(u+\lambda)+ 2 \sigma \Jb\, \Tb^0(u), \qquad \sigma = \ir^{-N(p'-p)},
\end{equation}
as in (2.22) of \cite{BazhMang2007} and analogous to (7.28) of \cite{MDPR}. In the six-vertex model, the matrix $\Jb$, related to the braid limit $u\to \ir \infty$, is a diagonal matrix given by
\be
\Jb = \frac12(\omega^{p'} \ir^{-2p\svec S^z}+ \omega^{-p'} \ir^{2p\svec S^z}),\qquad [\vec T^n(u),\vec J]=0,
\ee
where the matrix $\vec S^z$ is the spin-$\half$ magnetization with eigenvalues $S^z\in \half \mathbb Z$ and the complex parameter $\omega$ is the twist. In the loop model, $\Jb$ is an element of $\eptl_N(\alpha,\beta)$ that is independent of $u$. {We recall that the definition of this algebra is given in \cref{app:diag.algebras}.} On the standard module $\mathsf W_{N,d}$, $\Jb$ is proportional to the identity matrix, with an eigenvalue $J_d$ that is independent of $N$ and given by
\be
J_d = \frac12(\omega^{p'} \ir^{-pd}+ \omega^{-p'} \ir^{pd}).
\ee
As explained in \cref{app:diag.algebras}, in the context of  the loop model, $\omega$ is a free parameter that appears in the definition of the standard modules and measures the winding of the defects around the cylinder.

The fused transfer matrices with $n=yp'+j> p'$ satisfy the generalized closure relations
\be
\label{eq:genclosure}
\Tb^{yp'+j}(u) = \sigma^y U_{y}(\Jb)\,\Tb^j(u)+\sigma^{y-1} U_{y-1}(\Jb)\, \Tb^{p'-2-j}(u+(j+1)\lambda),\qquad y \in \mathbb Z,\qquad j=0, 1, \dots, p'-1,
\ee
where $U_k(x)$ is the $k$-th Chebyshev polynomial of the second kind. In the following, we write
\be
\Jb = \frac12\big(\eE^{\ir \boldsymbol\Lambda} + \eE^{-\ir \boldsymbol\Lambda}\big)
\ee
where $\eE^{\ir \boldsymbol\Lambda} = \omega^{p'} \ir^{-2p\svec S^z}$ for the vertex model. In the loop model, $\eE^{\ir \boldsymbol\Lambda}$ is an element of $\eptl_N(\alpha,\beta)$ that acts as the identity on the standard modules, with the eigenvalue $\omega^{p'} \ir^{-pd}$.

In a given magnetization sector, the infinite fusion limits $n \to \pm\infty$ of the eigenvalues $T^{n}(u)$ exist provided the limit is taken through suitable subsequences. Explicitly, setting $n = yp'+j$, we find
\be
\lim_{y\to\pm \infty} \frac{ T^{yp'+j}(u)}{\sigma^{y-1}U_{y-1}(J)}=\begin{cases}
\calT^{j,\pm}(u),&\mbox{$|\eE^{\ir \Lambda}|> 1$ or $\eE^{\ir \Lambda}=\pm 1$}\\[0.15cm]
\calT^{j,\mp}(u),&|\eE^{\ir \Lambda}|<1
\end{cases}
\ee
where $\calT^{j,\pm}(u)$ are the eigenvalues of the extended $Q$ matrices
\be
\label{eq:calT}
\vec\calT^{j,\pm}(u)=\sigma\,\eE^{\pm \ir \boldsymbol\Lambda}\,\vec T^j(u)+\vec T^{p'-j-2}(u+(j+1)\lambda).
\ee
Taking the limit through these subsequences is the proper way to make sense of the infinite fusion limit \eqref{fuselimit}. 
This limit exists in sectors with $p S^z\in \mathbb Z$ for periodic boundary conditions ($\omega=1$) but it does not exist for general complex twists $\omega$ on the unit circle.

\subsection[Extended $T$-system and extended $T$-$Q$ relations]{Extended $\boldsymbol T$-system and extended $\boldsymbol T$-$\boldsymbol Q$ relations}
\label{SecBilinear}

Motivated by the previous section, let us put aside questions of the existence of the infinite fusion limit and work with the extended $Q$ matices $\vec\calT^{j,\pm}(u)$ directly as defined in \eqref{eq:calT}. 
From \eqref{eq:genclosure}, we find that the $\vec\calT^{j,\pm}(u)$ satisfy the periodicity properties
\be \label{eq:per.con.periodic}
\vec\calT^{p'+j,\pm}(u) = \sigma\, \eE^{\pm \ir \boldsymbol\Lambda} \vec\calT^{j,\pm}(u), \qquad \vec\calT^{j,\pm}(u + \pi) = (-1)^N\vec\calT^{j,\pm}(u)
\ee
and the conjugacy property
\be
\vec\calT^{p'-j-2,\pm}(u) = \sigma\, \eE^{\pm \ir \bold \Lambda} \vec\calT^{j,\mp}\big(u-(j+1)\lambda\big).
\ee
They also satisfy the bilinear factorization identities
\be
\label{eq:bilinear.in.T}
\vec\calT^{j,+}(u)\vec\calT^{j,-}(u)=\vec T^{p'-1}(u)\vec T^{p'-1}\big(u+(j+1)\lambda\big),\qquad j\in \mathbb Z.
\ee
The right side of \eqref{eq:bilinear.in.T} can equivalently be written as $\sigma^2 \vec\calT^{p'-1,+}(u)\vec\calT^{p'-1,-}(u)$ because $\vec\calT^{p'-1,\pm}(u) = \sigma\, \eE^{\pm \ir \boldsymbol\Lambda} \vec T^{p'-1}(u)$. After a shift of $-j \lambda$, the bilinear factorization identities take the explicit form
\be
\label{eq:bilinear.take.2}
\big(\sigma\,\eE^{\ir \boldsymbol \Lambda}\vec T^j(u-j\lambda)+\vec T^{p'-j-2}(u+\lambda)\big)\big(\sigma\,\eE^{-\ir \boldsymbol \Lambda}\vec T^j(u-j\lambda)+\vec T^{p'-j-2}(u+\lambda)\big)=\vec T^{p'-1}(u-j\lambda)\vec T^{p'-1}(u+\lambda).
\ee
For the case $j=0$,
this agrees with (2.24) of \cite{BazhMang2007}. 
In fact, the identity \eqref{eq:bilinear.take.2} is the special case $k=p'\!-\!j\!-\!1$ of the more general two-index extended $T$-system of bilinear identities
\be
\label{eq:quad.identity}
\Tb^{j+k}_{-j}\Tb^{p'-1}_1=\Tb^j_{-j} \Tb^{p'-1-k}_{k+1} + 2 \sigma \Jb\, \Tb^j_{-j} \Tb^{k-1}_{1} + \Tb^{k-1}_{1}\Tb^{p'-2-j}_{1}, \qquad j,k \in \mathbb Z.
\ee
These bilinear identities are proved in \cref{AppA.periodic}. 

Using the fusion hierarchy \eqref{eq:hierarchy}, we separately obtain the set of {\it extended} $T$-$Q$ relations
\begin{subequations}
\label{extTQ}
\begin{alignat}{2}
\vec T(u)\vec\calT^{j,\pm}(u-j\lambda)&=f(u) \vec\calT^{j-1,\pm}(u-j\lambda)+f(u-\lambda)\vec\calT^{j+1,\pm}(u-j\lambda),\label{extTQa}\\[0.1cm]
\vec T(u)\vec\calT^{j,\pm}(u+\lambda)&=f(u-\lambda) \vec\calT^{j-1,\pm}(u+2\lambda)+f(u)\vec\calT^{j+1,\pm}(u),\label{extTQb}
\end{alignat}
\end{subequations}
where $\vec\calT^{j,\pm}(u)$ is defined by \eqref{eq:calT}. 
Indeed, setting $\vec\calT_k^{j,\pm}=\vec\calT^{j,\pm}(u+k\lambda)$ and expanding the left side using the fusion hierarchies \eqref{eq:hierarchy} gives
\begin{align}
\vec T_0^1\vec\calT_{-j}^{j,\pm}&=\vec T_0^1\big[\sigma\,\eE^{\pm \ir \boldsymbol\Lambda}\,\vec T_{-j}^j+\vec T_1^{p'-j-2}\big]
=\sigma\,\eE^{\pm \ir \boldsymbol\Lambda}(f_0\vec T_{-j}^{j-1}+f_{-1}\vec T_{-j}^{j+1})+
(f_{-1}\Tb^{p'-j-3}_2 + f_0 \Tb^{p'-j-1}_0)
\nonumber\\[4pt]
&
=f_0\big[\sigma\,\eE^{\pm \ir \boldsymbol\Lambda}\,\vec T_{-j}^{j-1}+\vec T_0^{p'-j-1}\big]+f_{-1}\big[\sigma\,\eE^{\pm \ir \boldsymbol\Lambda}\,\vec T_{-j}^{j+1}+\vec T_2^{p'-j-3}\big]
=f_0 \vec\calT_{-j}^{j-1,\pm}+f_{-1}\vec\calT_{-j}^{j+1,\pm}.
\end{align}
{The relation \eqref{extTQb} is obtained via a similar derivation. In the extended $T$-$Q$ relations \eqref{extTQ}, the extended $Q$ matrices $\vec Q^{j,\pm}(u)$ are no longer auxiliary quantities but are defined as polynomials in $\vec T(u)$ through \eqref{eq:calT} and \eqref{eq:hierarchy}.}

\subsection[Baxter $T$-$Q$ eigenvalue relations and eigenvalue decompositions]{Baxter $\boldsymbol T$-$\boldsymbol Q$ eigenvalue relations and eigenvalue decompositions}\label{sec:decomp.diag}

In this section, we analyse the extended $T$-$Q$ eigenvalue relations to deduce the usual Baxter $T$-$Q$ scalar relation. 
The eigenvalues $\calT^{j,\pm}(u)$ and $T^{p'-1}(u)$ of the corresponding transfer matrices also satisfy the bilinear factorization identities \eqref{eq:bilinear.in.T}. Each such eigenvalue is a Laurent polynomial in $z=\eE^{\ir u}$ with maximal and minimal powers $\pm N$. Let us denote by $\{v_k\}$ the set of zeros of $T^{p'-1}(u+\lambda)$ in the complex $u$-plane in a given vertical strip of width $\pi$, and likewise by $\{w_\ell^{j,\pm}\}$ the sets of zeros of $\calT^{j,\pm}(u-j \lambda)$. The bilinear factorization identities \eqref{eq:bilinear.in.T} imply that
\be
\label{eq:samezeros}
\prod_k \sin(u-v_k)\sin\!\big(u-(j+1)\lambda -v_k\big) = \prod_\ell \sin(u-w^{j,+}_\ell)\sin(u-w^{j,-}_\ell).
\ee
For generic $\omega$ with $e^{\ir \Lambda}\ne \pm 1$, we observe empirically that the zeros are all simple and that there are no complete $p'$-strings. In contrast, at $e^{\ir \Lambda}= \pm 1$, complete $p'$ strings can and do occur; this case is discussed further at the end of this subsection. Each zero $u=u_m$ of \eqref{eq:samezeros} belongs to $\{v_k\}$ or $\{v_k+(j+1)\lambda\}$, and likewise belongs to $\{w_\ell^{j,+}\}$ or $\{w_\ell^{j,-}\}$. {Let us write 
\be 
E^\pm = E^{j,\pm} =  \{v_k\} \cap \{w_\ell^{j,\pm}\}, \qquad j = 0, \dots, p'-2.
\ee 
For each $j$, the sets $E^{j,+}$ and $E^{j,-}$ comprise all the zeros of $T^{p'-1}(u+\lambda)$. Crucially, we observe that these sets {\em do not depend on $j$}. This implies that the division of the zeros of $T^{p'-1}(u+\lambda)$ between $\calT^{j,+}(u-j\lambda)$ and $\calT^{j,-}(u-j\lambda)$ is identical for all $j$. This non trivial statement justifies our choice to write these sets as $E^\pm$, with no reference to $j$.}

The proposition below gives a partial proof of this claim. It is based on the assumption that the functions $f_k$ and $T^1_k$ are non zero when specialized to values of $u$ in $E^{j,\pm}$, for each $j = 0, \dots, p'-2$ and $k \in \mathbb Z$. Our explorations with small $N$ using a computer program support this assumption.
\begin{Proposition}\label{prop:Ej}
With the above assumption, the sets $E^{j,+}$ and $E^{j,-}$ are independent of $j$.
\end{Proposition}
{\scshape Proof.} The proof uses the relation
\be\label{eq:secondTQ}
T^1_{-j-1}\calT^{j,\pm}_{-j} = f_{-2-j} \calT^{j-1,\pm}_{1-j} + f_{-1-j} \calT^{j+1,\pm}_{-1-j},
\ee
which is obtained from \eqref{extTQb} by a shift by $-(j+1)\lambda$. We prove the statement for $E^{j,+}$; the proof for $E^{j,-}$ is identical. Let $u_0$ be an element of $\in E^{0,+}$. From the definition of $E^{0,+}$, both $Q^{0,+}_{0}$ and $Q^{-1,+}_{1}$ vanish at $u=u_0$. Setting $j = 0$ and $u=u_0$ in \eqref{eq:secondTQ}, we find that two of the three terms are zero. This implies that the last term $f_{-1}Q^{1,+}_{-1}$ also vanishes at $u=u_0$. From the assumption that $f_{-1} \neq 0$ for $u = u_0$, we thus find that $u_0\in E^{1,+}$ and therefore $E^{0,+} \subseteq E^{1,+}$.

Let us now consider a value $u = u_1 \in E^{1,+}$. The relation \eqref{eq:secondTQ} with $j = 0$ and $u=u_1$ has two terms which vanish at $u=u_1$, implying that the last term $T^1_{-1} \calT^{0,\pm}_{0}$ is also zero at this value.
From the assumption that $T^1_{-1} \neq 0$ for $u=u_1$, we find that $u_1 \in E^{0,+}$ and therefore that $E^{1,+} \subseteq E^{0,+}$. We thus deduce that $E^{0,+} = E^{1,+}$.

By repeating the argument using \eqref{eq:secondTQ} with $j=1$, one finds $E^{1,+} = E^{2,+}$. The proof that $E^{j,+} = E^{j+1,+}$ on increasing values of $j$ is obtained with the same arguments.
\hfill $\square$
\medskip

From explorations using our computer program, we find that for the vertex model, the cardinalities of the sets $E^+$ and $E^-$ are $|E^\pm | = \frac N2 \mp S^z$. Likewise for the standard module $\mathsf W_{N,d}$ of the loop model, these sets have cardinalities $|E^\pm| = \frac {N\mp d}2$. We then define 
\be 
\label{eq:Qpm.periodic}
Q^+(u) = \prod_{u_m \in E^+} \sin(u-u_m), \qquad Q^-(u) = \prod_{u_m \in E^-} \sin(u-u_m).
\ee
The values $u_m \in E^{\pm}$ are therefore the Bethe roots.
Moreover, the Bazhanov-Lukyanov-Zamolodchikov
eigenvalue decompositions \cite{BLZ97,BazhMang2007} follow:
\begin{subequations}\label{eq:TQQ}
\begin{alignat}{2}
\calT^{j,+}(u-j\lambda) &= R^{j,+}(u)Q^+(u){Q^-\big(u-(j+1)\lambda\big)},\label{eq:TQQp}\\  
\calT^{j,-}(u-j\lambda) &=  R^{j,-}(u)Q^-(u){Q^+\big(u-(j+1)\lambda\big)},\label{eq:TQQm}\\
T^{p'-1}(u+\lambda) &= \phi(u) Q^+(u)Q^-(u),\label{eq:TQQpp}\\  
T^{p'-1}(u-j\lambda) &= \phi(u) Q^+\big(u-(j+1)\lambda\big)Q^-\big(u-(j+1)\lambda\big).\label{eq:TQQppv2}
\end{alignat}
\end{subequations}
{The division of the factors of $Q^\pm\big(u-(j+1)\lambda\big)$ between $\calT^{j,+}(u-j\lambda)$ and $\calT^{j,-}(u-j\lambda)$ is imposed by the conjugacy property in \eqref{eq:per.con.periodic}.} For $e^{\ir \Lambda}\ne \pm 1$, $R^{j,\pm}(u)$ and $\phi(u)$ are constants satisfying $R^{j,+}(u)R^{j,-}(u)=\phi(u)^2$. 
They can be computed by comparing the coefficient of $z^{N}$ of the left and right sides of \eqref{eq:TQQp} and \eqref{eq:TQQm}. For the right sides, these maximal coefficients are obtained directly from \eqref{eq:Qpm.periodic}. For the left sides, they are obtained from the braid limit of \eqref{eq:calT}. The eigenvalues of the braid fused transfer matrices are indeed known, see \cite{Zhou96b} for the vertex model and \cite{MDPR} for the loop model.
This calculation yields
\be
\frac{R^{j+1,\pm}(u)}{R^{j,\pm}(u)} = {\tilde \omega}^{\pm 1}, \qquad \tilde \omega = \omega\, \times
\left\{\begin{array}{ll}
\eE^{-\ir \pi S^z}& \textrm{for the vertex model},\\[0.15cm]
\eE^{-\ir \pi d/2}& \textrm{for the loop model}.
\end{array}\right.
\ee
The decompositions \eqref{eq:TQQ} are in fact related by a shift of argument to the decompositions of \cite{BazhMang2007} in their equations (2.29) and (2.31). 
The extended $T$-$Q$ and Baxter $T$-$Q$ relations now take the equivalent scalar forms:
\begin{subequations}
\begin{alignat}{2}
T(u)\calT^{j,\pm}(u-j\lambda)&=f(u) \calT^{j-1,\pm}(u-j\lambda)+f(u-\lambda)\calT^{j+1,\pm}(u-j\lambda),\\[0.1cm]
T(u) Q^\pm(u)&={\tilde \omega^{\mp 1}}f(u)  Q^\pm(u-\lambda)+{\tilde \omega^{\pm 1}}f(u-\lambda) Q^\pm(u+\lambda).\label{eq:tqfinal}
\end{alignat}
\end{subequations}
The first identity is the eigenvalue form of the extended $T$-$Q$ relation \eqref{extTQa} and holds for all $j$. In the second equation, we used \eqref{eq:TQQp} and \eqref{eq:TQQm} and divided by ${R^{j,\pm}(u)}Q^\pm\big(u-(j+1)\lambda\big)$. Remarkably, the dependence on $j$ disappears in \eqref{eq:tqfinal}. Substituting $u=u_m$ in \eqref{eq:tqfinal}, we recover the Bethe ansatz equations for periodic twisted boundary conditions on the cylinder. We note the presence of factors $\eE^{\pm\ir \pi S^z}$ and $\eE^{\pm\ir \pi d/2}$ in the $T$-$Q$ relation, which are absent in Baxter's usual $T$-$Q$ equation \eqref{BaxterTQ} and are due to our choice of gauge for the $R$-matrix.

An interesting feature occurs in the sector with magnetization $S^z = m$ if the twist is such that $e^{\ir \Lambda} = \pm 1$. In these special cases, at the level of the eigenvalues, \eqref{eq:bilinear.take.2} reads
\be
\label{eq:bilinear.take.3}
\big(\sigma\,T^j(u-j\lambda) \mp T^{p'-j-2}(u+\lambda)\big)^2=T^{p'-1}(u-j\lambda) T^{p'-1}(u+\lambda).
\ee
All the zeros of the left side are double, so the same must hold for the right side. Numerically, we observe that each of these zeros is twice degenerate, and never more. Depending on how the zeros are split between the two factors on the right side, the zeros of $T^{p'-1}(u+\lambda)$ can either be single or double. If $u=u_0$ is a single zero of $T^{p'-1}(u+\lambda)$, the double zero of the left side of \eqref{eq:bilinear.take.2} is evenly split between the two factors of the right side, implying that $T^{p'-1}(u_0-j\lambda) = 0$. This holds true for $j \in \mathbb Z$, so that $T^{p'-1}(u+\lambda) = 0 $ for $u = u_0, u_0 + \lambda, \dots, u_0 + (p'-1)\lambda$. The zeros of $T^{p'-1}(u+\lambda)$ are therefore either double zeros or single zeros forming complete $p'$-strings. In the decompositions \eqref{eq:TQQ}, the zeros of the complete $p'$-strings are encoded in the functions ${R^{j,\pm}(u)}$ and $\phi(u)$, which in this case are not constants. Clearly, $\phi(u) = \phi(u+\lambda)$. The sets $E^+$ and $E^-$ are equal, implying that $Q^+(u)=Q^-(u)=Q(u)$, $Q^+(u)Q^-(u)=Q(u)^2$ and {$R^{j,+}(u)=R^{j,-}(u) = R^j(u)$}. This last function satisfies ${R^j(u)^2} = \phi(u)^2$ and therefore ${R^j(u) = R^j(u+\lambda)}$.

\section{$\boldsymbol{U_q(s\ell(2))}$ invariant/Kac vacuum boundary conditions on the strip}\label{sec:strip}

In this section, we consider the six-vertex model at roots of unity with diagonal $U_q(s\ell(2))$ invariant boundary conditions and the logarithmic minimal models ${\cal LM}(p,p')$ with Kac vacuum boundary conditions. The functional equations satisfied by the fused transfer matrices of these two models coincide. More general diagonal and off-diagonal boundary conditions on the strip are considered in \cref{sec:Robin}.

\subsection{Fusion hierarchies}

The double-row transfer matrices $\vec D^n(u)$ of the $s\ell(2)$ models with $U_q(s\ell(2))$ invariant/Kac vacuum boundary conditions on the strip satisfy the fusion hierarchy relations~\cite{Zhou96b,MDPR}
\begin{subequations}
\label{eq:hierarchyD}
\begin{alignat}{2}
\hspace{-0.2cm}s_{n-2}s_{2n-1} \vec D^n(u)\vec D^1(u+n\lambda)&=s_{n-3}s_{2n}f(u+n\lambda)\vec D^{n-1}(u)+s_{n-1}s_{2n-2}f(u+(n-1)\lambda)\vec D^{n+1}(u)\label{eq:hierarchy1D}\\[0.1cm]
s_{n}s_{-1} \vec D^1(u)\vec D^n(u+\lambda)&=s_{n+1}s_{-2}f(u-\lambda)\vec D^{n-1}(u+2\lambda)+s_{n-1}s_0 f(u)\vec D^{n+1}(u)\label{eq:hierarchy2D}
\end{alignat}
\end{subequations}
where 
\be\label{eq:Kac.convention}
\vec D^{-1}(u)=0,\qquad \vec D^0(u)=f(u-\lambda) \vec I, \qquad f(u) = \prod_{i=1}^N \frac{\sin (u + \zeta_i)\sin (u - \zeta_i)}{\sin^2 \lambda} 
, \qquad s_k = \frac{\sin(2u+k \lambda)}{\sin \lambda},
\ee
and $\zeta_i$ are bulk inhomogeneities. The extra $s_i$ functions appearing in \eqref{eq:hierarchyD} are so-called {\it surface terms}. In the following, we sometimes use the short-hand notation
\be\label{eq:Df}
\vec D^j_k=\vec D^j(u+k\lambda), \qquad f_k = f(u+k \lambda).\ee
Moreover, the convention
\be
\label{eq:neg.D}
\Db^n(u) := -\Db^{-2-n}(u+(n+1)\lambda)\qquad n\le -2
\ee
allows us to extend the definition of $\Db^n(u)$ to $n\le -2$.

After applying a shift of $-n\lambda$ to the arguments, \eqref{eq:hierarchy1D} takes the form
\be
s_{-n-2}s_{-1} \vec D^1(u)\vec D^n(u-n\lambda)=s_{-n-3}s_0f(u)\vec D^{n-1}(u-n\lambda)+s_{-n-1}s_{-2}f(u-\lambda)\vec D^{n+1}(u-n\lambda).
\ee
Formally, this can be written as a {\it generalized} $T$-$Q$ relation
\begin{align}
s_{-1}\vec D^1(u)\vec Q^n(u)=s_0f(u)\vec Q^{n-1}(u-\lambda)+s_{-2}f(u-\lambda)\vec Q^{n+1}(u+\lambda),\qquad [\vec D(u),\vec Q^j(u)]=0
\end{align}
where
\bea
\vec Q^n(u)=s_{-n-2}\vec D^n(u-n\lambda).
\eea
As observed in \cite{YNZ2006}, this would lead to a $T$-$Q$ equation if the limit $\lim_{n\to\infty} \vec Q^n(u)$ exists. But again, the existence of this simple limit is problematic.

\subsection[Closure, the infinite fusion limit and extended $Q$-matrices]{Closure, infinite fusion limit and extended $\boldsymbol Q$-matrices}

For the root of unity cases $\lambda = \lambda_{p,p'}$, the fused transfer matrices satisfy the closure relation 
\begin{equation}\label{eq:closure.D}
\Db^{p'}(u)=2 \sigma \Db^0(u)+ \Db^{p'-2}(u+\lambda), \qquad \sigma = (-1)^{p'-p},
\end{equation}
analogous to (7.9) of \cite{MDPR}. Setting $n=yp'+j$, this closure relation generalizes to the higher fused transfer matrices
\be
\label{eq:closure.gen.D}
\Db^{yp'+j}(u) = \sigma^y (y+1)\Db^j(u)+ \sigma^{y-1} y \Db^{p'-2-j}(u+(j+1)\lambda),\qquad y \in \mathbb Z,\qquad j = 0, 1,\dots, p'-1.
\ee
We take the limit $n \to \infty$ of $\Db^n(u)$ through subsequences, by setting $n = yp' + j$ and sending $y$ to infinity with $j$ finite:
\be
\lim_{y\to\infty} \frac{\vec D^{yp'+j}(u-n\lambda)}{y \sigma^{y-1}}=\vec\calD^j(u).
\ee
This limit exists and defines the extended $Q$ matrices as the linear combinations
\be
\vec\calD^j(u)=\sigma \vec D^j(u)+ \vec D^{p'-j-2}(u+(j+1)\lambda) \qquad j \in \mathbb Z.
\ee
In contrast to the case of twisted boundary conditions on the cylinder, the infinite fusion limit taken through subsequences always exist on the strip. Moreover, the same extended $Q$ matrices are obtained by taking the limit $y \to -\infty$.
The matrices $\vec\calD^{j}(u)$ satisfy the following periodicity and conjugacy properties:
\be
{\vec\calD^{j}(u+\pi) = \sigma\, \vec\calD^{j}(u)},\qquad \vec\calD^{p'+j}(u) = \sigma\, \vec\calD^{j}(u), \qquad \vec\calD^{p'-j-2}(u) = \sigma\, \vec\calD^{j}\big(u-(j+1)\lambda\big).
\ee

\subsection[Extended $T$-system and extended $T$-$Q$ relations]{Extended $\boldsymbol T$-system and extended $\boldsymbol T$-$\boldsymbol Q$ relations}

The extended $Q$ matrices satisfy the bilinear factorization identities
\be
\label{eq:bilinear.in.D}
\big(s_{j-2}\vec\calD^{j}(u)\big)^2=s_{-3}s_{2j-1}\vec D^{p'-1}(u)\vec D^{p'-1}\big(u+(j+1)\lambda\big),\qquad j\in \mathbb Z.
\ee
Indeed, after a shift of $-j \lambda$, \eqref{eq:bilinear.in.D} explicitly reads
\be
\label{eq:bilinear.in.D.take.2}
(s_{-j-2})^2\big(\sigma\vec D^j(u-j\lambda)+ \vec D^{p'-j-2}(u+\lambda)\big)^2=s_{-2j-3}s_{-1}\vec D^{p'-1}(u-j\lambda)\vec D^{p'-1}(u+\lambda).
\ee
This is the special case $k=p'-1-j$ of the more general two-index extended $T$-system of bilinear identities 
\be
\Db^{j+k}_{-j}\Db^{p'-1}_1=\frac{s_{k-1}s_{-2-j}}{s_{k-j-2}s_{-1}}\Big(\Db^j_{-j} \Db^{p'-1-k}_{k+1} + 2 \sigma \Db^j_{-j} \Db^{k-1}_{1} + \Db^{k-1}_{1}\Db^{p'-2-j}_{1}\Big), \qquad j,k \in \mathbb Z\label{KacTsystem}.
\ee
These identities are proved in \cref{AppA.boundary}.

From the fusion hierarchy relations \eqref{eq:hierarchy1D} and \eqref{eq:hierarchy2D}, we obtain the {\it extended} $T$-$Q$ relations: 
{
\begin{subequations}
\begin{alignat}{2}
s_{-j-2}s_{-1}\vec D(u)\vec\calD^{j}(u-j\lambda)&=s_{-j-3}s_{0}f(u) \vec\calD^{j-1}(u-j\lambda)+s_{-j-1}s_{-2}f(u-\lambda)\vec\calD^{j+1}(u-j\lambda),
\label{KacExtTQ}\\[0.15cm]
s_{j}s_{-1}\vec D(u)\vec\calD^{j}(u+\lambda)&=s_{j+1}s_{-2}f(u-\lambda) \vec\calD^{j-1}(u+2\lambda)+s_{j-1}s_{0}f(u)\vec\calD^{j+1}(u).
\end{alignat}
\end{subequations}
}
Indeed, setting $\vec \calD^j_k=\vec\calD^j(u+k\lambda)$ and expanding the left side gives
\begin{align}
s_{-j-2}s_{-1}\vec D_0^1\vec\calT_{-j}^{j}&=s_{-j-2}s_{-1}\vec D_0^1\big(\sigma \vec D_{-j}^j+\vec D_1^{p'-j-2}\big)\nonumber\\[4pt]
&\hspace{-1.5cm}
=s_{-j-3}s_0\sigma f_0\vec D_{-j}^{j-1}+\sigma s_{-j-1}s_{-2} f_{-1}\vec D_{-j}^{j+1}+s_{-j-1}s_{-2} f_{-1} \vec D_2^{p'-j-3}+s_{-j-3}s_0\sigma f_0\vec D_0^{p'-j-1}\nonumber\\[4pt]
&\hspace{-1.5cm}
=s_{-j-3}s_{0} f_0\big(\sigma\vec D_{-j}^{j-1}+\vec D_0^{p'-j-1}\big)+s_{-j-1}s_{-2} f_{-1}\big(\sigma\vec D_{-j}^{j+1}+\vec D_2^{p'-j-3}\big)\nonumber \\[4pt]
&\hspace{-1.5cm}
=s_{-j-3}s_{0} f_0 \vec\calT_{-j}^{j-1}+s_{-j-1}s_{-2} f_{-1}\vec\calT_{-j}^{j+1}.
\end{align}
{The second relation is proven similarly.}

\subsection{Polynomial reductions}

The transfer matrices $\Db^{p'-1}(u)$ and extended $Q$ matrices $\Dbm^{j}(u)$ appearing in \eqref{eq:bilinear.in.D} in fact have a number of simple overall factors arising from the surface terms in the fusion hierarchy. Setting $k=-j$ in \eqref{KacTsystem} gives 
\be
f_{-j-1}\Db^{p'-1}_1=\frac{s_{-j-1}s_{-j-2}}{s_{-2j-2}s_{-1}}\Big(\Db^j_{-j} \Db^{p'-1+j}_{-j+1} + 2 \sigma \Db^j_{-j} \Db^{-j-1}_{1} + \Db^{-j-1}_{1}\Db^{p'-2-j}_{1}\Big).
\ee
For $j = -1$ and $j=0$, the trigonometric prefactor on the right side equals $1$. For the other values of $j$, this prefactor is not $1$. This implies that the transfer matrix $\Db^{p'-1}_1$ vanishes if $\prod_{i=0}^{p'-2} s_{i} = 0$. We therefore have the factorization
\be
\Db^{p'-1}_0 = \Big(\prod_{i=0}^{p'-2} s_{i-2}\Big) \Dbh^{p'-1}_0,
\ee
where the matrix entries of the reduced transfer matrix $\Dbh^{p'-1}_0$ are Laurent polynomials in $\eE^{\ir u}$. 

The bilinear factorization identities \eqref{eq:bilinear.in.D} thus become
\be
\big(s_{j-2} \Dbm^{j}_0\big)^2=\Big(\prod_{i=1}^{p'} s_i^2\Big)\Dbh^{p'-1}_0\Dbh^{p'-1}_{j+1}.
\ee
As a result, we deduce that $\Dbm^j_0$ factorizes as
\be
\Dbm^j_0 = \Big(\prod_{\substack{i=1\\i \neq j-2}}^{p'} \! s_i\, \Big)\, \Dbmh^j_0
\ee
where the matrix entries of the reduced $Q$ matrix $\Dbmh^j_0$ are also Laurent polynomials in $\eE^{\ir u}$. After applying a shift of $-j \lambda$, the bilinear factorization identities \eqref{eq:bilinear.in.D} take the reduced form
\be\label{eq:simpler.bilinear}
(\Dbmh^j_{-j})^2 = \Dbh^{p'-1}_{-j}\Dbh^{p'-1}_{1}.
\ee

\subsection[Baxter $T$-$Q$ eigenvalue relations and eigenvalue decompositions]{Baxter $\boldsymbol T$-$\boldsymbol Q$ eigenvalue relations and eigenvalue decompositions}\label{sec:TQ.Kac}

The left side of \eqref{eq:simpler.bilinear} is a perfect square and is satisfied eigenvalue by eigenvalue. This allows us to make a number of deductions regarding the zeros of $\hat D^{p'-1}(u)$.

{First, we note that, as a polynomial in $z = \eE^{\ir u}$, the maximal degree width of $\hat Q^j(u)$ and $\hat D^{p'-1}(u)$ is $4(N-p'+1)$. For $p'-1>N$, this number is negative. In these cases, the matrices $\Dbm^j(u)$ and $\Db^{p'-1}(u)$ are exactly zero. Indeed, all the representations that come into play for $p'-1>N$ also appear in the RSOS model, for which the fusion hierarchy is known to truncate at fusion level $p'-1$: $\hat D^{p'-1}(u) = 0$. The extended $T$-system identities \eqref{KacTsystem} still hold but are trivially satisfied, with each side equal to zero.}

Here we are interested in $p'$ fixed and $N$ large. For $N \ge p'-1$, the matrices $\Dbm^j(u)$ and $\Db^{p'-1}(u)$ are non-zero, but there are still eigenstates with $D^{p'-1}(u)  = Q^{j}(u) = 0$. These belong to representations of the Temperley-Lieb algebra that also appear in the RSOS model. Our method to construct $Q(u)$ via the extended $T$-system does not apply to these special eigenvalues. From here, we avoid these special representations and restrict our attention to states for which the eigenvalues $D^{p'-1}(u)$ and $Q^{p'-1}(u)$ are both nonzero.

{In this case,}
arguing as in the case of twisted boundary conditions, it follows that $\hat D^{p'-1}(u+\lambda)$ can have double zeros as well as single zeros organized into complete $p'$-strings. For a given eigenstate, {we denote by $E^j$ the set of zeros common to $\hat \calD^j(u-j\lambda)$ and $\hat D^{p'-1}(u+\lambda)$, excluding the complete $p'$ strings. Extending the arguments used in \cref{prop:Ej} to this case, we again find that $E^{j+1}= E^j$. We therefore write $E = E^j$ and define
\be
Q(u) = \prod_{u_m \in\, E} \sin(u-u_m).
\ee
The cardinality of the set $E$ in this case is specific to each eigenvalue, depending on its content in $p'$-strings.} We then have
\be
\hat D^{p'-1}(u+\lambda) = \phi(u) Q(u)^2, \qquad \hat D^{p'-1}(u-j\lambda) = \phi(u) Q(u-(j+1)\lambda)^2,
\ee
where $\phi(u) = \phi(u+\lambda)$ encodes the zeros of the complete $p'$-strings. The bilinear factorization identities then become
\be
\hat \calD^j(u-j\lambda)^2 = \big(\phi(u)Q(u-(j+1)\lambda)Q(u)\big)^2.
\ee
Taking the square root yields the decomposition
\be
\hat \calD^j(u-j\lambda) = R^j\phi(u)Q(u)Q(u-(j+1)\lambda)
\ee
where the $R^j$ are constants satisfying $(R^j)^2=1$. One can again compute the ratios $R^{j+1}/R^j$ by comparing the coefficient of the maximal term in $z = \eE^{\ir u}$. {In this case, because the coefficient of $z^{2(N-p'-1)}$ of each $\hat \calD^j(u-j\lambda)$ is zero, one has to compare the coefficients of the next leading term. Alternatively, this ratio can be obtained from the limit $\eta, \xi \to \infty$ of the same ratios \eqref{eq:R.ratios.Robin} for the Robin boundary conditions, for which the coefficient of the maximal power is non vanishing. We find
\be
\label{eq:R.ratios.diag}
\frac{R^{j+1}}{R^j}=1
\ee
in all cases. We have also confirmed that \eqref{eq:R.ratios.diag} holds for small values of $N$ using our computer implementation.}

Finally, we rewrite the scalar extended $T$-$Q$ relation \eqref{KacExtTQ}  in terms of $\hat \calD^j(u)$ and then $Q(u)$. Dividing the resulting equation throughout by $R^j\phi(u)Q(u-(j+1)\lambda)\prod_{i=1}^{p'}s_i$ gives
\begin{align}
s_{-1} D(u) Q(u)=s_0f(u)Q(u-\lambda)+s_{-2}f(u-\lambda) Q(u+\lambda)
\label{KacScalarTQ}
\end{align}
which is Baxter's $T$-$Q$ relation for the $U_q(s\ell(2))$ invariant/Kac vacuum boundary case. Assuming that the zeros appear in pairs under the crossing symmetry $Q(u)=Q(-u-\lambda)$ gives 
\bea
Q(u) = \prod_{m=1}^M \sin(u-u_m)\sin(u+u_m+\lambda).
\eea
Substituting this into \eqref{KacScalarTQ} and setting $u=u_m$ so that the left side vanishes then gives the Bethe ansatz equations previously obtained in \cite{AlcarazEtAl87,Sklyanin,Nepomechie2002}.

\section{Off-diagonal/Robin vacuum boundary conditions on the strip}\label{sec:Robin}

In this section, we apply our functional equation techniques to the more difficult problem of $s\ell(2)$ models at roots of unity with off-diagonal/Robin vacuum boundary conditions on both the left and right edges of the strip. This very general case reduces to the $U_q(s\ell(2))$ invariant/Kac vacuum boundary conditions of the previous section in the special limit $\eta,\xi \to \ir \infty$.

\subsection{Fusion hierarchies}\label{sec:FH.Robin.2}

For off-diagonal/Robin vacuum boundary conditions, the fusion hierarchy relations are given by
\begin{subequations}
\label{eq:FH.Robin.2}
\begin{alignat}{2}
s_{n-2}s_{2n-1}\Db^n_0\Db^1_n &= s_{n-1}s_{2n-2}f_{n-1}\Db^{n+1}_0 + s_{n-3}s_{2n}f_n g_{n-1} \Db^{n-1}_0,\\[0.15cm]
s_{n}s_{-1}\Db^1_0\Db^n_1 &= s_{n-1}s_{0}f_{0}\Db^{n+1}_0 + s_{n+1}s_{-2} f_{-1} g_0 \Db^{n-1}_2,
\end{alignat}
\end{subequations}
where the functions $f_k$ and $s_k$ are defined in \eqref{eq:Kac.convention} and \eqref{eq:Df}. The function $g_k$ is defined by
\be
g_k = \Gamma_L(u+k\lambda)\Gamma_L(-u-k\lambda)\Gamma_R(u+k\lambda)\Gamma_R(-u-k\lambda),
\ee
with $\Gamma_{L,R}(u)$ given respectively, for the vertex and loop models, by \eqref{eq:Gamma.6V} and \eqref{eq:Gamma.loops}.
We use the notation
\be
\Db^{-1}_k=0, \qquad \Db^0_k=f_{k-1} \Ib,\qquad \Db^n_k = \Db^n(u+k \lambda),
\ee
as well as the convention 
\be
\label{eq:neg.D.Robin2}
\Db^n_0 := - \Big(\prod_{j=n}^{-2}g_j^{-1}\Big)\Db^{-2-n}_{n+1},\qquad n\le -2
\ee
for matrices with negative fusion labels. The fusion hierarchy relations \eqref{eq:FH.Robin.2} hold for arbitrary values of~$\lambda$. The proof is obtained by extending the arguments given in Appendix~E.1 of \cite{MDPR} to include the surface functions~$g_j$.

\subsection[Closure, the infinite fusion limit and extended $Q$ matrices]{Closure, the infinite fusion limit and extended $\boldsymbol Q$ matrices}\label{sec:T.Robin.2}

For the roots of unity cases $\lambda = \lambda_{p,p'}$, we conjecture that the fused transfer matrices satisfy the closure relation 
\begin{subequations}
\label{eq:closure.Robin.2}
\be
\Db^{p'}_0=(a_+ + a_- - a_0) \Db^0_0+ g_{-1}\Db^{p'-2}_1
\ee
where 
\be
a_{\pm} = (-1)^{p'-p}\prod_{j=1}^{p'} \Gamma_L\big(\!\pm (u+j\lambda)\big)\Gamma_R\big(\!\pm (u+j\lambda)\big), \qquad a_0 = \Kb \prod_{j=1}^{p'} s_j^2.
\ee
\end{subequations}
For the six-vertex model, $\Kb$ is proportional to the identity matrix, with the overall constant $K$ given by
\begin{equation}
  \label{eq:K.6V}
  K = (-1)^{Np}\,\left((\mu_2\nu_2)^{p'} + \frac{1}{(\mu_2\nu_2)^{p'}}\right)
    - (-1)^{p}\,\left((\mu_1\nu_1)^{p'} + \frac{1}{(\mu_1\nu_1)^{p'}}\right)\,.
\end{equation}
For the loop model, $\Kb$ is an element of $\mathsf{TL}^{\textrm{\tiny$(2)$}}_N(\beta, \alpha_1, \alpha_2, \beta_1, \beta_2, \gamma)$ that is independent of $u$. On the standard modules $\mathsf {V}^{\textrm{\tiny $(2)$}}_{N,d}$, $\Kb$ is proportional to the identity matrix, with an eigenvalue $K_d$ that is independent of $N$ and given by
\be
\label{eq:Kd.def}
K_0 = \prod_{\ell=0}^{p'-1} \Big(\gamma^2 + (q^\ell - q^{-\ell})(q^\ell b_+ - q^{-\ell}b_-) \Big), \quad \qquad K_{d} = 0 \qquad \textrm{for } d > 0,
\ee 
where $q = \eE^{\ir \lambda}$ and 
\be
b_\pm = \bigg(\frac{\alpha_1 q^{\pm1/2}-\alpha_2 q^{\mp1/2}}{q-q^{-1}}\bigg)\bigg(\frac{\beta_1 q^{\pm1/2}-\beta_2 q^{\mp1/2}}{q-q^{-1}}\bigg).
\ee
We do not currently have a proof of this closure relation. As evidence of the conjecture \eqref{eq:closure.Robin.2}, we have confirmed that (i) it holds for small values of $N$ using a computer program, (ii) it holds in the braid limit, (iii) it reduces to \eqref{eq:closure.D} in the limits $\eta,\xi \to \ir \infty$, and (iv) it is consistent with the crossing and periodicity symmetries. For $p' = p+1$, the closure condition \eqref{eq:closure.Robin.2} is equivalent to the similar relations obtained in \cite{Nepomechie2002}.
 
More generally, the fused transfer matrices $\Db^{yp' + j}_0$, for $j = 0, \dots, p'-1$, satisfy the closure relations
\be
\Db^{yp'+j}_0 = \mathcal U_y \Db^j_0+ \mathcal U_{y-1} \Big(\prod_{i=0}^j g_{i-1}\Big) \Db^{p'-2-j}_{j+1},
\ee
where $\mathcal U_\ell$ is related to Chebyshev polynomials  of the second kind:
{
\be
\mathcal U_\ell = (a_+a_-)^{\ell/2}\, U_\ell\bigg(\frac{a_+ + a_- - a_0}{2(a_+a_-)^{1/2}}\bigg).
\ee
}
This can be rewritten as
\be
\mathcal U_\ell = \frac{c_+^{\ell+1}-c_-^{\ell+1}}{c_+ - c_-}
\ee
where
\be
c_\pm = \frac12\Big(a_+ + a_- - a_0 \pm \sqrt{(a_+ + a_- - a_0)^2-4 a_+ a_-}\Big).
\ee
Because of the presence of the square root singularity, $c_+$ and $c_-$ are
not in general analytic functions of $\eE^{\ir u}$. One exception is the
special case $a_0=0$, for which the argument of the square root is a perfect
square and
\be
c_\pm\big|_{a_0=0} = a_\pm. 
\ee

For $\lambda = \lambda_{p,p'}$, we define the extended $Q$ matrices $\Dbm^{j,\pm}_0$ by
\be\label{eq:def.Dbm.Robin.2}
\Dbm^{j,\pm}_0 = c_\pm \Db^j_0 + \Big(\prod_{i = 0}^j g_{i-1}\Big) \Db^{p'-2-j}_{j+1}.
\ee
These share the properties of $c_\pm$ and $c_0$ in being non-analytic functions of $\eE^{\ir u}$ for $a_0\ne 0$. They satisfy the following periodicity, crossing and conjugacy properties:
\begin{subequations}\label{eq:symmetries.Robin.2}
{
\begin{alignat}{2}
&\Dbm^{j,\pm}(u+\pi) = \Dbm^{j,\pm}(u), \qquad \Dbm^{p'+j,\pm}(u) = c_\pm \Dbm^{j,\pm}(u), \qquad \Dbm^{j,+}\big((2-j)\lambda - u\big) = \Dbm^{j,-}(u), 
\label{eq:symmetries.a.Robin.2}\\[0.2cm]
&\Dbm^{p'-j-2,+}(u) = \Dbm^{j,-}(u-(j+1)\lambda)\, \frac{\prod_{k = 0}^{p'-j-2}\Gamma_{L}(u+k \lambda)\Gamma_{R}(u+k \lambda)}{\prod_{k = p'-j-1}^{p'-1}\Gamma_{L}(-u-k\lambda)\Gamma_{R}(-u-k\lambda)}.
\label{eq:symmetries.b.Robin.2}
\end{alignat}
} 
\end{subequations}

If $|c_+/c_-|> 1$,  
then the extended $Q$ matrices $\Dbm^{j,\pm}_0$ are given by taking the infinite fusion limit $n\to \infty$ of $\Db^n_0$ through suitable subsequences:
\be
\Dbm^{j,\pm}_0=\lim_{y\to \pm \infty} \frac{\Db^{yp'+j}_0}{\mathcal U_{y-1}}.
\ee
If $|c_+/c_-|< 1$, 
the same results hold but with $\Dbm^{j,+}_0$ and $\Dbm^{j,-}_0$ interchanged.

\subsection[Extended $T$-system]{Extended $\boldsymbol T$-system}\label{sec:T.Robin.3}

The extended $Q$ matrices $\Dbm^{j,\pm}_0$ together satisfy the bilinear factorization identities
\be
\label{eq:bilinear.Robin.2}
s_{j-2}^2 \Dbm^{j,+}_0\Dbm^{j,-}_0 = s_{-3}s_{2j-1} \Big(\prod_{i=0}^j g_{i-1}\Big) \Db^{p'-1}_{0}\Db^{p'-1}_{j+1}.
\ee
This is a special case of the extended $T$-system
\be
\label{eq:extended.T.Robin.2}
\Db^{j+k}_{-j}\Db^{p'-1}_1=\frac{s_{k-1}s_{-2-j}}{s_{k-j-2}s_{-1}}\bigg(\Big(\prod_{i=0}^{k-1}g_i \Big)\Db^j_{-j} \Db^{p'-1-k}_{k+1} + (a_+ + a_- -a_0) \Db^j_{-j} \Db^{k-1}_{1} + \Big(\prod_{i=p'-j-1}^{p'-1}\hspace{-0.2cm}g_i\, \Big)\Db^{k-1}_{1}\Db^{p'-2-j}_{1}\bigg).
\ee
Indeed, for $k = p'-1-j$, this reduces to \eqref{eq:bilinear.Robin.2} using the identities
\be
c_+ c_- = a_+ a_- =  \prod_{i=1}^{p'} g_i, \qquad c_+ + c_- = a_+ + a_- - a_0.
\ee
We do not give the explicit proof of the extended $T$-system of bilinear identities \eqref{eq:extended.T.Robin.2} since they involve a straightforward extension of the arguments of \cref{AppA.boundary} to include the surface functions~$g_i$. We also note that \eqref{eq:extended.T.Robin.2} holds for $j,k \in \mathbb Z$ provided that the improper products appearing there are interpreted in accord with the conventions:
\be
\prod_{i = 0}^{-1} g_i := 1,\qquad\prod_{i = 0}^\ell g_i := \prod_{i = \ell + 1}^{-1} g_i^{-1} \quad \textrm{for}\quad \ell \le -2.
\ee

\subsection[Factorizations and extended $T$-$Q$ relations]{Factorizations and extended $\boldsymbol T$-$\boldsymbol Q$ relations}\label{sec:factors.Robin.2}

The matrices $\Dbm^{j,\pm}_0$ also satisfy the following {\em generalized} $T$-$Q$ relations:
\begin{subequations}
\begin{alignat}{2}
s_{-j-2}s_{-1}\Db_0^1 \Dbm^{j,\pm}_{-j} &= s_{-j-3}s_0 f_0 g_{-1} \Dbm^{j-1,\pm}_{-j} + s_{-j-1}s_{-2} f_{-1} \Dbm^{j+1,\pm}_{-j},\label{eq:DD.Robin.2}\\[0.15cm]
s_{j}s_{-1}\Db_0^1 \Dbm^{j,\pm}_{1} &= s_{j+1}s_{-2} f_{-1} g_{0} \Dbm^{j-1,\pm}_{2} + s_{j-1}s_{0} f_{0} \Dbm^{j+1,\pm}_{0}.
\end{alignat}
\end{subequations}
As in previous cases, these follow simply from the fusion hierarchy relations \eqref{eq:FH.Robin.2}. These relations are not the {\em extended} 
$T$-$Q$ relations we seek since the prefactors depend on $j$. This will be remedied by removing simple factors in $\Dbm^{j,\pm}(u)$.

Indeed, the matrices $\Db^{p'-1}(u)$ and $\Dbm^{j,\pm}(u)$ have a number of simple overall factors arising from the surface terms in the fusion hierarchy. By setting $k=-j$ in \eqref{eq:extended.T.Robin.2}, we find that $f_{-j-1}\Db^{p'-1}_1$ is proportional to $\frac{s_{-j-1}s_{-j-2}}{s_{-2j-2}s_{-1}}$, for $j = 1, \dots, p'$, as in the $U_q(s\ell(2))$ invariant/Kac vacuum case. We conclude that $\Db^{p'-1}_0$ can be written as 
\be
\Db^{p'-1}_0 = \Big(\prod_{i=0}^{p'-2} s_{i-2}\Big) \Dbh^{p'-1}_0,
\ee
where the matrix entries of $\Dbh^{p'-1}_0$ are Laurent polynomials in $\eE^{\ir u}$. The bilinear factorization identities \eqref{eq:bilinear.Robin.2} then become
\be
s_{j-2}^2\Dbm^{j,+}_0\Dbm^{j,-}_0=\Big(\prod_{i=1}^{p'} s_i^2\Big)\Big(\prod_{i=0}^j g_{i-1}\Big)\Dbh^{p'-1}_0\Dbh^{p'-1}_{j+1}.\label{LHSRHS}
\ee

The trigonometric factors appearing on the right side of \eqref{LHSRHS} must be divided between the matrices $\Dbm^{j,+}_0$ and $\Dbm^{j,-}_0$ on the left side. To understand how this division occurs, we first note that $\Dbm^{j,+}_0$ has an overall factor of $\prod_{i=0}^j \Gamma_L(u_{i-1})\Gamma_R(u_{i-1})$, {where we use the abbreviation $u_k = u+k\lambda$}. This happens because, in the definition \eqref{eq:def.Dbm.Robin.2}, both $c_+$ and $\prod_{i=0}^j g_{i-1}$ vanish under the condition $\prod_{i=0}^j \Gamma_L(u_{i-1})\Gamma_R(u_{i-1}) = 0$. Likewise, $\Dbm^{j,-}_0$ has an overall factor of $\prod_{i=0}^j \Gamma_L(-u_{i-1})\Gamma_R(-u_{i-1})$. 

The other trigonometric terms are divided evenly between $\Dbm^{j,+}_0$ and $\Dbm^{j,-}_0$. To see this, suppose that $\Dbm^{j,-}_0$ vanishes for $s_i=0$, for some $i \in \{j-1, \dots, p'+j-3\}$. From the definition of the extended $Q$ matrices, we arrive at the equality
\be
\label{eq:QminusQ}
\Dbm^{j,+}_0 - \Dbm^{j,-}_0 = (c_+ - c_-)\Db^j_0.
\ee
It follows that $(a_+ - a_-) = 0$ and $a_0 = 0$ for $s_i = 0$. Under this condition, $c_+ = c_-$ and the right side of \eqref{eq:QminusQ} vanishes. With the assumption that $\Dbm^{j,-}_0$ vanishes for $s_i=0$, the left side reduces to $\Dbm^{j,+}_0$ and therefore vanishes as well. This confirms that a factor of $s_i$ can also be factorized from $\Dbm^{j,+}_0$. The argument is identical if we initially assume that $\Dbm^{j,+}_0$ vanishes for $s_i=0$. We therefore have
\be\label{eq:renormalizeQ}
\Dbm^{j,\pm}_0 = \Big(\prod_{i=j-1}^{p'+j-3} \! s_i\, \Big)\Big(\prod_{i=0}^j \Gamma_L(\pm u_{i-1})\Gamma_R(\pm u_{i-1})\Big)\, \Dbmh^{j,\pm}_0,
\ee
where, in the general case, $\Dbmh^{j,+}_0$ and $\Dbmh^{j,-}_0$ are non-analytic functions in $\eE^{\ir u}$ that are finite in the complex $u$-plane. Applying a shift of $-j \lambda$, the bilinear factorization identities \eqref{LHSRHS} simplify to
\be\label{eq:simpler.bilinear.Robin.2}
\Dbmh^{j,+}_{-j}\Dbmh^{j,-}_{-j} = \Dbh^{p'-1}_{-j}\Dbh^{p'-1}_{1}.
\ee
Similarly, \eqref{eq:DD.Robin.2} simplifies to give the sought-after extended $T$-$Q$ relations:
\be\label{eq:extended.TQ.Robin.2}
s_{-1}\Db_0^1 \Dbmh^{j,\pm}_{-j} = s_0 f_0 \Gamma_L\big(\!\pm\!(\lambda-u)\big)\Gamma_R\big(\!\pm\!(\lambda-u)\big)\Dbmh^{j-1,\pm}_{-j} + s_{-2} f_{-1} \Gamma_L(\pm u)\Gamma_R(\pm u)\Dbmh^{j+1,\pm}_{-j}.
\ee

\subsection[Baxter $T$-$Q$ eigenvalue relations and eigenvalue decompositions for $a_0=0$]{Baxter $\boldsymbol{T}$-$\boldsymbol{Q}$ eigenvalue relations and eigenvalue decompositions for $\boldsymbol{a_0=0}$}

The bilinear factorization identities \eqref{eq:simpler.bilinear.Robin.2} hold at the level of the eigenvalues of the corresponding transfer matrices. For $a_0 = 0$, the eigenvalues of $\Dbmh^{j,\pm}(u)$ are Laurent polynomials in $\eE^{\ir u}$ and the standard analysis of $T$-$Q$ relations applies. This does not apply for $a_0\ne 0$. This case is discussed separately in the next section.

{For the vertex model, $a_0$ is equal to zero if the parameters $\mu_1$, $\mu_2$ and $\nu_1$, $\nu_2$ are fixed such that the right side of \eqref{eq:K.6V} is zero.} For the loop model, $a_0$ vanishes (i) for eigenvalues in a standard module $\mathsf V_{N,d}$ with $d>0$, and (ii) for eigenvalues in the standard module $\mathsf V_{N,0}$ with $\gamma$ specialized such that $K_0$, defined in \eqref{eq:Kd.def}, vanishes. For both the vertex and loop models, it also applies in the limiting case $\eta \to \ir \infty$ and/or $\xi \to \ir \infty$, corresponding to $U_q(s\ell(2))$ invariant/Kac vacuum boundary conditions on one or both edges of the strip. In this last case, it is readily verified that the $a_0$ term in \eqref{eq:closure.Robin.2} is absent in the limit. 

For the specialization $a_0=0$, we now make a number of deductions about the eigenvalues of $\Dmh^{j,\pm}(u)$ and $\hat D^{p'-1}(u)$. As Laurent polynomials in $\eE^{\ir u}$, their degree width is $4N$ which is thus reduced compared to $\Dm^{j,\pm}(u)$ and $D^{p'-1}(u)$. Therefore, \eqref{eq:bilinear.Robin.2} is an equality between two Laurent polynomials of degree width $8N$. Let us denote by $\{v_k\}$ the set of zeros of $\hat D^{p'-1}(u+\lambda)$ in the complex $u$-plane, and likewise by $\{w_\ell^{j,\pm}\}$ the sets of zeros of $\hat \calD^{j,\pm}(u-j \lambda)$. The bilinear factorization identities imply that
\be
\label{eq:samezeros.Robin.2}
\prod_\ell \sin(u-w^{j,+}_\ell)\sin(u-w^{j,-}_\ell) = \prod_k \sin(u-v_k)\sin\!\big(u-(j+1)\lambda -v_k\big).
\ee
For generic values of the parameters, we conjecture that the zeros are all simple and that there are no complete $p'$-strings. Each zero $u=u_m$ of \eqref{eq:samezeros.Robin.2} belongs to $\{v_k\}$ or $\{v_k+(j+1)\lambda\}$, and likewise belongs to $\{w_\ell^{j,+}\}$ or $\{w_\ell^{j,-}\}$. So we define the {sets $E^{j,\pm} = \{v_k\} \cap \{w_\ell^{j,\pm}\}$. Repeating the proof of \cref{prop:Ej} for this case, we again find that $E^{j+1,\pm} = E^{j,\pm}$. We thus write $E^\pm = E^{j,\pm}$. We find from our computer implementation that, for generic values of the parameters, these sets have cardinalities $|E^\pm| = 2N$.} The corresponding $Q^\pm(u)$ functions are defined as
\be
Q^+(u) = \prod_{u_m \in E^+} \sin(u-u_m), \qquad Q^-(u) = \prod_{u_m \in E^-} \sin(u-u_m),
\ee
where the $u_m$ are the Bethe roots.
With these definitions, we have the decompositions
\begin{subequations}
\begin{alignat}{2}
\hat \calD^{j,+}(u-j\lambda) &= R^{j,+}(u)Q^+(u)Q^-\big(u-(j+1)\lambda\big), \label{eq:DQQp.2}\\  
\hat \calD^{j,-}(u-j\lambda) &=  R^{j,-}(u)Q^-\big(u\big)Q^+(u-(j+1)\lambda), \label{eq:DQQm.2}\\
\hat D^{p'-1}(u+\lambda) &= \phi(u) Q^+(u)Q^-(u),\\  
\hat D^{p'-1}(u-j\lambda) &= \phi(u)Q^+\big(u-(j+1)\lambda\big)Q^-\big(u-(j+1)\lambda\big).
\end{alignat}
\end{subequations}
{The division of the $Q^\pm\big(u-(j+1)\lambda\big)$ factors between $\hat \calD^{j,+}(u-j\lambda)$ and $\hat \calD^{j,0}(u-j\lambda)$ is dictated by the conjugacy property \eqref{eq:symmetries.b.Robin.2}.}
In the generic case where $a_+ \neq a_-$, there are no complete $p'$-strings, so $R^{j,\pm}(u)$ and $\phi(u)$ are constants and satisfy $R^{j,+}(u)R^{j,-}(u) = \phi^2(u)$. {In this case, the braid limits of $\hat \calD^{j,\pm}(u-j\lambda)$ are non-zero, allowing us to compute the ratios of consecutive $R^{j,\pm}(u)$ constants. We find
\be
\label{eq:R.ratios.Robin}
\frac{R^{j+1,\pm}(u)}{R^{j,\pm}(u)}=1.
\ee
}
Baxter's $T$-$Q$ eigenvalue relation takes the form
\be
s_{-1}D(u) Q^\pm(u)=s_{0}f_0 \Gamma_L\big(\!\pm\!(\lambda-u)\big)\Gamma_R\big(\!\pm\!(\lambda-u)\big) Q^\pm(u-\lambda)+s_{-2}f_{-1} \Gamma_L(\pm u)\Gamma_R(\pm u) Q^\pm(u+\lambda).\label{eq:tqfinal.Robin.2}
\ee
This is obtained by considering \eqref{eq:extended.TQ.Robin.2} at the level of the eigenvalues, rewriting the relation in terms of the functions $Q^\pm(u)$ using \eqref{eq:DQQp.2} and \eqref{eq:DQQm.2} and dividing throughout by the factor $R^{j,\pm}(u)Q^\pm\big(u-(j+1)\lambda\big)$.
Remarkably, the dependence on $j$ again disappears in \eqref{eq:tqfinal.Robin.2}. The only difference between the $T$-$Q$ relations for this $a_0=0$ case and the simpler case studied in \cref{sec:strip} are the extra factors of $\Gamma_L\big(\!\pm\!(\lambda-u)\big)\Gamma_R\big(\!\pm\!(\lambda-u)\big)$ and $\Gamma_L(\pm u)\Gamma_R(\pm u)$ on the right side. {Setting $u=u_m$ in \eqref{eq:tqfinal.Robin.2} yields the Bethe ansatz equations for the Robin boundary case.}

The case $a_+ = a_-$ is special and gives rise to double zeros and complete $p'$-strings, as in this case $\Dbmh^{j,+}(u) = \Dbmh^{j,-}(u)$ and the left side of \eqref{eq:bilinear.Robin.2} is a perfect square. In this case, the functions $R^{j,\pm}(u)$ and $\phi(u)$ are not constants and account for the zeros of the complete $p'$-strings. The analysis of this subcase is identical to the $U_q(s\ell(2))$ invariant/Kac vacuum boundary case presented in \cref{sec:TQ.Kac}.

\subsection[Extended $T$-$Q$ relations for $a_0 \neq 0$]{Extended $\boldsymbol{T}$-$\boldsymbol{Q}$ relations for $\boldsymbol{a_0\neq 0}$}

The results of the previous section are consistent with those of \cite{Caoetal03,Nepo04,YNZ2006,YZ2006}, where it was found that the transfer matrix eigenvalues satisfy Baxter's $T$-$Q$ relations provided that the parameters of the $K$ matrices satisfy a specific identity, here written as $a_0=0$. In this section, we investigate the case  $a_0 \neq 0$, where the eigenvalues $\Dmh^{j,+}(u)$ and $\Dmh^{j,-}(u)$ are not analytic functions of $\eE^{\ir u}$. The bilinear factorization identity \eqref{eq:simpler.bilinear.Robin.2} holds in this case as well. At the level of the eigenvalues of the corresponding transfer matrices, the right side is a Laurent polynomial of degree $8N$. The left side, although it takes the form of a product of two functions that are not analytic in $z = \eE^{\ir u}$, is also a Laurent polynomial of degree $8N$. Because of their non-analyticity, we are no longer able to factorize $\Dmh^{j,+}(u)$ and $\Dmh^{j,-}(u)$ as products of two Laurent polynomials, as in \eqref{eq:DQQp.2} and \eqref{eq:DQQm.2}.

Let us therefore return to the functional equations at the level of matrices. Following \cite{MurganNS2006}, the relations \eqref{eq:DD.Robin.2}, involving non analytic functions, can be replaced by a set of similar relations involving only analytic functions. To this end, one defines two new objects $\Dbm^{j,0}(u)$ and $\Dbm^{j,1}(u)$ satisfying
\be
\label{eq:QDelta}
\Dbm^{j,\pm}(u) = \Dbm^{j,0}(u) \pm \frac12\sqrt{\Delta} \Dbm^{j,1}(u), \qquad \Delta = (a_+ +a_- - a_0)^2 - 4 a_+ a_-.
\ee
Their explicit definition is
\be
\Dbm^{j,0}(u) = \frac12(a_+ + a_- - a_0)\Db^j(u) + \Big(\prod_{i=0}^j g_{i-1}\Big)\Db^{p'-2-j}\big(u+(j+1)\lambda\big), \qquad \Dbm^{j,1}(u) = \Db^j(u).
\ee
Both are Laurent polynomials in the variable $\eE^{\ir u}$. The non-analyticity of the problem is captured by the function $\sqrt\Delta$. From the periodicity property given in \eqref{eq:symmetries.a.Robin.2}, we can restrict our focus to the functions $\Dbm^{j,\pm}(u)$ for $j = 1, \dots, p'$. The similar periodicity properties for $\Dbm^{j,0}(u)$ and $\Dbm^{j,1}(u)$ are
\begin{subequations}
\begin{alignat}{2}
\Dbm^{p'+j,0}(u) &= \frac12(a_+ + a_- - a_0) \Dbm^{j,0}(u) + \frac \Delta 4 \Dbm^{j,1}(u),\\[0.15cm]
\Dbm^{p'+j,1}(u) &= \Dbm^{j,0}(u) +  \frac12(a_+ + a_- - a_0) \Dbm^{j,1}(u).
\end{alignat}
\end{subequations}
We obtain generalized $T$-$Q$ relations for $\Dbm^{j,0}(u)$ and $\Dbm^{j,1}(u)$ by equating separately the terms  in \eqref{eq:DD.Robin.2} that are analytic and those that have the prefactor $\sqrt \Delta$. We find
\begin{subequations}
\be
s_{-j-2}s_{-1}\Db_0^1 \Dbm^{j,\ell}_{-j} = s_{-j-3}s_0 f_0 g_{-1} \Dbm^{j-1,\ell}_{-j} + s_{-j-1}s_{-2} f_{-1} \Dbm^{j+1,\ell}_{-j}, \qquad \ell = 0, 1, \quad j = 2, \dots, p'-1,
\ee
and
\begin{alignat}{2}
s_{-3}s_{-1}\Db^1_0 \Dbm^{1,0}_{-1} &= \frac{s_{-4}s_0 f_0 g_{-1}}{a_+ a_-}\Big(\frac12(a_+ + a_- - a_0)\Dbm^{p',0}_{-1} - \frac\Delta4\Dbm^{p',1}_{-1}\Big) + s_{-2}^2 f_{-1} \Dbm^{2,0}_{-1},\\[0.15cm]
s_{-3}s_{-1}\Db^1_0 \Dbm^{1,1}_{-1} &= \frac{s_{-4}s_0 f_0 g_{-1}}{a_+ a_-}\Big(\frac12(a_+ + a_- - a_0)\Dbm^{p',1}_{-1} - \Dbm^{p',0}_{-1}\Big) + s_{-2}^2 f_{-1} \Dbm^{2,1}_{-1},\\[0.15cm]
s_{-2}s_{-1}\Db^1_0 \Dbm^{p',0}_{0} &= s_{-3}s_0 f_0 g_{-1}\Dbm^{p'-1,0}_0 + s_{-1}s_{-2} f_{-1} \Big(\frac12(a_+ + a_- - a_0) \Dbm^{1,0}_0 + \frac \Delta 4 \Dbm^{1,1}_0\Big),\\[0.15cm]
s_{-2}s_{-1}\Db^1_0 \Dbm^{p',1}_{0} &= s_{-3}s_0 f_0 g_{-1}\Dbm^{p'-1,1}_0 + s_{-1}s_{-2} f_{-1} \Big(\frac12(a_+ + a_- - a_0) \Dbm^{1,1}_0 + \Dbm^{1,0}_0\Big).
\end{alignat}
\end{subequations}

As argued in \cref{sec:factors.Robin.2}, $\Dbm^{j,\pm}(u) = 0$ if $u$ is specialized to values for which $\prod_{\substack{i=1, i \neq j-2}}^{p'} s_i$ vanishes. The same property is transferred to $\Dbm^{j,0}(u)$ and $\Dbm^{j,1}(u)$. However, the renormalized factors involving the functions $\Gamma_L(u)$ and $\Gamma_R(u)$ in \eqref{eq:renormalizeQ} are different for $\Dbm^{j,+}(u)$ and $\Dbm^{j,-}(u)$, so there is no such factorization for $\Dbm^{j,0}(u)$ and $\Dbm^{j,1}(u)$. We thus define
\be
\Dbm^{j,\ell}_0 = \Big(\prod_{\substack{i=1\\[0.05cm]i \neq j-2}}^{p'} \! s_i\, \Big) \Dbmh^{j,\ell}_0,\qquad \ell = 0,1.
\ee
The matrices $\Dbmh^{j,0}_0$ and $\Dbmh^{j,1}_0$ are analytic, finite in the complex $u$-plane and satisfy the relations
\begin{subequations}
\begin{alignat}{2}
s_{-1}\Db_0^1 \Dbmh^{j,\ell}_{-j} &= s_0 f_0 g_{-1} \Dbmh^{j-1,\ell}_{-j} + s_{-2} f_{-1} \Dbmh^{j+1,\ell}_{-j}, \qquad \ell = 0, 1, \quad j = 2, \dots, p'-1,\\[0.15cm]
s_{-1}\Db^1_0 \Dbmh^{1,0}_{-1} &= \frac{s_0 f_0 g_{-1}}{a_+ a_-}\Big(\frac12(a_+ + a_- - a_0)\Dbmh^{p',0}_{-1} - \frac\Delta4\Dbmh^{p',1}_{-1}\Big) + s_{-2} f_{-1} \Dbmh^{2,0}_{-1},\\[0.15cm]
s_{-1}\Db^1_0 \Dbmh^{1,1}_{-1} &= \frac{s_0 f_0 g_{-1}}{a_+ a_-}\Big(\frac12(a_+ + a_- - a_0)\Dbmh^{p',1}_{-1} - \Dbmh^{p',0}_{-1}\Big) + s_{-2} f_{-1} \Dbmh^{2,1}_{-1},\\[0.15cm]
s_{-1}\Db^1_0 \Dbmh^{p',0}_{0} &= s_0 f_0 g_{-1}\Dbmh^{p'-1,0}_0 + s_{-2} f_{-1} \Big(\frac12(a_+ + a_- - a_0) \Dbmh^{1,0}_0 + \frac \Delta 4 \Dbmh^{1,1}_0\Big),\\[0.15cm]
s_{-1}\Db^1_0 \Dbmh^{p',1}_{0} &= s_0 f_0 g_{-1}\Dbmh^{p'-1,1}_0 + s_{-2} f_{-1} \Big(\frac12(a_+ + a_- - a_0) \Dbmh^{1,1}_0 + \Dbmh^{1,0}_0\Big).
\end{alignat}
\end{subequations}
These hold at the level of the eigenvalues as well. We therefore recover the {\it generalized} $T$-$Q$ relations (3.40) and (3.41) of Murgan, Nepomechie and Shi \cite{MurganNS2006}. Their results were obtained via an ansatz (see their equation (3.37)) which, in the present context, is simply the decomposition \eqref{eq:QDelta}.

We note that the results of Murgan, Nepomechie and Shi \cite{MurganNS2006} that were restricted to the discrete set of values $\lambda=\pi/(p+1)$ with $p=1,2,\ldots$ are now extended to all roots of unity $\lambda=\lambda_{p,p'}$.

\section{Conclusion}\label{sec:conclusion}

In this paper, we have considered the mutually commuting $1\times n$ fused single and double-row transfer matrices of the critical six-vertex and dense loop models at roots of unity $q=\eE^{\ir\lambda}$ with $\lambda$ given by \eqref{rational}. 
These points represent a countable dense set of points on the critical line of the six-vertex model. The dense loop models at roots of unity coincide with the logarithmic minimal models ${\cal LM}(p,p')$. 
For these $s\ell(2)$ models we used extended $T$-systems of bilinear identities to obtain the extended $T$-$Q$ relations \eqref{extTQ}, \eqref{KacExtTQ} and \eqref{eq:extended.TQ.Robin.2}. 
The extended $Q$ matrices $\vec\calT^{j,\pm}(u)$ and $\vec\calT^{j}(u)$ appearing in these equations are unambiguously identified as explicit linear combinations of standard fused transfer matrices $\vec T^j(u)$ with locally defined Boltzmann face weights.
We also deduce the usual Baxter $T$-$Q$ relations for eigenvalues.
An obvious advantage of our approach, compared to some other approaches, is that no reference state is needed. 
We also argue that, together, the extended $T$-systems of bilinear identities and the extended $T$-$Q$ relations reflect the extended $s\ell(2)$ loop algebra symmetries present at roots of unity.

It could be said that, in some measure, Baxter's $Q$ matrix has remained somewhat mysterious for more than four decades. 
Indeed the standard approach, as in \cite{BazhMang2007}, is to assume the existence of a $Q$ matrix satisfying Baxter's $T$-$Q$ relation and subject to certain analyticity properties. The usual results then follow but the analyticity assumptions are difficult to justify. 
In our approach, we instead use the extended $T$-system bilinear identities to derive extended $T$-$Q$ relations.
In stark contrast, our uniquely defined extended $Q$ matrices are no longer auxiliary quantities and there is nothing mysterious about them. The analyticity properties of the extended $Q$ matrices are directly accessible.
Moreover, at the level of spectra, we have shown that the extended $T$-$Q$ relations imply the usual Baxter $T$-$Q$ relation. 
These results are of particular interest for the dense loop models since the construction of $Q(u)$ and a derivation of the $T$-$Q$ relations were previously unknown for these models.

A remarkable consequence of our approach is that, for modest system sizes $N$, the eigenvalues of the extended $Q$ matrices and their zeros can be explicitly and directly obtained numerically on a computer, for both the six-vertex and loop model. In particular, all of the Bethe roots appear among these zeros! These zeros may also include complete $p'$ strings but, in this case, the zeros of the complete $p'$ strings appear in distinguished positions determined by the transfer matrix eigenvalues on the right side of the bilinear factorization identities \eqref{eq:bilinear.in.T}, \eqref{eq:bilinear.in.D} and \eqref{eq:bilinear.Robin.2} and not in continuous positions as allowed~\cite{GainutdinovNepomechie2016} by the Baxter $T$-$Q$ relation. In this way the patterns of zeros of the eigenvalues $\calT^{j,\pm}(u)$ and $Q^\pm(u)$ can be studied, for small system sizes $N$, without worrying about missing some eigenvalues~\cite{GHNS2015} or dealing with unphysical solutions~\cite{HaoNepomechieS2014,NepomechieWang2014} to the Bethe ansatz equations. Indeed, this opens up the possibility that the patterns of zeros of $\calT^{j,\pm}(u)$,  and $Q^\pm(u)$ can be completely classified in the same way that the patterns of zeros of $T^j(u)$ with $j=1,2$ are classified~\cite{MDKP2017} for critical percolation. 
Another advantage of the current approach is that all the $T$-systems considered here share a common universal $D$-type $Y$-system that closes finitely and, at least in principle, can be solved analytically for the conformal spectra using the nonlinear integral equation and dilogarithm techniques of \cite{KlumperP,MDKP2017}. 
The $Y$-system is universal~\cite{universal} in the sense that it holds for all boundary conditions and all topologies. 
The disadvantage of the methods of this paper, which are built on functional equations, is that they only give access to eigenvalue spectra in comparison with the algebraic Bethe ansatz which also gives access to eigenstates.

Lastly, the approach of this paper is also expected to generalize to more general $(r,s)$ type boundary conditions~\cite{rsBoundaries1,rsBoundaries2} and to other models at roots of unity including the higher level fused $s\ell(2)$ systems~\cite{PRTsuper2014,MDPR} with transfer matrices $\vec T^{m,n}(u)$, the symmetric eight-vertex model~\cite{BaxterQ,Baxter73} and the $A_2^{(1)}$ or $s\ell(3)$ models~\cite{MDPR2018}.

\goodbreak

\section*{Acknowledgments} 

AMD and PAP acknowledge the hospitality of the University of Hannover where this work was initiated.
AMD is an FNRS Postdoctoral Researcher under the project CR28075116. He acknowledges support from the EOS-contract O013018F and from the University of Melbourne, where parts of this work were carried out.
HF is a member of the DFG funded research unit \emph{Correlations in Integrable Quantum Many-Body Systems} (FOR2316). This paper was finished while PAP was visiting the APCTP.

\goodbreak
\appendix

\section{Diagrammatic Temperley-Lieb algebras}\label{app:diag.algebras}

In this section, we review the definition of the diagrammatic algebras $\tl_N(\beta)$, $\eptl_N(\alpha,\beta)$ and $\tl^{\textrm{\tiny$(2)$}}_N(\beta, \alpha_1, \alpha_2, \beta_1, \beta_2, \gamma)$ used to describe the dense loop model on the cylinder and strip. 
These ``linear" algebras correspond to planar algebras~\cite{JonesPlanar} in which the action is restricted to a fixed direction. 
The analogous planar diagrammatic algebra for the vertex models is just the planar algebra of local tensor contractions on local spin states.

\paragraph{Temperley-Lieb algebra.} On the strip, the dense loop models with Kac vacuum boundary conditions are described by the Temperley-Lieb algebra $\tl_N(\beta)$ \cite{TL71,Jones83,M91,GW93,W95,RSA14}. This is a diagrammatic algebra defined by
\be
 \tl_N(\beta)=\big\langle I,\,e_j ;\,j=1,\ldots,N-1\big\rangle,
 \ee
 where
\be
\label{eq:TLdiag}
I=\,
\begin{pspicture}[shift=-0.55](0.0,-0.65)(2.0,0.45)
\pspolygon[fillstyle=solid,fillcolor=lightlightblue,linewidth=0pt](0,-0.35)(2.0,-0.35)(2.0,0.35)(0,0.35)
\rput(1.4,0.0){\small$...$}
\psline[linecolor=blue,linewidth=1.5pt]{-}(0.2,0.35)(0.2,-0.35)\rput(0.2,-0.55){$_1$}
\psline[linecolor=blue,linewidth=1.5pt]{-}(0.6,0.35)(0.6,-0.35)\rput(0.6,-0.55){$_2$}
\psline[linecolor=blue,linewidth=1.5pt]{-}(1.0,0.35)(1.0,-0.35)\rput(1.0,-0.55){$_3$}
\psline[linecolor=blue,linewidth=1.5pt]{-}(1.8,0.35)(1.8,-0.35)\rput(1.8,-0.55){$_N$}
\end{pspicture} 
\ \ ,\qquad
 e_j=\,
 \begin{pspicture}[shift=-0.55](0.0,-0.65)(3.2,0.45)
\pspolygon[fillstyle=solid,fillcolor=lightlightblue,linewidth=0pt](0,-0.35)(3.2,-0.35)(3.2,0.35)(0,0.35)
\rput(0.6,0.0){\small$...$}
\rput(2.6,0.0){\small$...$}
\psline[linecolor=blue,linewidth=1.5pt]{-}(0.2,0.35)(0.2,-0.35)\rput(0.2,-0.55){$_1$}
\psline[linecolor=blue,linewidth=1.5pt]{-}(1.0,0.35)(1.0,-0.35)
\psline[linecolor=blue,linewidth=1.5pt]{-}(2.2,0.35)(2.2,-0.35)
\psline[linecolor=blue,linewidth=1.5pt]{-}(3.0,0.35)(3.0,-0.35)\rput(3.0,-0.55){$_{N}$}
\psarc[linecolor=blue,linewidth=1.5pt]{-}(1.6,0.35){0.2}{180}{0}\rput(1.35,-0.55){$_j$}
\psarc[linecolor=blue,linewidth=1.5pt]{-}(1.6,-0.35){0.2}{0}{180}\rput(1.85,-0.55){$_{j+1}$}
\end{pspicture} \ \ .
\ee
The diagrams on the right sides are referred to as {\it connectivities}. To multiply two connectivities $a_1$ and $a_2$, one draws $a_2$ above $a_1$ and obtains a connectivity $a_3$ by ``following" the loop segments connecting the top and bottom $N$ nodes. The product is $a_3$ times a multiplicative factor of $\beta$ for each closed loop appearing in the resulting diagram. For instance, for $N=4$, we have 
\be
\begin{pspicture}[shift=-0.6](0.0,-0.35)(1.6,1.05)
\pspolygon[fillstyle=solid,fillcolor=lightlightblue,linewidth=0pt](0,-0.35)(1.6,-0.35)(1.6,0.35)(0,0.35)
\psline[linecolor=blue,linewidth=1.5pt]{-}(0.2,0.35)(0.2,-0.35)
\psline[linecolor=blue,linewidth=1.5pt]{-}(1.4,0.35)(1.4,-0.35)
\psarc[linecolor=blue,linewidth=1.5pt]{-}(0.8,0.35){0.2}{180}{0}
\psarc[linecolor=blue,linewidth=1.5pt]{-}(0.8,-0.35){0.2}{0}{180}
\rput(0,0.7){
\pspolygon[fillstyle=solid,fillcolor=lightlightblue,linewidth=0pt](0,-0.35)(1.6,-0.35)(1.6,0.35)(0,0.35)
\psline[linecolor=blue,linewidth=1.5pt]{-}(1.0,0.35)(1.0,-0.35)
\psline[linecolor=blue,linewidth=1.5pt]{-}(1.4,0.35)(1.4,-0.35)
\psarc[linecolor=blue,linewidth=1.5pt]{-}(0.4,0.35){0.2}{180}{0}
\psarc[linecolor=blue,linewidth=1.5pt]{-}(0.4,-0.35){0.2}{0}{180}
}
\end{pspicture}\ \ = \ \
\begin{pspicture}[shift=-0.25](0.0,-0.35)(1.6,0.35)
\pspolygon[fillstyle=solid,fillcolor=lightlightblue,linewidth=0pt](0,-0.35)(1.6,-0.35)(1.6,0.35)(0,0.35)
\psarc[linecolor=blue,linewidth=1.5pt]{-}(0.4,0.35){0.2}{180}{0}
\psarc[linecolor=blue,linewidth=1.5pt]{-}(0.8,-0.35){0.2}{0}{180}
\psbezier[linecolor=blue,linewidth=1.5pt]{-}(0.2,-0.35)(0.2,0)(1.0,0)(1.0,0.35)
\psline[linecolor=blue,linewidth=1.5pt]{-}(1.4,-0.35)(1.4,0.35)
\end{pspicture}\ \ ,
\qquad 
\begin{pspicture}[shift=-0.6](0.0,-0.35)(1.6,1.05)
\pspolygon[fillstyle=solid,fillcolor=lightlightblue,linewidth=0pt](0,-0.35)(1.6,-0.35)(1.6,0.35)(0,0.35)
\psbezier[linecolor=blue,linewidth=1.5pt]{-}(0.2,0.35)(0.2,-0.0)(1.0,0.0)(1.0,-0.35)
\psbezier[linecolor=blue,linewidth=1.5pt]{-}(0.6,0.35)(0.6,-0.0)(1.4,0.0)(1.4,-0.35)
\psarc[linecolor=blue,linewidth=1.5pt]{-}(1.2,0.35){0.2}{180}{0}
\psarc[linecolor=blue,linewidth=1.5pt]{-}(0.4,-0.35){0.2}{0}{180}
\rput(0,0.7){
\pspolygon[fillstyle=solid,fillcolor=lightlightblue,linewidth=0pt](0,-0.35)(1.6,-0.35)(1.6,0.35)(0,0.35)
\psline[linecolor=blue,linewidth=1.5pt]{-}(0.2,0.35)(0.2,-0.35)
\psbezier[linecolor=blue,linewidth=1.5pt]{-}(0.6,-0.35)(0.6,0.0)(1.4,0.0)(1.4,0.35)
\psarc[linecolor=blue,linewidth=1.5pt]{-}(1.2,-0.35){0.2}{0}{180}
\psarc[linecolor=blue,linewidth=1.5pt]{-}(0.8,0.35){0.2}{180}{0}
}
\end{pspicture}\ \ = \beta \ \
\begin{pspicture}[shift=-0.55](0.0,-0.65)(1.6,0.45)
\pspolygon[fillstyle=solid,fillcolor=lightlightblue,linewidth=0pt](0,-0.35)(1.6,-0.35)(1.6,0.35)(0,0.35)
\psbezier[linecolor=blue,linewidth=1.5pt]{-}(0.2,0.35)(0.2,-0.0)(1.0,0.0)(1.0,-0.35)
\psarc[linecolor=blue,linewidth=1.5pt]{-}(0.8,0.35){0.2}{180}{0}
\psarc[linecolor=blue,linewidth=1.5pt]{-}(0.4,-0.35){0.2}{0}{180}
\psline[linecolor=blue,linewidth=1.5pt]{-}(1.4,0.35)(1.4,-0.35)
\end{pspicture}\ \ .
\ee
As is well known, the diagrammatic rules for the products of connectivities follow from the relations imposed between the generators:
\be
e_j^2=\beta e_j, \qquad e_j e_{j\pm1} e_j = e_j, \qquad e_i e_j = e_j e_i \qquad (|i-j|>1).
\label{eq:TLdef}
\ee
We illustrate these relations with two examples for $N=4$:
\be
(e_2)^2 = \ \ \begin{pspicture}[shift=-0.6](0.0,-0.35)(1.6,1.05)
\pspolygon[fillstyle=solid,fillcolor=lightlightblue,linewidth=0pt](0,-0.35)(1.6,-0.35)(1.6,0.35)(0,0.35)
\pspolygon[fillstyle=solid,fillcolor=lightlightblue,linewidth=0pt](0,-0.35)(1.6,-0.35)(1.6,0.35)(0,0.35)
\psline[linecolor=blue,linewidth=1.5pt]{-}(0.2,0.35)(0.2,-0.35)
\psline[linecolor=blue,linewidth=1.5pt]{-}(1.4,0.35)(1.4,-0.35)
\psarc[linecolor=blue,linewidth=1.5pt]{-}(0.8,0.35){0.2}{180}{0}
\psarc[linecolor=blue,linewidth=1.5pt]{-}(0.8,-0.35){0.2}{0}{180}
\rput(0,0.7){
\pspolygon[fillstyle=solid,fillcolor=lightlightblue,linewidth=0pt](0,-0.35)(1.6,-0.35)(1.6,0.35)(0,0.35)
\pspolygon[fillstyle=solid,fillcolor=lightlightblue,linewidth=0pt](0,-0.35)(1.6,-0.35)(1.6,0.35)(0,0.35)
\psline[linecolor=blue,linewidth=1.5pt]{-}(0.2,0.35)(0.2,-0.35)
\psline[linecolor=blue,linewidth=1.5pt]{-}(1.4,0.35)(1.4,-0.35)
\psarc[linecolor=blue,linewidth=1.5pt]{-}(0.8,0.35){0.2}{180}{0}
\psarc[linecolor=blue,linewidth=1.5pt]{-}(0.8,-0.35){0.2}{0}{180}
}
\end{pspicture}\ \ = \beta \ \
\begin{pspicture}[shift=-0.55](0.0,-0.65)(1.6,0.45)
\pspolygon[fillstyle=solid,fillcolor=lightlightblue,linewidth=0pt](0,-0.35)(1.6,-0.35)(1.6,0.35)(0,0.35)
\pspolygon[fillstyle=solid,fillcolor=lightlightblue,linewidth=0pt](0,-0.35)(1.6,-0.35)(1.6,0.35)(0,0.35)
\psline[linecolor=blue,linewidth=1.5pt]{-}(0.2,0.35)(0.2,-0.35)
\psline[linecolor=blue,linewidth=1.5pt]{-}(1.4,0.35)(1.4,-0.35)
\psarc[linecolor=blue,linewidth=1.5pt]{-}(0.8,0.35){0.2}{180}{0}
\psarc[linecolor=blue,linewidth=1.5pt]{-}(0.8,-0.35){0.2}{0}{180}
\end{pspicture}\ \ ,
\qquad 
e_2 e_3 e_2 = \ \
\begin{pspicture}[shift=-0.95](0.0,-0.35)(1.6,1.75)
\pspolygon[fillstyle=solid,fillcolor=lightlightblue,linewidth=0pt](0,-0.35)(1.6,-0.35)(1.6,0.35)(0,0.35)
\psline[linecolor=blue,linewidth=1.5pt]{-}(0.2,0.35)(0.2,-0.35)
\psline[linecolor=blue,linewidth=1.5pt]{-}(1.4,0.35)(1.4,-0.35)
\psarc[linecolor=blue,linewidth=1.5pt]{-}(0.8,0.35){0.2}{180}{0}
\psarc[linecolor=blue,linewidth=1.5pt]{-}(0.8,-0.35){0.2}{0}{180}
\rput(0,0.7){
\pspolygon[fillstyle=solid,fillcolor=lightlightblue,linewidth=0pt](0,-0.35)(1.6,-0.35)(1.6,0.35)(0,0.35)
\psline[linecolor=blue,linewidth=1.5pt]{-}(0.2,0.35)(0.2,-0.35)
\psline[linecolor=blue,linewidth=1.5pt]{-}(0.6,0.35)(0.6,-0.35)
\psarc[linecolor=blue,linewidth=1.5pt]{-}(1.2,0.35){0.2}{180}{0}
\psarc[linecolor=blue,linewidth=1.5pt]{-}(1.2,-0.35){0.2}{0}{180}
}
\rput(0,1.4){
\pspolygon[fillstyle=solid,fillcolor=lightlightblue,linewidth=0pt](0,-0.35)(1.6,-0.35)(1.6,0.35)(0,0.35)
\psline[linecolor=blue,linewidth=1.5pt]{-}(0.2,0.35)(0.2,-0.35)
\psline[linecolor=blue,linewidth=1.5pt]{-}(1.4,0.35)(1.4,-0.35)
\psarc[linecolor=blue,linewidth=1.5pt]{-}(0.8,0.35){0.2}{180}{0}
\psarc[linecolor=blue,linewidth=1.5pt]{-}(0.8,-0.35){0.2}{0}{180}
}
\end{pspicture}\ \ = \ \
\begin{pspicture}[shift=-0.25](0.0,-0.35)(1.6,0.35)
\pspolygon[fillstyle=solid,fillcolor=lightlightblue,linewidth=0pt](0,-0.35)(1.6,-0.35)(1.6,0.35)(0,0.35)
\psline[linecolor=blue,linewidth=1.5pt]{-}(0.2,0.35)(0.2,-0.35)
\psline[linecolor=blue,linewidth=1.5pt]{-}(1.4,0.35)(1.4,-0.35)
\psarc[linecolor=blue,linewidth=1.5pt]{-}(0.8,0.35){0.2}{180}{0}
\psarc[linecolor=blue,linewidth=1.5pt]{-}(0.8,-0.35){0.2}{0}{180}
\end{pspicture}\ \ .
\ee

\paragraph{Enlarged periodic Temperley-Lieb algebra.}
On the cylinder, dense loop models are described by the periodic Temperley-Lieb algebra \cite{L91,MS93,GL98,G98,EG99}. Here, we work with the so-called {\it enlarged} 
periodic Temperley-Lieb algebra $\eptl_N(\alpha, \beta)$ defined in \cite{PRVcyl2010}: 
\be
\eptl_N(\alpha, \beta) = \big\langle \Omega, \Omega^{-1},\,e_j;\,j=1,\ldots,N\big\rangle.\qquad  
\ee
The identity $I$ is an element of the algebra and is obtained from the product of $\Omega$ and $\Omega^{-1}$. The generators $e_j$, for $1\le j\le N-1$, are drawn as in \eqref{eq:TLdiag}, but on a rectangle with periodic boundary conditions in the horizontal direction. The connectivities for the other generators are:
\be
e_N= \
\begin{pspicture}[shift=-0.45](0,-0.55)(2.4,0.35)
\pspolygon[fillstyle=solid,fillcolor=lightlightblue,linewidth=0pt](0,-0.35)(2.4,-0.35)(2.4,0.35)(0,0.35)
\rput(0.2,-0.55){$_1$}\rput(0.6,-0.55){$_2$}\rput(1.0,-0.55){$_3$}\rput(1.8,-0.55){\small$...$}\rput(2.2,-0.55){$_N$}
\rput(1.4,0.0){\small$...$}
\psarc[linecolor=blue,linewidth=1.5pt]{-}(0.0,0.35){0.2}{-90}{0}
\psarc[linecolor=blue,linewidth=1.5pt]{-}(0.0,-0.35){0.2}{0}{90}
\psline[linecolor=blue,linewidth=1.5pt]{-}(0.6,0.35)(0.6,-0.35)
\psline[linecolor=blue,linewidth=1.5pt]{-}(1.0,0.35)(1.0,-0.35)
\psline[linecolor=blue,linewidth=1.5pt]{-}(1.8,0.35)(1.8,-0.35)
\psarc[linecolor=blue,linewidth=1.5pt]{-}(2.4,-0.35){0.2}{90}{180}
\psarc[linecolor=blue,linewidth=1.5pt]{-}(2.4,0.35){0.2}{180}{-90}
\psframe[fillstyle=solid,linecolor=white,linewidth=0pt](-0.1,-0.4)(0,0.4)
\psframe[fillstyle=solid,linecolor=white,linewidth=0pt](2.4,-0.4)(2.5,0.4)
\end{pspicture} \ ,
\qquad
\begin{pspicture}[shift=-0.45](-0.7,-0.55)(2.0,0.35)
\rput(0.2,-0.55){$_1$}\rput(0.6,-0.55){$_2$}\rput(1.0,-0.55){$_3$}\rput(1.4,-0.55){\small$...$}\rput(1.8,-0.55){$_N$}
\pspolygon[fillstyle=solid,fillcolor=lightlightblue,linewidth=0pt](0,-0.35)(2.0,-0.35)(2.0,0.35)(0,0.35)
\multiput(0,0)(0.4,0){6}{\psbezier[linecolor=blue,linewidth=1.5pt]{-}(-0.2,-0.35)(-0.2,-0.0)(0.2,0.0)(0.2,0.35)}
\psframe[fillstyle=solid,linecolor=white,linewidth=0pt](-0.3,-0.4)(0,0.4)
\psframe[fillstyle=solid,linecolor=white,linewidth=0pt](2.0,-0.4)(2.4,0.4)
\rput(-0.55,0.042){$\Omega=$}
\end{pspicture} \ ,
\qquad
\begin{pspicture}[shift=-0.45](-1.1,-0.55)(2.0,0.35)
\rput(0.2,-0.55){$_1$}\rput(0.6,-0.55){$_2$}\rput(1.0,-0.55){$_3$}\rput(1.4,-0.55){\small$...$}\rput(1.8,-0.55){$_N$}
\pspolygon[fillstyle=solid,fillcolor=lightlightblue,linewidth=0pt](0,-0.35)(2.0,-0.35)(2.0,0.35)(0,0.35)
\multiput(0,0)(0.4,0){6}{\psbezier[linecolor=blue,linewidth=1.5pt]{-}(-0.2,0.35)(-0.2,-0.0)(0.2,0.0)(0.2,-0.35)}
\psframe[fillstyle=solid,linecolor=white,linewidth=0pt](-0.3,-0.4)(0,0.4)
\psframe[fillstyle=solid,linecolor=white,linewidth=0pt](2.0,-0.4)(2.4,0.4)
\rput(-0.75,0.07){$\Omega^{-1}=$}
\end{pspicture} \ .
\ee
The defining relations of $\eptl_N(\alpha, \beta)$ include \eqref{eq:TLdef}, where the indices in the set $\{1, \dots, N\}$ are taken modulo $N$. They also include the relations
\be
\Omega \Omega^{-1} = \Omega^{-1} \Omega = I, \qquad
\Omega e_i \Omega^{-1} = e_{i-1}, \qquad
\Omega^{N} e_N = e_N \Omega^{N}, \qquad
e_{N-1}\cdots e_2 e_1 = \Omega^2 e_1.
\ee
The ``generators" $e_j$ are not actually independent and are related by translations. For $N$ even, there are two extra relations:
\be
E \Omega^{\pm 1} E \,=\, \alpha E,\qquad E=\, e_2e_4\ldots e_{N-2}e_N.
\ee
In the diagrammatic setting, these relations replace each non-contractible loop with a weight or fugacity~$\alpha$. For $N=4$, this is depicted as
\be 
\begin{pspicture}[shift=-0.95](0.0,-0.35)(1.6,1.75)
\pspolygon[fillstyle=solid,fillcolor=lightlightblue,linewidth=0pt](0,-0.35)(1.6,-0.35)(1.6,0.35)(0,0.35)
\psarc[linecolor=blue,linewidth=1.5pt]{-}(0.4,0.35){0.2}{180}{0}
\psarc[linecolor=blue,linewidth=1.5pt]{-}(0.4,-0.35){0.2}{0}{180}
\psarc[linecolor=blue,linewidth=1.5pt]{-}(1.2,0.35){0.2}{180}{0}
\psarc[linecolor=blue,linewidth=1.5pt]{-}(1.2,-0.35){0.2}{0}{180}
\rput(0,0.7){
\pspolygon[fillstyle=solid,fillcolor=lightlightblue,linewidth=0pt](0,-0.35)(1.6,-0.35)(1.6,0.35)(0,0.35)
\multiput(0,0)(0.4,0){5}{\psbezier[linecolor=blue,linewidth=1.5pt]{-}(-0.2,-0.35)(-0.2,-0.0)(0.2,0.0)(0.2,0.35)}
\psframe[fillstyle=solid,linecolor=white,linewidth=0pt](-0.3,-0.35)(-0.003,0.35)
\psframe[fillstyle=solid,linecolor=white,linewidth=0pt](1.603,-0.35)(2.0,0.35)
}
\rput(0,1.4){
\pspolygon[fillstyle=solid,fillcolor=lightlightblue,linewidth=0pt](0,-0.35)(1.6,-0.35)(1.6,0.35)(0,0.35)
\psarc[linecolor=blue,linewidth=1.5pt]{-}(0.4,0.35){0.2}{180}{0}
\psarc[linecolor=blue,linewidth=1.5pt]{-}(0.4,-0.35){0.2}{0}{180}
\psarc[linecolor=blue,linewidth=1.5pt]{-}(1.2,0.35){0.2}{180}{0}
\psarc[linecolor=blue,linewidth=1.5pt]{-}(1.2,-0.35){0.2}{0}{180}
}
\end{pspicture}\ \ = \alpha \ \
\begin{pspicture}[shift=-0.25](0.0,-0.35)(1.6,0.35)
\pspolygon[fillstyle=solid,fillcolor=lightlightblue,linewidth=0pt](0,-0.35)(1.6,-0.35)(1.6,0.35)(0,0.35)
\psarc[linecolor=blue,linewidth=1.5pt]{-}(0.4,0.35){0.2}{180}{0}
\psarc[linecolor=blue,linewidth=1.5pt]{-}(0.4,-0.35){0.2}{0}{180}
\psarc[linecolor=blue,linewidth=1.5pt]{-}(1.2,0.35){0.2}{180}{0}
\psarc[linecolor=blue,linewidth=1.5pt]{-}(1.2,-0.35){0.2}{0}{180}
\end{pspicture}\ \ = \ \ 
\begin{pspicture}[shift=-0.95](0.0,-0.35)(1.6,1.75)
\pspolygon[fillstyle=solid,fillcolor=lightlightblue,linewidth=0pt](0,-0.35)(1.6,-0.35)(1.6,0.35)(0,0.35)
\psarc[linecolor=blue,linewidth=1.5pt]{-}(0.4,0.35){0.2}{180}{0}
\psarc[linecolor=blue,linewidth=1.5pt]{-}(0.4,-0.35){0.2}{0}{180}
\psarc[linecolor=blue,linewidth=1.5pt]{-}(1.2,0.35){0.2}{180}{0}
\psarc[linecolor=blue,linewidth=1.5pt]{-}(1.2,-0.35){0.2}{0}{180}
\rput(0,0.7){
\pspolygon[fillstyle=solid,fillcolor=lightlightblue,linewidth=0pt](0,-0.35)(1.6,-0.35)(1.6,0.35)(0,0.35)
\multiput(0,0)(0.4,0){5}{\psbezier[linecolor=blue,linewidth=1.5pt]{-}(-0.2,0.35)(-0.2,0.0)(0.2,0.0)(0.2,-0.35)}
\psframe[fillstyle=solid,linecolor=white,linewidth=0pt](-0.3,-0.35)(-0.003,0.35)
\psframe[fillstyle=solid,linecolor=white,linewidth=0pt](1.603,-0.35)(2.0,0.35)
}
\rput(0,1.4){
\pspolygon[fillstyle=solid,fillcolor=lightlightblue,linewidth=0pt](0,-0.35)(1.6,-0.35)(1.6,0.35)(0,0.35)
\psarc[linecolor=blue,linewidth=1.5pt]{-}(0.4,0.35){0.2}{180}{0}
\psarc[linecolor=blue,linewidth=1.5pt]{-}(0.4,-0.35){0.2}{0}{180}
\psarc[linecolor=blue,linewidth=1.5pt]{-}(1.2,0.35){0.2}{180}{0}
\psarc[linecolor=blue,linewidth=1.5pt]{-}(1.2,-0.35){0.2}{0}{180}
}
\end{pspicture}\ \ .
\ee

The standard modules for $\eptl_N(\alpha,\beta)$ are denoted by $\mathsf W_{N,d}$. These are constructed from link states on $N$ nodes with $d$ defects. As an example, here are the link states for $\mathsf W_{4,2}$:
\begin{equation}
\psset{unit=0.9}
\begin{pspicture}[shift=-0.08](0.0,0)(1.6,0.5)
\psline[linewidth=0.5pt](0,0)(1.6,0)
\psarc[linecolor=blue,linewidth=1.5pt]{-}(0.4,0){0.2}{0}{180}
\psline[linecolor=blue,linewidth=1.5pt]{-}(1.0,0)(1.0,0.5)
\psline[linecolor=blue,linewidth=1.5pt]{-}(1.4,0)(1.4,0.5)
\end{pspicture} \ \ \ 
\begin{pspicture}[shift=-0.08](0.0,0)(1.6,0.5)
\psline[linewidth=0.5pt](0,0)(1.6,0)
\psline[linecolor=blue,linewidth=1.5pt]{-}(0.2,0)(0.2,0.5)
\psline[linecolor=blue,linewidth=1.5pt]{-}(1.4,0)(1.4,0.5)
\psarc[linecolor=blue,linewidth=1.5pt]{-}(0.8,0){0.2}{0}{180}
\end{pspicture} \ \ \ 
\begin{pspicture}[shift=-0.08](0.0,0)(1.6,0.5)
\psline[linewidth=0.5pt](0,0)(1.6,0)
\psline[linecolor=blue,linewidth=1.5pt]{-}(0.2,0)(0.2,0.5)
\psline[linecolor=blue,linewidth=1.5pt]{-}(0.6,0)(0.6,0.5)
\psarc[linecolor=blue,linewidth=1.5pt]{-}(1.2,0){0.2}{0}{180}
\end{pspicture} \ \ \
\begin{pspicture}[shift=-0.08](0.0,0)(1.6,0.5)
\psline[linewidth=0.5pt](0,0)(1.6,0)
\psarc[linecolor=blue,linewidth=1.5pt]{-}(0,0){0.2}{0}{90}
\psline[linecolor=blue,linewidth=1.5pt]{-}(0.6,0)(0.6,0.5)
\psline[linecolor=blue,linewidth=1.5pt]{-}(1.0,0)(1.0,0.5)
\psarc[linecolor=blue,linewidth=1.5pt]{-}(1.6,0){0.2}{90}{180}
\end{pspicture} \ \ .
\end{equation}
For $d>0$, the standard modules depend on a twist parameter $\omega$. For $d=0$, the standard module depends on the weight $\alpha$ of the non contractible loops, which for convenience we express as $\alpha = \omega + \omega^{-1}$. In stating results of \cref{sec:periodic} about standard modules, we follow the convention used in \cite{MDKP2017} for their definition.

\paragraph{Two-boundary Temperley-Lieb algebra.}
If loop segments are permitted to be attached to the left and right boundaries, the natural algebraic structure is the two-boundary Temperley-Lieb algebra \cite{NRG2005,dGN09,GMP17}. We denote this algebra by $\tl^{\textrm{\tiny$(2)$}}_N(\beta, \alpha_1, \alpha_2, \beta_1, \beta_2, \gamma)$. It is defined as 
\be
\tl^{\textrm{\tiny$(2)$}}_N(\beta, \alpha_1, \alpha_2, \beta_1, \beta_2, \gamma) = \big\langle I, e_j;\,j=0,\ldots,N\big\rangle.\qquad  
\ee
The identity and the generators $e_j$ for $j = 1, \dots, N-1$ are identified to connectivities as in \eqref{eq:TLdiag} whereas, for the other two generators, we have
\be
e_0= \
\begin{pspicture}[shift=-0.45](0,-0.55)(2.4,0.35)
\pspolygon[fillstyle=solid,fillcolor=lightlightblue,linewidth=0pt](0,-0.35)(2.4,-0.35)(2.4,0.35)(0,0.35)
\rput(0.2,-0.55){$_1$}\rput(0.6,-0.55){$_2$}\rput(1.0,-0.55){$_3$}\rput(1.8,-0.55){\small$...$}\rput(2.2,-0.55){$_N$}
\rput(1.4,0.0){\small$...$}
\psarc[linecolor=blue,linewidth=1.5pt]{-}(0.0,0.35){0.2}{-90}{0}
\psarc[linecolor=blue,linewidth=1.5pt]{-}(0.0,-0.35){0.2}{0}{90}
\psline[linecolor=blue,linewidth=1.5pt]{-}(0.6,0.35)(0.6,-0.35)
\psline[linecolor=blue,linewidth=1.5pt]{-}(1.0,0.35)(1.0,-0.35)
\psline[linecolor=blue,linewidth=1.5pt]{-}(1.8,0.35)(1.8,-0.35)
\psline[linecolor=blue,linewidth=1.5pt]{-}(2.2,0.35)(2.2,-0.35)
\psframe[fillstyle=solid,linecolor=white,linewidth=0pt](-0.1,-0.4)(0,0.4)
\psframe[fillstyle=solid,linecolor=white,linewidth=0pt](2.4,-0.4)(2.5,0.4)
\end{pspicture}\ \ , \qquad \ \
e_N= \
\begin{pspicture}[shift=-0.45](0,-0.55)(2.4,0.35)
\pspolygon[fillstyle=solid,fillcolor=lightlightblue,linewidth=0pt](0,-0.35)(2.4,-0.35)(2.4,0.35)(0,0.35)
\rput(0.2,-0.55){$_1$}\rput(0.6,-0.55){$_2$}\rput(1.0,-0.55){$_3$}\rput(1.8,-0.55){\small$...$}\rput(2.2,-0.55){$_N$}
\rput(1.4,0.0){\small$...$}
\psline[linecolor=blue,linewidth=1.5pt]{-}(0.2,0.35)(0.2,-0.35)
\psline[linecolor=blue,linewidth=1.5pt]{-}(0.6,0.35)(0.6,-0.35)
\psline[linecolor=blue,linewidth=1.5pt]{-}(1.0,0.35)(1.0,-0.35)
\psline[linecolor=blue,linewidth=1.5pt]{-}(1.8,0.35)(1.8,-0.35)
\psarc[linecolor=blue,linewidth=1.5pt]{-}(2.4,-0.35){0.2}{90}{180}
\psarc[linecolor=blue,linewidth=1.5pt]{-}(2.4,0.35){0.2}{180}{-90}
\psframe[fillstyle=solid,linecolor=white,linewidth=0pt](-0.1,-0.4)(0,0.4)
\psframe[fillstyle=solid,linecolor=white,linewidth=0pt](2.4,-0.4)(2.5,0.4)
\end{pspicture}\ \ .
\ee
The relations defining this algebra include those given in \eqref{eq:TLdef}, with $i,j, j+1$ and $j-1$ in the range $1, \dots, N-1$. There are extra relations involving the extra generator for the left and right boundaries:
\begin{subequations}
\begin{alignat}{5}
e_1 e_0 e_1 &= \alpha_1 e_1, \qquad  &&e_0^2 &&= \alpha_2 e_N, \qquad &&[e_0, e_j] = 0 \qquad &&(j \ge 2),\\[0.10cm]
e_{N-1} e_N e_{N-1} &= \beta_1 e_{N-1}, \qquad &&e_N^2 &&= \beta_2 e_N, \qquad &&[e_j, e_N] = 0 \qquad &&(j \le N-1).\label{eq:bdy.relations}
\end{alignat}
\end{subequations}
For the right boundary, the first two relations in \eqref{eq:bdy.relations} are depicted as follows for $N=4$:
\be 
\begin{pspicture}[shift=-0.95](0.0,-0.35)(1.6,1.75)
\pspolygon[fillstyle=solid,fillcolor=lightlightblue,linewidth=0pt](0,-0.35)(1.6,-0.35)(1.6,0.35)(0,0.35)
\psline[linecolor=blue,linewidth=1.5pt]{-}(0.2,-0.35)(0.2,0.35)
\psline[linecolor=blue,linewidth=1.5pt]{-}(0.6,-0.35)(0.6,0.35)
\psarc[linecolor=blue,linewidth=1.5pt]{-}(1.2,0.35){0.2}{180}{0}
\psarc[linecolor=blue,linewidth=1.5pt]{-}(1.2,-0.35){0.2}{0}{180}
\rput(0,0.7){
\pspolygon[fillstyle=solid,fillcolor=lightlightblue,linewidth=0pt](0,-0.35)(1.6,-0.35)(1.6,0.35)(0,0.35)
\psline[linecolor=blue,linewidth=1.5pt]{-}(0.2,-0.35)(0.2,0.35)
\psline[linecolor=blue,linewidth=1.5pt]{-}(0.6,-0.35)(0.6,0.35)
\psline[linecolor=blue,linewidth=1.5pt]{-}(1.0,-0.35)(1.0,0.35)
\psarc[linecolor=blue,linewidth=1.5pt]{-}(1.6,-0.35){0.2}{90}{180}
\psarc[linecolor=blue,linewidth=1.5pt]{-}(1.6,0.35){0.2}{180}{-90}
}
\rput(0,1.4){
\pspolygon[fillstyle=solid,fillcolor=lightlightblue,linewidth=0pt](0,-0.35)(1.6,-0.35)(1.6,0.35)(0,0.35)
\psline[linecolor=blue,linewidth=1.5pt]{-}(0.2,-0.35)(0.2,0.35)
\psline[linecolor=blue,linewidth=1.5pt]{-}(0.6,-0.35)(0.6,0.35)
\psarc[linecolor=blue,linewidth=1.5pt]{-}(1.2,0.35){0.2}{180}{0}
\psarc[linecolor=blue,linewidth=1.5pt]{-}(1.2,-0.35){0.2}{0}{180}
}
\end{pspicture}\ \ = \beta_1 \ \
\begin{pspicture}[shift=-0.25](0.0,-0.35)(1.6,0.35)
\pspolygon[fillstyle=solid,fillcolor=lightlightblue,linewidth=0pt](0,-0.35)(1.6,-0.35)(1.6,0.35)(0,0.35)
\psline[linecolor=blue,linewidth=1.5pt]{-}(0.2,-0.35)(0.2,0.35)
\psline[linecolor=blue,linewidth=1.5pt]{-}(0.6,-0.35)(0.6,0.35)
\psarc[linecolor=blue,linewidth=1.5pt]{-}(1.2,0.35){0.2}{180}{0}
\psarc[linecolor=blue,linewidth=1.5pt]{-}(1.2,-0.35){0.2}{0}{180}
\end{pspicture}\ \ , \qquad
\begin{pspicture}[shift=-0.6](0.0,-0.35)(1.6,1.05)
\pspolygon[fillstyle=solid,fillcolor=lightlightblue,linewidth=0pt](0,-0.35)(1.6,-0.35)(1.6,0.35)(0,0.35)
\psline[linecolor=blue,linewidth=1.5pt]{-}(0.2,-0.35)(0.2,0.35)
\psline[linecolor=blue,linewidth=1.5pt]{-}(0.6,-0.35)(0.6,0.35)
\psline[linecolor=blue,linewidth=1.5pt]{-}(1.0,-0.35)(1.0,0.35)
\psarc[linecolor=blue,linewidth=1.5pt]{-}(1.6,-0.35){0.2}{90}{180}
\psarc[linecolor=blue,linewidth=1.5pt]{-}(1.6,0.35){0.2}{180}{-90}
\rput(0,0.7){
\pspolygon[fillstyle=solid,fillcolor=lightlightblue,linewidth=0pt](0,-0.35)(1.6,-0.35)(1.6,0.35)(0,0.35)
\psline[linecolor=blue,linewidth=1.5pt]{-}(0.2,-0.35)(0.2,0.35)
\psline[linecolor=blue,linewidth=1.5pt]{-}(0.6,-0.35)(0.6,0.35)
\psline[linecolor=blue,linewidth=1.5pt]{-}(1.0,-0.35)(1.0,0.35)
\psarc[linecolor=blue,linewidth=1.5pt]{-}(1.6,-0.35){0.2}{90}{180}
\psarc[linecolor=blue,linewidth=1.5pt]{-}(1.6,0.35){0.2}{180}{-90}
}
\end{pspicture}\ \ = \beta_2 \ \
\begin{pspicture}[shift=-0.25](0.0,-0.35)(1.6,0.35)
\pspolygon[fillstyle=solid,fillcolor=lightlightblue,linewidth=0pt](0,-0.35)(1.6,-0.35)(1.6,0.35)(0,0.35)
\psline[linecolor=blue,linewidth=1.5pt]{-}(0.2,-0.35)(0.2,0.35)
\psline[linecolor=blue,linewidth=1.5pt]{-}(0.6,-0.35)(0.6,0.35)
\psline[linecolor=blue,linewidth=1.5pt]{-}(1.0,-0.35)(1.0,0.35)
\psarc[linecolor=blue,linewidth=1.5pt]{-}(1.6,-0.35){0.2}{90}{180}
\psarc[linecolor=blue,linewidth=1.5pt]{-}(1.6,0.35){0.2}{180}{-90}
\end{pspicture}\ \ .
\ee
Assigning the integers $1, 2, \dots$ to the loop segments attached to the boundary starting from the bottom, we see that the fugacity of a boundary loop is $\beta_1$ if the parity of the lowest loop segment is odd, and $\beta_2$ if it is even. The same applies for $\alpha_1$ and $\alpha_2$ for the left boundary. Finally, there are two more relations for loops that are tied to both boundaries:
\be\label{eq:EF}
EFE = \gamma^2 E, \qquad FEF = \gamma^2 F,
\ee
where
\be
E = \left\{\begin{array}{cl}
e_1e_3 \cdots e_{N-1} & N \textrm{ even},\\[0.15cm]
e_1e_3 \cdots e_{N} & N \textrm{ odd},
\end{array}\right.
\qquad
F = \left\{\begin{array}{cl}
e_0e_2 \cdots e_{N} & N \textrm{ even},\\[0.15cm]
e_0e_2 \cdots e_{N-1} & N \textrm{ odd}.
\end{array}\right. 
\ee
For $N = 4$ and for $N = 5$, the leftmost relation in \eqref{eq:EF} is depicted as follows:
\be 
\begin{pspicture}[shift=-0.95](0.0,-0.35)(1.6,1.75)
\pspolygon[fillstyle=solid,fillcolor=lightlightblue,linewidth=0pt](0,-0.35)(1.6,-0.35)(1.6,0.35)(0,0.35)
\psarc[linecolor=blue,linewidth=1.5pt]{-}(0.4,0.35){0.2}{180}{0}
\psarc[linecolor=blue,linewidth=1.5pt]{-}(0.4,-0.35){0.2}{0}{180}
\psarc[linecolor=blue,linewidth=1.5pt]{-}(1.2,0.35){0.2}{180}{0}
\psarc[linecolor=blue,linewidth=1.5pt]{-}(1.2,-0.35){0.2}{0}{180}
\rput(0,0.7){
\pspolygon[fillstyle=solid,fillcolor=lightlightblue,linewidth=0pt](0,-0.35)(1.6,-0.35)(1.6,0.35)(0,0.35)
\psarc[linecolor=blue,linewidth=1.5pt]{-}(0.0,0.35){0.2}{-90}{0}
\psarc[linecolor=blue,linewidth=1.5pt]{-}(0.0,-0.35){0.2}{0}{90}
\psarc[linecolor=blue,linewidth=1.5pt]{-}(0.8,0.35){0.2}{180}{0}
\psarc[linecolor=blue,linewidth=1.5pt]{-}(0.8,-0.35){0.2}{0}{180}
\psarc[linecolor=blue,linewidth=1.5pt]{-}(1.6,-0.35){0.2}{90}{180}
\psarc[linecolor=blue,linewidth=1.5pt]{-}(1.6,0.35){0.2}{180}{-90}
}
\rput(0,1.4){
\pspolygon[fillstyle=solid,fillcolor=lightlightblue,linewidth=0pt](0,-0.35)(1.6,-0.35)(1.6,0.35)(0,0.35)
\psarc[linecolor=blue,linewidth=1.5pt]{-}(0.4,0.35){0.2}{180}{0}
\psarc[linecolor=blue,linewidth=1.5pt]{-}(0.4,-0.35){0.2}{0}{180}
\psarc[linecolor=blue,linewidth=1.5pt]{-}(1.2,0.35){0.2}{180}{0}
\psarc[linecolor=blue,linewidth=1.5pt]{-}(1.2,-0.35){0.2}{0}{180}
}
\end{pspicture}\ \ = \gamma^2 \ \
\begin{pspicture}[shift=-0.25](0.0,-0.35)(1.6,0.35)
\pspolygon[fillstyle=solid,fillcolor=lightlightblue,linewidth=0pt](0,-0.35)(1.6,-0.35)(1.6,0.35)(0,0.35)
\psarc[linecolor=blue,linewidth=1.5pt]{-}(0.4,0.35){0.2}{180}{0}
\psarc[linecolor=blue,linewidth=1.5pt]{-}(0.4,-0.35){0.2}{0}{180}
\psarc[linecolor=blue,linewidth=1.5pt]{-}(1.2,0.35){0.2}{180}{0}
\psarc[linecolor=blue,linewidth=1.5pt]{-}(1.2,-0.35){0.2}{0}{180}
\end{pspicture}
\ \ ,\qquad \ \ 
\begin{pspicture}[shift=-0.95](0.0,-0.35)(2.0,1.75)
\pspolygon[fillstyle=solid,fillcolor=lightlightblue,linewidth=0pt](0,-0.35)(2.0,-0.35)(2.0,0.35)(0,0.35)
\psarc[linecolor=blue,linewidth=1.5pt]{-}(0.4,0.35){0.2}{180}{0}
\psarc[linecolor=blue,linewidth=1.5pt]{-}(0.4,-0.35){0.2}{0}{180}
\psarc[linecolor=blue,linewidth=1.5pt]{-}(1.2,0.35){0.2}{180}{0}
\psarc[linecolor=blue,linewidth=1.5pt]{-}(1.2,-0.35){0.2}{0}{180}
\psarc[linecolor=blue,linewidth=1.5pt]{-}(2.0,-0.35){0.2}{90}{180}
\psarc[linecolor=blue,linewidth=1.5pt]{-}(2.0,0.35){0.2}{180}{-90}
\rput(0,0.7){
\pspolygon[fillstyle=solid,fillcolor=lightlightblue,linewidth=0pt](0,-0.35)(2.0,-0.35)(2.0,0.35)(0,0.35)
\psarc[linecolor=blue,linewidth=1.5pt]{-}(0.0,0.35){0.2}{-90}{0}
\psarc[linecolor=blue,linewidth=1.5pt]{-}(0.0,-0.35){0.2}{0}{90}
\psarc[linecolor=blue,linewidth=1.5pt]{-}(0.8,0.35){0.2}{180}{0}
\psarc[linecolor=blue,linewidth=1.5pt]{-}(0.8,-0.35){0.2}{0}{180}
\psarc[linecolor=blue,linewidth=1.5pt]{-}(1.6,0.35){0.2}{180}{0}
\psarc[linecolor=blue,linewidth=1.5pt]{-}(1.6,-0.35){0.2}{0}{180}
}
\rput(0,1.4){
\pspolygon[fillstyle=solid,fillcolor=lightlightblue,linewidth=0pt](0,-0.35)(2.0,-0.35)(2.0,0.35)(0,0.35)
\psarc[linecolor=blue,linewidth=1.5pt]{-}(0.4,0.35){0.2}{180}{0}
\psarc[linecolor=blue,linewidth=1.5pt]{-}(0.4,-0.35){0.2}{0}{180}
\psarc[linecolor=blue,linewidth=1.5pt]{-}(1.2,0.35){0.2}{180}{0}
\psarc[linecolor=blue,linewidth=1.5pt]{-}(1.2,-0.35){0.2}{0}{180}
\psarc[linecolor=blue,linewidth=1.5pt]{-}(2.0,-0.35){0.2}{90}{180}
\psarc[linecolor=blue,linewidth=1.5pt]{-}(2.0,0.35){0.2}{180}{-90}
}
\end{pspicture}\ \ = \gamma^2 \ \
\begin{pspicture}[shift=-0.25](0.0,-0.35)(2.0,0.35)
\pspolygon[fillstyle=solid,fillcolor=lightlightblue,linewidth=0pt](0,-0.35)(2.0,-0.35)(2.0,0.35)(0,0.35)
\psarc[linecolor=blue,linewidth=1.5pt]{-}(0.4,0.35){0.2}{180}{0}
\psarc[linecolor=blue,linewidth=1.5pt]{-}(0.4,-0.35){0.2}{0}{180}
\psarc[linecolor=blue,linewidth=1.5pt]{-}(1.2,0.35){0.2}{180}{0}
\psarc[linecolor=blue,linewidth=1.5pt]{-}(1.2,-0.35){0.2}{0}{180}
\psarc[linecolor=blue,linewidth=1.5pt]{-}(2.0,-0.35){0.2}{90}{180}
\psarc[linecolor=blue,linewidth=1.5pt]{-}(2.0,0.35){0.2}{180}{-90}

\end{pspicture}\ \ .
\ee

We denote by $\mathsf {V}^{\textrm{\tiny $(2)$}}_{N,d}$ the standard modules over $\tl^{\textrm{\tiny$(2)$}}_N(\beta, \alpha_1, \alpha_2, \beta_1, \beta_2, \gamma)$. These are defined in terms of link states on $N$ nodes with $d$ defects. The loop segments in these link states connect nodes pairwise or connect a single node to one of the two boundaries in a planar fashion such that the loop segments do not cross. As an example, here are the link states for $\mathsf {V}^{\textrm{\tiny $(2)$}}_{4,2}$:
\be
\psset{unit=0.9}
\begin{pspicture}[shift=-0.08](0.0,0)(1.6,0.5)
\psline[linewidth=0.5pt](0,0)(1.6,0)
\psarc[linecolor=blue,linewidth=1.5pt]{-}(0.4,0){0.2}{0}{180}
\psline[linecolor=blue,linewidth=1.5pt]{-}(1.0,0)(1.0,0.5)
\psline[linecolor=blue,linewidth=1.5pt]{-}(1.4,0)(1.4,0.5)
\end{pspicture} \ \ \ 
\begin{pspicture}[shift=-0.08](0.0,0)(1.6,0.5)
\psline[linewidth=0.5pt](0,0)(1.6,0)
\psline[linecolor=blue,linewidth=1.5pt]{-}(0.2,0)(0.2,0.5)
\psline[linecolor=blue,linewidth=1.5pt]{-}(1.4,0)(1.4,0.5)
\psarc[linecolor=blue,linewidth=1.5pt]{-}(0.8,0){0.2}{0}{180}
\end{pspicture} \ \ \ 
\begin{pspicture}[shift=-0.08](0.0,0)(1.6,0.5)
\psline[linewidth=0.5pt](0,0)(1.6,0)
\psline[linecolor=blue,linewidth=1.5pt]{-}(0.2,0)(0.2,0.5)
\psline[linecolor=blue,linewidth=1.5pt]{-}(0.6,0)(0.6,0.5)
\psarc[linecolor=blue,linewidth=1.5pt]{-}(1.2,0){0.2}{0}{180}
\end{pspicture} \ \ \
\begin{pspicture}[shift=-0.08](0.0,0)(1.6,0.5)
\psline[linewidth=0.5pt](0,0)(1.6,0)
\psarc[linecolor=blue,linewidth=1.5pt]{-}(0,0){0.2}{0}{90}
\psline[linecolor=blue,linewidth=1.5pt]{-}(0.6,0)(0.6,0.5)
\psline[linecolor=blue,linewidth=1.5pt]{-}(1.0,0)(1.0,0.5)
\psarc[linecolor=blue,linewidth=1.5pt]{-}(1.6,0){0.2}{90}{180}
\end{pspicture} \ \ \
\begin{pspicture}[shift=-0.08](0.0,0)(1.6,0.5)
\psline[linewidth=0.5pt](0,0)(1.6,0)
\psarc[linecolor=blue,linewidth=1.5pt]{-}(0,0){0.2}{0}{90}
\psbezier[linecolor=blue,linewidth=1.5pt]{-}(0.6,0)(0.6,0.5)(0.1,0.5)(0,0.5)
\psline[linecolor=blue,linewidth=1.5pt]{-}(1.0,0)(1.0,0.5)
\psline[linecolor=blue,linewidth=1.5pt]{-}(1.4,0)(1.4,0.5)
\end{pspicture} \ \ \
\begin{pspicture}[shift=-0.08](0.0,0)(1.6,0.5)
\psline[linewidth=0.5pt](0,0)(1.6,0)
\psline[linecolor=blue,linewidth=1.5pt]{-}(0.2,0)(0.2,0.5)
\psline[linecolor=blue,linewidth=1.5pt]{-}(0.6,0)(0.6,0.5)
\psarc[linecolor=blue,linewidth=1.5pt]{-}(1.6,0){0.2}{90}{180}
\psbezier[linecolor=blue,linewidth=1.5pt]{-}(1.0,0)(1.0,0.5)(1.5,0.5)(1.6,0.5)
\end{pspicture}
\ \ .
\ee
In this case, the weight $\gamma$ of loops touching both boundaries appears only in the standard module $\mathsf {V}^{\textrm{\tiny $(2)$}}_{N,0}$.

\section{Proofs of extended $\boldsymbol T$-system bilinear identities}\label{AppA}

\subsection{Twisted boundary conditions on the cylinder}\label{AppA.periodic}

In the compact notations \eqref{eq:compact.cyl}, the fusion hierarchy equations for twisted boundary conditions on the cylinder are
\begin{subequations}
\label{eq:2fushier}
\begin{alignat}{2}
\Tb^n_k \Tb^1_{n+k} &= f_{n+k} \Tb^{n-1}_k +  f_{n+k-1} \Tb^{n+1}_k\,,\\[0.15cm]
\Tb^1_k \Tb^n_{1+k} &= f_{k-1} \Tb^{n-1}_{k+2} +  f_{k} \Tb^{n+1}_k\,,
\end{alignat}
\end{subequations}
where $n,k \in \mathbb Z$. 
The fusion index $n$ is extended to negative values by implementing the convention
\be
\label{eq:Tnegativen}
\Tb^n_k = - \Tb^{-2-n}_{n+1+k}, \qquad \Tb^{-1}_k = 0.
\ee
The periodicity and closure relations \eqref{eq:closure.cyl} are
\be
\label{eq:periodicity.cyl}
\Tb^{n}_{k+p'} = \sigma^2 \Tb^{n}_{k}, \qquad f_{k+p'} = \sigma^2 f_k,
\ee
\be
\label{eq:closure}
\Tb^{p'}_k = \Tb^{p'-2}_{k+1} + 2 \sigma \Jb\, \Tb^0_k, \qquad \sigma = \ir^{-N(p'-p)}.
\ee

\begin{Proposition}
\label{prop:TTrel}
The following extended $T$-system bilinear identities hold:
\be
\label{eq:quad.identity}
\Tb^{j+k}_{-j}\Tb^{p'-1}_1=\Tb^j_{-j} \Tb^{p'-1-k}_{k+1} + 2 \sigma \Jb\, \Tb^j_{-j} \Tb^{k-1}_{1} + \Tb^{k-1}_{1}\Tb^{p'-2-j}_{1}, \qquad j,k \in \mathbb Z.
\ee
\end{Proposition}
{\scshape Proof.} Let us respectively denote by $\Lb_{j,k}$ and $\Rb_{j,k}$ the left and right sides of \eqref{eq:quad.identity}:
\be
\Lb_{j,k} = \Tb^{j+k}_{-j}\Tb^{p'-1}_1, \qquad \Rb_{j,k} = \Tb^j_{-j} \Tb^{p'-1-k}_{k+1} + 2 \sigma \Jb\, \Tb^j_{-j} \Tb^{k-1}_{1} + \Tb^{k-1}_{1}\Tb^{p'-2-j}_{1}. 
\ee
These satisfy the same two recursive identities:
\begin{subequations}
\label{eq:recursiveT}
\begin{alignat}{2}
\Tb^1_{-1-j} \Lb_{j,k} &= f_{-2-j} \Lb_{j-1,k} + f_{-1-j} \Lb_{j+1,k}\, , \label{eq:recursiveT.a}\\[0.15cm]
\Tb^1_{-1-j} \Rb_{j,k} &= f_{-2-j} \Rb_{j-1,k} + f_{-1-j} \Rb_{j+1,k}\, , \label{eq:recursiveT.b}\\[0.15cm] 
\Tb^1_k \Lb_{j,k} &= f_k \Lb_{j,k-1} + f_{k-1} \Lb_{j,k+1}\, , \label{eq:recursiveT.c}\\[0.15cm]
\Tb^1_k \Rb_{j,k} &= f_k \Rb_{j,k-1} + f_{k-1} \Rb_{j,k+1}\, . \label{eq:recursiveT.d}
\end{alignat}
\end{subequations}
Using the fusion hierarchy, the first relation is proven as follows:
\begin{alignat}{2}
\Tb^1_{-1-j} \Lb_{j,k} &= (\Tb^1_{-1-j}\Tb^{j+k}_{-j})\Tb^{p'-1}_1 = (f_{-2-j}\Tb^{j+k-1}_{1-j}+ f_{-1-j}\Tb^{j+k+1}_{-1-j})\Tb^{p'-1}_1 
\nonumber\\[0.15cm]&
= f_{-2-j} \Lb_{j-1,k} + f_{-1-j} \Lb_{j+1,k}\,.
\end{alignat}
Similarly,
\begin{alignat}{2}
&\Tb^1_{-1-j} \Rb_{j,k} = (\Tb^1_{-1-j}\Tb^{j}_{-j})(\Tb^{p'-1-k}_{k+1} + 2 \sigma \Jb\, \Tb^{k-1}_{1}) + \sigma^2 \Tb^{k-1}_{1} (\Tb^{p'-2-j}_{1}\Tb^1_{p'-1-j})\nonumber\\[0.15cm]
&\quad = (f_{-2-j}\Tb^{j-1}_{1-j}+f_{-1-j}\Tb^{j+1}_{-1-j})(\Tb^{p'-1-k}_{k+1} + 2 \sigma \Jb\, \Tb^{k-1}_{1})+ \sigma^2\Tb^{k-1}_{1} (f_{p'-1-j}\Tb^{p'-3-j}_1+f_{p'-2-j}\Tb^{p'-1-j}_1)\nonumber\\[0.15cm]
&\quad = f_{-2-j} \Rb_{j-1,k} + f_{-1-j} \Rb_{j+1,k}\,.
\end{alignat}
The identities \eqref{eq:recursiveT.c} and \eqref{eq:recursiveT.d} are derived with the same arguments.

We note that \eqref{eq:quad.identity} holds trivially for $k=0$, and likewise for $j=-1$. For $(j,k) = (0,-1)$, it reads
\be
0= \Tb^0_0\Tb^{p'}_0 + 2 \sigma \Jb\,\Tb^0_0 \Tb^{-2}_{1} + \Tb^{-2}_{1}  \Tb^{p-2}_{1}
\ee
which holds by virtue of \eqref{eq:Tnegativen} and \eqref{eq:closure}. From the points $(0,0)$, $(0,-1)$, $(-1,0)$ and $(-1,-1)$ where $\Lb_{j,k}=\Rb_{j,k}$, the equality is proved for $j,k \in \mathbb Z$ inductively using the recursion relations \eqref{eq:recursiveT}.
\hfill $\square$

\subsection[$U_q(s\ell(2))$ invariant/Kac vacuum boundary conditions on the strip]{$\boldsymbol{U_q(s\ell(2))}$ invariant/Kac vacuum boundary conditions on the strip}\label{AppA.boundary}

In the compact notation \eqref{eq:Df}, the fusion hierarchy equations for $U_q(s\ell(2))$/Kac vacuum boundary conditions are 
\begin{subequations}
\label{eq:2fushierD}
\begin{alignat}{2}
\Db^n_k \Db^1_{n+k} &= \frac{s_{n+2k-3}\,s_{2n+2k}}{s_{n+2k-2}\,s_{2n+2k-1}} f_{n+k}\Db^{n-1}_k + \frac{s_{n+2k-1}\,s_{2n+2k-2}}{s_{n+2k-2}\,s_{2n+2k-1}} f_{n+k-1} \Db^{n+1}_k\, ,\\[0.15cm]
\Db^1_k \Db^n_{1+k} &= \frac{s_{n+2k+1}\,s_{2k-2}}{s_{n+2k}\,s_{2k-1}} f_{k-1} \Db^{n-1}_{k+2} + \frac{s_{n+2k-1}\,s_{2k}}{s_{n+2k}\,s_{2k-1}} f_{k} \Db^{n+1}_k\, .
\end{alignat}
\end{subequations}
The fusion index $n$ is extended to negative values by implementing the convention
\be
\label{eq:Dnegativen}
\Db^n_k = - \Db^{-2-n}_{n+1+k}, \qquad \Db^{-1}_k = 0. 
\ee
The periodicity and closure relations \eqref{eq:closure.D} are
\be
\label{eq:periodicity.D}
\Db^{n}_{k+p'} = \Db^{n}_{k}\,, \qquad f_{k+p'} = f_k, \qquad s_{k+p'} = \sigma s_{k},
\ee
\be
\label{eq:closureD}
\Db^{p'}_k = \Db^{p'-2}_{k+1} + 2 \sigma \Db^0_k\,, \qquad \sigma = (-1)^{p'-p}.
\ee

\begin{Proposition}
The following extended $T$-system bilinear identities hold:
\be
\label{eq:quad.identity.D}
\Db^{j+k}_{-j}\Db^{p'-1}_1=\frac{s_{k-1}s_{-2-j}}{s_{k-j-2}s_{-1}}\Big(\Db^j_{-j} \Db^{p'-1-k}_{k+1} + 2 \sigma \Db^j_{-j} \Db^{k-1}_{1} + \Db^{k-1}_{1}\Db^{p'-2-j}_{1}\Big), \qquad j,k \in \mathbb Z.
\ee
\end{Proposition}
{\scshape Proof.} The proof uses the same arguments as in \cref{prop:TTrel}.
We denote by $\Lb_{j,k}$ and $\Rb_{j,k}$ the left- and right sides of \eqref{eq:quad.identity.D}:
\be
\Lb_{j,k} = \Db^{j+k}_{-j}\Db^{p'-1}_1, \qquad \Rb_{j,k} = \frac{s_{k-1}s_{-2-j}}{s_{k-j-2}s_{-1}}\Big(\Db^j_{-j} \Db^{p'-1-k}_{k+1} + 2 \sigma \Db^j_{-j} \Db^{k-1}_{1} + \Db^{k-1}_{1}\Db^{p'-2-j}_{1}\Big). 
\ee
These satisfy the same two recursive identities:
\begin{subequations}
\label{eq:recursiveD}
\begin{alignat}{2}
\Db^1_{-1-j} \Lb_{j,k} &= \frac{s_{k-j-1}\,s_{-4-2j}}{s_{k-j-2}\,s_{-3-2j}} f_{-2-j}\Lb_{j-1,k} + \frac{s_{k-j-3}\,s_{-2-2j}}{s_{k-j-2}\,s_{-3-2j}} f_{-1-j} \Lb_{j+1,k}\, ,\label{eq:recursiveD.a}\\[0.15cm]
\Db^1_{-1-j} \Rb_{j,k} &= \frac{s_{k-j-1}\,s_{-4-2j}}{s_{k-j-2}\,s_{-3-2j}} f_{-2-j}\Rb_{j-1,k} + \frac{s_{k-j-3}\,s_{-2-2j}}{s_{k-j-2}\,s_{-3-2j}} f_{-1-j}\Rb_{j+1,k}\, ,\label{eq:recursiveD.b}\\[0.15cm] 
\Db^1_k \Lb_{j,k} &= \frac{s_{k-j-3}\,s_{2k}}{s_{k-j-2}\,s_{2k-1}} f_k \Lb_{j,k-1} + \frac{s_{k-j-1}\,s_{2k-2}}{s_{k-j-2}\,s_{2k-1}} f_{k-1} \Lb_{j,k+1}\, ,\label{eq:recursiveD.c}\\[0.15cm]
\Db^1_k \Rb_{j,k} &= \frac{s_{k-j-3}\,s_{2k}}{s_{k-j-2}\,s_{2k-1}} f_k \Rb_{j,k-1} + \frac{s_{k-j-1}\,s_{2k-2}}{s_{k-j-2}\,s_{2k-1}} f_{k-1} \Rb_{j,k+1}\,.\label{eq:recursiveD.d}
\end{alignat}
\end{subequations}
Using the fusion hierarchy, the first relation is proven as follows:
\begin{alignat}{2}
\Db^1_{-1-j} \Lb_{j,k} &= (\Db^1_{-1-j}\Db^{j+k}_{-j})\Db^{p'-1}_1 = \Big(\frac{s_{k-j-1}\,s_{-4-2j}}{s_{k-j-2}\,s_{-3-2j}}f_{-2-j}\Db^{j+k-1}_{1-j}+ \frac{s_{k-j-3}\,s_{-2-2j}}{s_{k-j-2}\,s_{-3-2j}}f_{-1-j}\Db^{j+k+1}_{-1-j}\Big)\Db^{p'-1}_1 \nonumber\\
&= \frac{s_{k-j-1}\,s_{-4-2j}}{s_{k-j-2}\,s_{-3-2j}} f_{-2-j} \Lb_{j-1,k} + \frac{s_{k-j-3}\,s_{-2-2j}}{s_{k-j-2}\,s_{-3-2j}} f_{-1-j} \Lb_{j+1,k}\,.
\end{alignat}
Similarly, 
\begin{alignat}{2}
\Db^1_{-1-j} \Rb_{j,k} &= \frac{s_{k-1}\,s_{-2-j}}{s_{k-j-2}\,s_{-1}} \Big((\Db^1_{-1-j}\Db^{j}_{-j})(\Db^{p'-1-k}_{k+1} + 2 \sigma \Db^{k-1}_{1}) + \Db^{k-1}_{1} (\Db^{p'-2-j}_{1}\Db^1_{p'-1-j})\Big)\nonumber\\[0.1cm]
& = \frac{s_{k-1}\,s_{-2-j}}{s_{k-j-2}\,s_{-1}}\Big(\frac{s_{-1-j}\,s_{-4-2j}}{s_{-2-j}\,s_{-3-2j}}f_{-2-j}\Db^{j-1}_{1-j}+\frac{s_{-3-j}\,s_{-2-2j}}{s_{-2-j}\,s_{-3-2j}}f_{-1-j}\Db^{j+1}_{-1-j}\Big)(\Db^{p'-1-k}_{k+1} + 2 \sigma \Db^{k-1}_{1}) \nonumber\\
& \hspace{1.3cm}+ \frac{s_{k-1}s_{-2-j}}{s_{k-j-2}s_{-1}}\Db^{k-1}_{1} \Big(\frac{s_{-3-j}\,s_{-2-2j}}{s_{-2-j}\,s_{-3-2j}}f_{-1-j}\Db^{p'-3-j}_1+\frac{s_{-1-j}\,s_{-4-2j}}{s_{-2-j}\,s_{-3-2j}}f_{-2-j}\Db^{p'-1-j}_1\Big)\nonumber\\[0.1cm]
& =  \frac{s_{k-j-1}\,s_{-4-2j}}{s_{k-j-2}\,s_{-3-2j}} f_{-2-j}\Rb_{j-1,k} +  \frac{s_{k-j-3}\,s_{-2-2j}}{s_{k-j-2}\,s_{-3-2j}} f_{-1-j}\Rb_{j+1,k}\,.
\end{alignat}
The identities \eqref{eq:recursiveD.c} and \eqref{eq:recursiveD.d} are derived with the same arguments.

We note that \eqref{eq:quad.identity.D} holds trivially for $k=0$, and likewise for $j=-1$. For $(j,k) = (0,-1)$, it reads
\be
0= \Db^0_0\Db^{p'}_0 + 2 \sigma \Db^0_0 \Db^{-2}_{1} + \Db^{-2}_{1}  \Db^{p-2}_{1}
\ee
which holds by virtue of \eqref{eq:Dnegativen} and \eqref{eq:closureD}. From the points $(0,0)$, $(0,-1)$, $(-1,0)$ and $(-1,-1)$ where $\Lb_{j,k}=\Rb_{j,k}$, the equality is proved for $j,k \in \mathbb Z$ inductively using the recursion relations \eqref{eq:recursiveD}. 
\hfill $\square$


\end{document}